%% file: ms.tex
\documentclass[manuscript]{aastex}

\def\arcs{\rlap{.}$^{\prime\prime}$}
\def\arcm{\rlap{.}$^{\prime}$}

\begin{document}


\title{Do Bars Drive Spiral Density Waves?}


\author{Ronald J. Buta}\affil{Department of Physics and Astronomy,
University of Alabama, Tuscaloosa, AL 35487, USA;
rbuta@bama.ua.edu}

\author{Johan H. Knapen}\affil{Instituto de Astrof\'\i sica de
Canarias, E-38200 La Laguna, Spain; jhk@iac.es}

\author{Bruce G. Elmegreen}\affil{IBM Research Division, T.J. Watson
Research Center, 1101 Kitchawan Road, Yorktown Heights, NY 10598,
USA; bge@watson.ibm.com}

\author{Heikki Salo \& Eija Laurikainen}\affil{Division of Astronomy,
Department of Physical Sciences, University of Oulu, Oulu FIN-90014,
Finland; hsalo@sun3.oulu.fi,eija@sun3.oulu.fi}

\author{Debra Meloy Elmegreen}\affil{Vassar College, Dept. of
Physics \& Astronomy, Box 745, Poughkeepsie, NY 12604, USA;
elmegreen@vassar.edu}

\author{Iv\^anio Puerari}\affil{Instituto Nacional de Astrof\'{\i}sica,
Optica y Electr\'onica, Tonantzintla, PUE 72840, Mexico;
puerari@inaoep.mx}

\author{David L. Block}\affil{Anglo American Cosmic Dust Laboratory,
School of Computational \& Applied Mathematics University of the
Witwatersrand P.O Box 60 Wits, 2050, South Africa;
David.Block@wits.ac.za}



\begin{abstract}
We present deep near-infrared $K_s$-band AAT IRIS2 observations of a
selected sample of nearby barred spiral galaxies, including some with
the strongest known bars. The sample covers a range of Hubble types
from SB0$^-$ to SBc. The goal is to determine if the torque strengths
of the spirals correlate with those of the bars, which might be
expected if the bars actually drive the spirals as has been predicted
by theoretical studies. This issue has implications for interpreting
bar and spiral fractions at high redshift. Analysis of previous samples
suggested that such a correlation exists in the near-infrared, where
effects of extinction and star formation are less important. However,
the earlier samples had only a few excessively strong bars. Our new
sample largely confirms our previous studies, but still any correlation
is relatively weak. We find two galaxies, NGC 7513 and UGC 10862, where
there is a only a weak spiral in the presence of a very strong bar. We
suggest that some spirals probably are driven by their bars at the same
pattern speed, but that this may be only when the bar is growing or if
there is abundant gas and dissipation.

\end{abstract}




\section{Introduction}

The bar phenomenon is a pervasive and complex aspect of disk galaxies.
A bar can be identified in $\sim$60\% or more of present-epoch disk
galaxies (Knapen et al. 2000; Laurikainen et al. 2004;
Menendez-Delmestre et al. 2007; Marinova \& Jogee 2007). Studies of
galaxies in the GEMS and GOODS fields suggest that this fraction has
been largely constant to at least $z$=1 (Elmegreen et al. 2004; Jogee
et al. 2004). Results from a larger sample in the COSMOS field
indicates that the bar fraction is approximately constant out to
$z$=0.84 for the most massive galaxies only, and that smaller and less
massive galaxies have a significantly declining bar fraction out to
that redshift (Sheth et al.  2008). There is also a slight correlation
between the presence of a bar and the presence of a prominent bulge
among the high redshift galaxies; this is consistent with the massive
galaxies having a constant bar fraction, since those galaxies tend to
have a bulge (Sheth et al. 2008). Another issue is the effect of
environment on bar fraction. Verley et al. (2007) showed that in a
sample of isolated galaxies, a comparable fraction is barred as in
samples not selected for isolation. Isolated barred galaxies also
were found to have a comparable distribution of bar strengths to a
non-isolation-selected sample.

An important question is how the strength of a bar impacts the features
seen in a barred galaxy. We are particularly interested in the relation
between the strength of a bar and the appearance or strength of a
spiral. Is there a correlation between bar strength and spiral arm
strength, as suggested by theoretical models? For example, Yuan \& Kuo
(1997; see also Kormendy \& Norman 1979; Elmegreen \& Elmegreen 1985)
showed that stronger bars excited sharper gaseous density waves than
weaker bars, although other parameters also affected the appearance of
the waves. The fact that some strong observed bars join to a strong
two-armed global spiral suggests that the bars and spirals are closely
connected and that a bar strength-spiral strength correlation may be
present. These global spirals are so tightly connected to the bar that
it would seem the two features have the same pattern speed. Two-armed
spirals around strong bars are rather common, representing
$\approx$70\% of typical field spirals, unlike non-barred field spirals
where only $\approx$30\% are two-armed (Elmegreen and Elmegreen 1982).
We consider this bar-spiral correlation as evidence for interaction
between the bar and the spiral, but do not know the nature of the
interaction. It could be through various resonances, for example, and
the exact resonances would determine the ratio of pattern speeds.

On the other hand, many bars are not connected to global two-armed
spirals. There are bars with flocculent blue arms around them, galaxies
with tiny bars and long irregular (swing amplified?) types of spirals
around them, multiple-armed patterns, and old bars (SB0) with no spiral
around them. It is clear that there is a wide variety in bar-disk
interactions that do not include driving. There are no complete
theoretical models which examine bar-driven density waves that consider
both gas and stars.

We suspect that bars may drive spirals only when (a) the bar is young
and growing in strength itself, or (b) there is ample gas in the
bar-spiral system. Each of these situations provides an ``arrow of
time" for the spiral to know whether to be leading or trailing
(Lynden-Bell \& Ostriker 1967). Dissipation, growth, and interactions
provide this but a steady state does not (e.g., Toomre 1969,
1981).  Elmegreen \& Elmegreen (1985) suggested that strong bars
can grow to extend all the way to corotation and organize the gas
clouds along strong outer spiral shocks. The issue of whether bars
drive spirals is fundamental to our understanding of galaxy evolution
because a close connection between bars and spirals should manifest
itself in the fractions of such features seen at high $z$.

Observationally, one way to evaluate these ideas is to use
near-infrared $K_s$-band images to infer the gravitational potential
due to the dominant stellar backbone of galaxies. With such potentials,
we can derive the relative importance of tangential forces due to bars
and spirals. Near-infrared imaging is a necessity because optical
images are confused by dust and star formation, whereas the $K_s$-band
emphasizes the mass distribution in the old disk (e.g., Block \&
Wainscoat 1991; Regan \& Elmegreen 1997; Block et al. 1999).

In a recent study, Block et al. (2004) analyzed 2.2$\mu$m images of 17
galaxies covering a range of bar strengths and Hubble types. A
Fourier-based technique was used to separate the bars from their
associated spirals and derive separate maximum relative torques (Buta,
Block, \& Knapen 2003). The bar and spiral strengths, $Q_b$
and $Q_s$, are derived from the maximum ratio of the tangential force
to the mean background radial force (Combes \& Sanders 1981), as
obtained after a Fourier decomposition of the galaxy image. To separate
the bar and spiral arm components, the empirical fact that the Fourier
amplitudes due to the bar increase with radius to a maximum in the same
way as they decline past the maximum was used (the ``symmetry
assumption").  Alternatively, some Fourier profiles can be fitted with
one or more gaussian components (Buta et al. 2005). With such
representations, the bar can be extrapolated into the spiral region,
and removed from the image. The net result is a ``bar+disk" image and a
``spiral+disk" image having the same total flux and identical $m$=0
backgrounds. These separated images can then be analyzed for maximum
relative torques due to the nonaxisymmetric features.

In Block et al. (2004), the sample was small and appeared to show a
correlation between bar strength $Q_b$ and spiral strength $Q_s$ for
galaxies having $Q_b$ $>$ 0.3. A hint of the same correlation was also
found by Buta et al. (2005) based on more than 100 galaxies in the Ohio
State University Bright Galaxy Survey (OSUBGS, Eskridge et al. 2002).
No galaxies in either study were found to have $Q_b$ $>$ 0.6 and $Q_s$
$<$ 0.2. The finding of such galaxies would seriously challenge the
idea that bars in general drive spirals.

In this paper, we complement the Block et al. (2004) study by analyzing
maximum relative torque strengths in 23 additional barred galaxies of
relatively normal morphology and high luminosity. The main selection
criterion was bar contrast, and we tried to include bars that were
likely to turn out excessively strong when the near-IR image is
converted into a gravitational potential. Because the new sample is not
statistical in nature, we cannot determine an average bar driving
strength for normal galaxies over the type range we have considered.
Instead, our goal is to determine if bars drive spirals at all, even if
only in a single case. For this we needed a highly selected sample,
such as the one we chose, to focus on the most important physical
processes that are involved.

The observations are described in section 2 while the images are
discussed in section 3. Our analysis of the images involved 2D
decompositions (section 4), image deprojection and bar-spiral
separation (section 5), estimates of bar radii (section 6), followed by
relative torque calculations (section 7). Descriptions of individual
galaxies are provided in section 8. A discussion is provided in section
9 and conclusions in section 10.

\section{Observations}

To examine whether there might be weak spirals in the presence of
extremely strong bars, we imaged the 23 galaxies in the $K$-short
($K_s$) band using the Infrared Imager and Spectrograph (IRIS2)
attached to the 3.9-m Anglo-Australian Telescope (AAT). The run took
place from 2004 June 28 to July 5. IRIS2 has a $1024\times 1024$ pixel
Rockwell HAWAII-1 HgCdTe detector mounted at the AAT's $f/8$ Cassegrain
focus, yielding a scale of 0.447 arcsec pix$^{-1}$ and a field of view
of 7.7\,arcmin square. For the larger of our sample galaxies we imaged
offset fields for subsequent sky subtraction of the galaxy images,
while for the majority of our targets which were small enough we took
series of 1\,min images of the galaxy in alternate quadrants of the
array. In either case, individual galaxy exposures were sky-subtracted
and then combined into the final images, which have total equivalent
on-source exposure times of around one hour in almost all cases and a
spatial resolution of typically 1.5\,arcsec (see Table~1 for details).
This gave exceptional depth to the images because of our interest in
measuring both bar and spiral torque strengths. The images were
calibrated using Two-Micron All-Sky Survey (2MASS) data (see below),
and depending on the size of the galaxy relative to the field of view,
reach a level of 22-24 mag arcsec$^{-2}$ in azimuthally-averaged
profiles. For the subsequent analysis, we cleaned the images of
foreground stars and subtracted any residual background, if the field
of view was large enough.

\section{Morphology}

Our sample is morphologically diverse and has a range of properties.
Tables 1 and 2 summarize some of these properties. Types are either
from the de Vaucouleurs Atlas of Galaxies (Buta, Corwin, \& Odewahn
2007, hereafter the deVA) or were estimated by RB from available
$B$-band images.
These are generally consistent with types listed in the Third Reference
Catalogue of Bright Galaxies (RC3, de Vaucouleurs et al. 1991). After
standard processing with IRAF\footnote{IRAF is distributed by the
National Optical Astronomy Observatories, which are operated by AURA,
Inc., under cooperative agreement with the National Science
Foundation.} routines, all images were placed into the same units as
the deVA, mag arcsec$^{-2}$. The $K_s$-band images were calibrated
using 2MASS photometry within a 14$^{\prime\prime}$ diameter aperture
compiled on the NASA/IPAC Extragalactic Database (NED) website. The
mean absolute blue magnitude is $<M_B^o>$ = $-$20.7$\pm$0.8 (s.d.) and
the mean absolute $K_s$-band magnitude is $<M_{K_s}>$ = $-$24.0$\pm$1.0
(s.d.). The galaxies are typical high luminosity systems.

Figures~\ref{images1}-~\ref{images17} show the de Vaucouleurs
Atlas-style images of the 23 galaxies. For seven of the galaxies, only
the $K_s$-band images are shown. For the remaining 16 galaxies, optical
images were used to derive color index maps which reveal the star
formation and dust distribution in an especially discriminating
manner. For several cases, we show the optical image and the color
index map on two scales, as well as the $K_s$-band image. The source
of each optical image is indicated in the caption to each figure.
These were calibrated using published photoelectric aperture
photometry as described in the deVA.

\section{Mean Orientation Parameters and Bulge-Disk Properties}

To facilitate deprojection of the galaxy images, mean orientation
parameters were derived from ellipse fits. The ellipticity and major
axis position angle of isophotes at different surface brightness levels
were derived by least-squares, and means at large radii were taken to
represent the tilt of the disk. The derived values are collected in
Table 1 as $<q>$, the mean disk axis ratio, and $<\phi>$, the mean
major axis position angle. The table also lists the mean position angle
of the bar, the range of radii used for $<q>$ and $<\phi>$, as well as
the full width at half maximum of the seeing profile.

Deprojection involved rotating the image by $-<\phi>$ and stretching
it along the $x$-axis by 1/$<q>$. However, before we could do this, it
was necessary to make some allowance for the shape of the bulge.
For this purpose, we used the two-dimensional
bulge/disk/bar decomposition technique of Laurikainen et al. (2004)
to allow for the likely less flattened shape of the bulge.
This uses an exponential disk defined by central surface brightness $\mu_{disk}$(0)
and radial scale length $h_r$; a spherical Sersic (1968) bulge model defined by
central surface brightness $\mu_{bulge}$(0), characteristic radius $h_b$,
and radial exponent $\beta$ = 1/$n$, where $n$ is the Sersic index; and a Ferrers bar defined
by maximum radii $a_{bar}$ and $b_{bar}$, central surface brightness $\mu_{bar}$(0),
the angle of the bar relative to the line of nodes $\phi_{bar}$, and the
bar exponent $n_{bar}$. The derived parameters are listed in Table 3, including
the fractional contributions of the bulge and bar to the total luminosity.
Figure~\ref{twodfits} shows the quality of the decompositions using azimuthally-averaged
surface brightness profiles. The actual fits were not made to such profiles, but
to individual pixels. Generally, it was necessary to fix the bar semi-major
axis radius, $a_{bar}$, to get a stable solution.

The most important parameters in these kinds of solutions are $\beta$,
$h_r$, and $B/T$.  The typical uncertainties in these parameters were
evaluated using synthetic data by Laurikainen et al. (2005), who showed
that the components can be recovered with an accuracy of nearly 5\%
when a bulge, disk, and bar are fitted simultaneously. Adding extra
components beyond these three could change the $B/T$ value further by
5\% (Laurikainen et al. 2006). Although additional components, such as
nuclear bars, could improve some of our solutions, we have preferred
the simpler three-component models since our goal with the
decompositions is mainly image deprojection.  Nevertheless, it is
interesting to examine the properties of the bulges in our sample.
Figure~\ref{pbulges} shows a graph of Sersic index $n$ versus the log
of the bulge-to-total luminosity ratio. The vertical and horizontal
dashed lines show limits considered by Kormendy and Kennicutt (2004) to
distinguish classical bulges from ``pseudobulges." Any bulge having
$B/T$ $>$ 0.5 was considered by them to be a classical bulge. However,
galaxies having $B/T$ $<$ 0.5 and $n$ $<$ 2 were considered most likely
to be pseudobulges. In our sample, 15 out of the 23 galaxies fall
within this domain. This significant fraction is consistent with the
findings of Laurikainen et al. (2007) for a much larger sample.

\section{Deprojected Images and Relative Fourier Intensity Amplitudes}

In order to derive bar strengths and other properties of the bars in
our sample, we need to deproject the images. To eliminate or minimize
bulge ``deprojection stretch," the artificial stretching of bulge
isophotes due to the less flattened shape of the bulge, we first
subtracted the bulge model from the total image as a spherical
component. Next, the disk was deprojected using IRAF routine IMLINTRAN
in flux-conserving mode, using the adopted orientation parameters in
Table 1. The bulge was then added back. In some cases, the assumption
of a spherical bulge is not a good approximation, and the process
oversubtracts bulge light along the galaxy minor axis. This leads to
``decomposition (or spherical bulge) pinch," where the inner isophotes
are pinched (rather than extended) perpendicular to the major axis.
Note that both deprojection stretch and decomposition pinch can lead to
artificial bar-like features, although these are generally weaker than
the actual bars and can be easily distinguished.

The relative Fourier intensity amplitudes of the galaxies were derived
as a function of radius. In each case, we analyzed the amplitudes in
the same manner as in Buta, Block, \& Knapen (2003) and Buta et al.
(2005). To separate the bars from the spirals, in some cases we used
the ``symmetry assumption" whereby the amplitudes are assumed to
decline past a maximum in the same manner as they rose to that
maximum.  The assumption is based on studies of galaxies where there is
little contamination of bar amplitudes by other features. In other
cases, we were able to fit one or more gaussians to the amplitudes to
allow for more complex features. The results of these analyses are
summarized in Figure~\ref{fourier}. Table 4 lists the maximum bar $m$=2
and 4 relative amplitudes, $A_{2b}$ and $A_{4b}$, based on the mappings
in Figure~\ref{fourier}. Table 5 summarizes gaussian fit parameters for
those galaxies where this representation provided a good mapping of the
bar. The separated bar and spiral images are shown in
Figure~\ref{separations}. No uncertainties are given for the
parameters in Tables 4 and 5 because the Fourier profiles are based on
averages over many pixels and their statistical uncertainty in the bar
regions is very small. Systematic uncertainties in the orientation
parameters due, for example, to inaccurate assumptions about intrinsic
disk shapes would generally lead to less than $\pm$10\% uncertainty in
the values of $A_{2b}$ and $A_{4b}$ for the average low inclination of
$\approx$35$^o$ for our sample. Uncertainties in $r_{2b}$ and $r_{4b}$
would also generally be less than $\pm$10\%.

\section{Bar Radii}

Erwin (2005) presented an analysis of bar radii and argued that bar
size must be important because it determines how much of a galaxy is
impacted by the bar itself. A variety of methods has been proposed to
measure this parameter (see, for example, Athanassoula \& Misiriotis
2007; Gadotti et al. 2007), but no single technique works consistently
for all possible bars. To estimate bar major axis radii, Erwin used
ellipse fits to isophotes which gave both lower and upper limits to bar
size. The lower limit was taken as the radius of maximum bar
ellipticity (see also Marinova \& Jogee 2007). Also, if a bar crossed a
ring or lens, he used the radius of the ring or lens as an estimate of
the bar size. Here we use another approach: the Fourier mappings of the
bars in Figure~\ref{fourier} show that bars do not simply end abruptly
but decline smoothly to zero intensity.  This suggests that bar radii
might be estimated from fractions of the maximum Fourier amplitudes. We
ask what fractions well-approximate what appears to be the bar radius
from visual inspection of our $K_s$-band images.

Table 6, column 6 summarizes our visual estimates, considered to be the
maximum extents of the features as seen on the images deprojected using
our Table 1 orientation parameters. A cursor was placed at each end of the
bar on a monitor screen and the coordinates read into a file. The
visual bar radius $r_{vis}$ is half the distance between these cursor
positions. Figure~\ref{rbar} shows comparisons between $r_{vis}$ and
four estimates based on fractions of the maximum Fourier amplitudes
$A_{2b}$ and $A_{4b}$ for the bar mappings in Figure~\ref{fourier}.
These are called ``Fourier amplitude fraction radii" $r_{FAF}$ in
Figure~\ref{rbar}. The bar mappings are well-defined for each of our
sample galaxies, but often the profile $I_2/I_0$ involves more
extrapolation than $I_4/I_0$. We found that $r(0.25A_{4b})$ provides a
very good approximation to the visual bar radius, matching over a range
of radii from 20$^{\prime\prime}$ to 100$^{\prime\prime}$. However,
$r(0.25A_{2b})$ gives radii somewhat larger than the visual bar radii.
Instead, we found that $r(0.40A_{2b})$ provided a comparably good match
to $r(0.25A_{4b})$, while $r(0.40A_{4b})$ tended to slightly
underestimate the bar radius.  We adopt $r_{bar}$ = $r(0.25A_{4b})$ as
our best estimates, listed in column 5 of Table 6.  Using the NED
Galactic standard of rest (GSR) distances, the linear bar radii in
column 7 of Table 6 were derived. The values are comparable to those
estimated by Erwin (2005) for galaxies in the type range of our
sample.

\section{Bar and Spiral Torque Strengths}

The strengths of the bars and spirals were derived by assuming that the $K_s$-band
light distribution traces the mass distribution. This seems reasonable for
bar strength, but one might question whether the near-IR is the best choice
for the spiral strength. One could argue that the $B$-band would be best for
the spiral because it is more sensitive to the cold component, which would be
more reactive to the bar forcing. Nevertheless, the old stellar component is
still the best tracer for the spiral arm strength. This would be the amplitude
of the actual mode in the stellar disk, and would measure its dynamical
significance. Note also that Elmegreen \& Elmegreen (1985) showed
that the arm-interarm contrast is essentially the same in the $B$ and 
$I$ bands in grand design spiral galaxies but is much stronger in $B$ than $I$ 
for flocculent galaxies. We expect any bar-driven spirals would be grand design.

As in Block et al. (2004), we used the Cartesian-coordinate method
described by Quillen, Frogel, \& Gonzalez (1994) to derive the
gravitational potentials. An exponential density distribution is used
for the vertical dimension with a type-dependent scale-height based on
the work of de Grijs (1998). From these potentials, the radial and
tangential forces were derived, and the bar and spiral strengths were
estimated from maps of the ratio $F_T(i,j)/F_{0R}(i,j)$, where
$F_T(i,j)$ is the tangential force and $F_{0R}(i,j)$ is the mean radial
(axisymmetric) force, all in the galaxy plane. From a quadrant
analysis, the maximum values $Q_T(r)$ = $|F_T(i,j)/F_{0R}(i,j)|_{max}$
were derived as a function of radius.  For the bar plus disk images,
the maximum of $Q_{Tb}(r)$ is $Q_b$ at radius $r_b$ while for the
spiral plus disk images, the maximum of $Q_{Ts}(r)$ is $Q_s$ at radius
$r_s$. The total maximum relative gravitational torque is given by
$Q_g$. 

Table 7 summarizes the derived maxima.  The uncertainties listed in
Table 7 are based on the analysis of Buta, Block, \& Knapen (2003) and
are estimated as percentages of the values. For $Q_s$ and $Q_b$, we
used 10\% for the orientation parameters (meaning
$\sigma(Q_b)$$\approx$0.1$Q_b$ due to this effect, etc.), 10\% for the
vertical scale height, 4\% for the bar extrapolations, and 10\% for the
spiral extrapolation, for an average sample galaxy inclined by
35$^{\circ}$. We also allowed for the scatter in the maxima in each
quadrant due to asymmetries in the spiral pattern. Since these
uncertainties are largely independent, we added them in quadrature. For
the error bars on the spiral contrasts $A_{2s}$ and $A_{4s}$, we used a
similar procedure but without any effect of vertical scale height. The
error bars are in any case only indicative. The uncertainties in the
Table 7 parameters tend to be larger than those for the Tables 4-6
parameters because $Q_b$ and $Q_s$ involve an uncertain vertical
scale-height and $Q_s$, $A_{2s}$, and $A_{4s}$ are especially sensitive
to the extrapolation of the bar. The bar in the present sample with the
largest value of $Q_b$ is found in UGC 10862, which has a very small
bulge and near-IR ansae. These conspire to make the bar strong.

Note that if the forcing due to the dark matter halo is more important
at larger radii than at smaller radii, then our $Q_s$ values are likely
to be more overestimated by ignoring the halo than would our $Q_b$
values, since $r_s \approx 2r_b$ on average. Our analysis assumes a
constant mass-to-light ratio. Buta, Laurikainen, \& Salo (2004) show
using a statistical approach how much the effect on $Q_b$ can be for
typical high luminosity galaxies. For galaxies of similar luminosities
to those in our AAT sample, the inclusion of a halo reduced $Q_b$
values by 6\% on average. Even if the effect on $Q_s$ is twice this
amount, it would still be small and have little impact on our results.

The spiral maximum at radius $r_s$ in Table 7 in some cases refers to
bright inner arms (e.g., NGC 175), while in others it refers to outer
arms (e.g., NGC 521). This depends on how the arms combine with the
declining background. In other cases, residual light of an extended
oval may contribute to what we call $Q_s$ (e.g., NGC 7155). The bar
radius $r(0.25A_{4b})$ from Table 6 correlates well with the $Q_b$
maximum radius $r_b$ in an impartial linear relation of the form
$r(0.25A_{4b}) = 1.583(\pm 0.031) r_b$, with a radius-dependent
dispersion of $\sigma$=0.082$r(0.25A_{4b})$.

As a check on these results, we also derived the potentials (and the
resulting $Q_b$,$Q_s$ values) using the polar grid method of
Laurikainen \& Salo (2002). Figure~\ref{compare} shows how well the two
methods agree for our 23 galaxies. In general, the agreement is good
but we find small average offsets: for $Q_b$, the Cartesian values are
on average larger than the polar grid values by 0.038 while for $Q_s$
they are larger by 0.033. These differences are not due to the
integration method (Cartesian versus polar grid) but are likely due to
slightly different treatments of the bulge and the vertical scale
height in the two independent sets of programs. However, because the
differences are virtually the same, they will have little or no effect
on a possible correlation between $Q_b$ and $Q_s$.

Figure~\ref{strengths} compares $Q_b$ and
the bar contrast parameters $A_{2b}$ and $A_{4b}$ with the relative radii of these
maximum parameters. The radii are normalized to $r_o(25)=D_o/2$, where $D_o$
is the extinction-corrected isophotal diameter at $\mu_B$=25.0 mag arcsec$^{-2}$
from RC3. The values of $A_{2b}$ and $A_{4b}$ (and $r_{2b}$ and $r_{4b}$) are from
Table 4, and are based on the same mappings (Figure~\ref{fourier}) 
used to derive $Q_b$.
There is little apparent correlation between $Q_b$ and $r_b/r_o(25)$,
but $A_{2b}$ and $A_{4b}$ do show some correlation. The correlation between 
$A_{2b}$ and
$r_{2b}/r(25)$ for this sample has already been discussed by Elmegreen et al.
(2007). These authors argue that this contrast correlation, in addition to a
correlation with central density, implies that bars grow in both length and
contrast over a Hubble time through angular momentum transfer to the disk
and halo.

Table 8 summarizes the mean values of several parameters from our analysis
for those galaxies where single or double gaussians well-represented the
relative Fourier profiles. Buta et al. (2006) showed that bars fitted
by double gaussians were stronger than those fitted with single gaussians.
In the present sample, the single and double gaussian bars have the same
relative torque strength on average, while the contrasts are higher for
the double gaussian features. The double gaussian features also have
higher average values of $r_{bar}/h_r$, bulge-to-total luminosity
ratio, and Sersic index, which may conspire to make the torque
strengths similar to those of the single gaussian bars.

Laurikainen et al. (2007) noted that strong bars with thin and thick
components have double-peaked profiles in all Fourier modes. But in
galaxies where only the $m$=2 profile (not the higher Fourier modes) of
the bar is double-peaked, the amplitude is most probably contaminated
by an inner or outer oval/lens.

One issue we can examine with our new dataset is the steepness with
which the bar declines near its ends. More evolved, stronger bars
should have more steeply falling amplitude profiles because their
orbits are pushed right up against the resonances. Non-circular orbits
and random stellar motions should broaden the bar's edge. We examine
this issue using $Q_{Tb}(r)$ forcing profiles, rather than the Fourier
luminosity profiles, because these most reliably trace how rapidly the
significance of the bar declines. Only the maxima $Q_b$ of the
$Q_{Tb}(r)$ profiles are compiled in Table 7, which also gives
$r_b=r(Q_b)$. To remove the effects of distance and scale, we normalize
the profiles as $S_{Tb}=Q_{Tb}(r)/Q_b$ versus $\rho=r/r_b$.
Figure~\ref{writeqb}a shows the normalized profiles for IC 1438 and IC
4290, the former having $Q_b$=0.12 and the latter having $Q_b$=0.52.
The curves show that the strong bar in IC 4290 has the more steeply
declining normalized bar torque profile. We measure the steepness of
the bar on its declining edge as the slope $S_b = dS_{Tb}/d\rho$ at the
point on the profile where $S_{Tb}$ drops to 0.5 (filled circles in
Figure~\ref{writeqb}a). Figure~\ref{writeqb}b shows that, for our 23
galaxies, $S_b$ generally declines with increasing $Q_b$, although with
considerable scatter.  On average, the bar torque profiles do decline
more steeply past the end of the bar for stronger bars than for weaker
bars, for our small sample.

We also define the relative bar-end drop-off distance as

$$f_b={(r(0.25Q_b)-r(0.75Q_b))\over r_b}$$ 

\noindent
where $r(0.75Q_b)$ is the radius where the $Q_{Tb}(r)$ profile drops to
75\% of the maximum value, $r(0.25Q_b)$ is the radius where the
$Q_{Tb}(r)$ profile drops to 25\% of its maximum value, and $r_b$ is again the
radius of the maximum from Table 7. This fraction is plotted versus
$Q_b$ in Figure~\ref{writeqb}c. The plot shows that like $S_b$, $f_b$ declines
slightly with increasing $Q_b$, which is also consistent with the stronger
bars having a steeper decline past the maximum, relative to the $Q_b$ bar
radius. 

We also examined whether $S_b$ and $f_b$ correlated with any other quantities,
such as $r_b/r_o(25)$, $r_b/h_r$, $M_B$, and $T$, but no other significant 
correlations were found.

\section{Description of Individual Galaxies}

In this section, bar and spiral strength classes are defined as in Buta \& Block
(2001) and Buta et al. (2005). Class 0 refers to values less than 0.05, class 1
to values ranging from 0.05 to 0.15, class 2 to values 0.15-0.25, etc. 

\begin{figure}
\figurenum{1}
\plotone{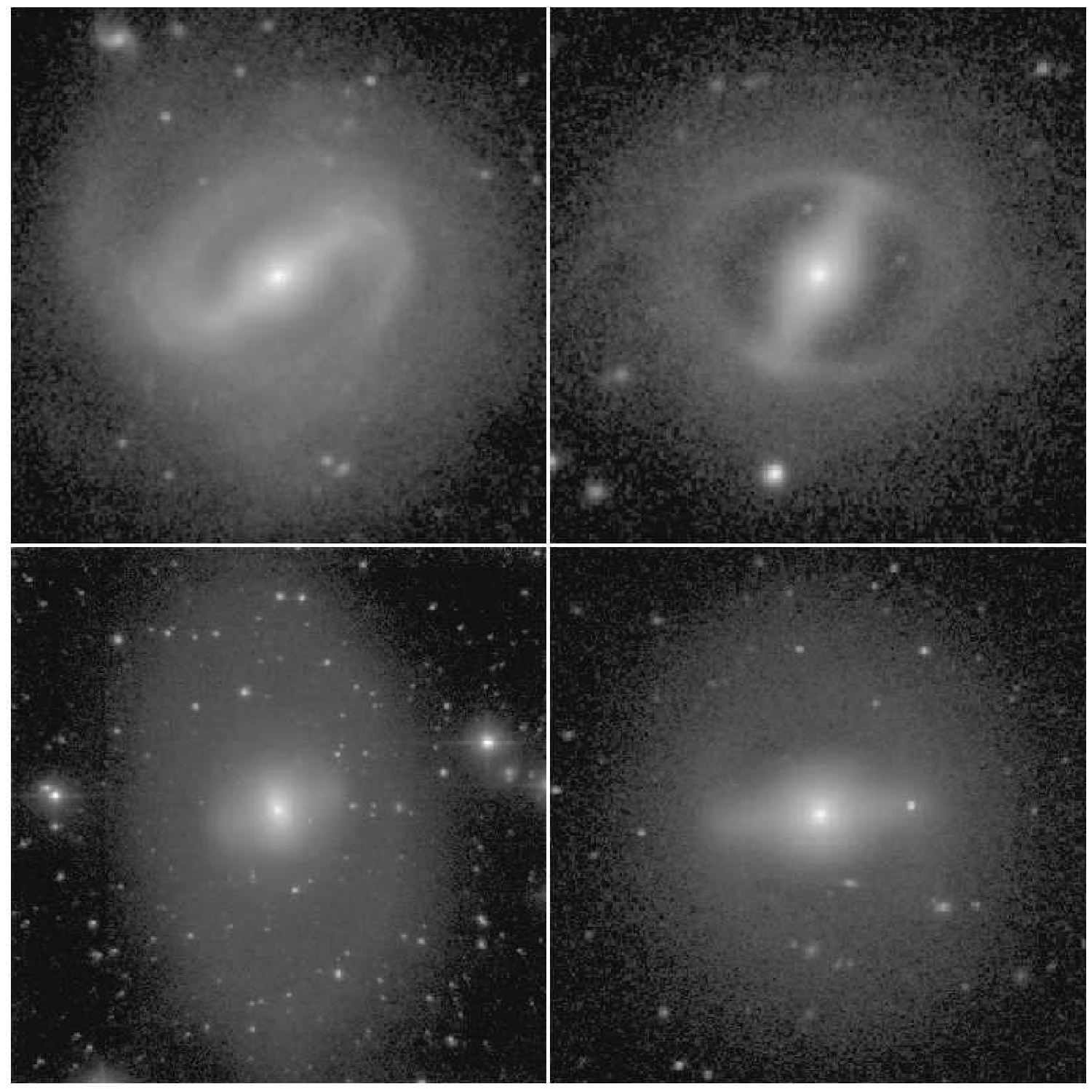}
\caption{$K_s$-band images of NGC 175 (upper left), NGC 5335 (upper right),
NGC 5365 (lower left), and NGC 7155 (lower right). The images are logarithmic
in units of mag arcsec$^{-2}$ and the square fields have side lengths
1\arcm 91 for
NGC 175 and NGC 5335, 3\arcm 73 for NGC 5365,
and 2\arcm 38 for NGC 7155. North is at the top and east
is to the left in each frame.}
\label{images1}
\end{figure}

{\it NGC 175} - The near-IR morphology of this galaxy
(Figure~\ref{images1}) is almost as structured as its blue light
morphology in the Hubble Atlas (Sandage 1961). The inner pseudoring has
a diameter of 13.6 kpc, comparable to the average for SB inner rings as
derived by de Vaucouleurs \& Buta (1980), after adjustment for distance
scale. The 2D decomposition gave an exponential bulge including 7.4\%
of the total luminosity.  Figures~\ref{fourier} and ~\ref{separations}
show that a single gaussian component well-represents the bar. Removal
of this bar representation from the image leaves an elongated inner
pseudoring that is slightly misaligned with the bar axis. This
suggests that the bar and spiral in this case have a different pattern
speed, because alignment is the normal rule for inner rings and
pseudorings (Buta 1995). The bar is strong and corresponds to bar
class 4, while the spiral class is 2.

{\it NGC 521} - Figure~\ref{images3} shows that the $B$- and $K_s$-band
morphologies are similar except that the outermost spiral features are
much weaker in $K_s$.  The bar is enveloped by a conspicuous inner
pseudoring in the $B$-band that is much weaker in $K_s$. The
deprojected $K_s$-band diameter of the ring is 14.9 kpc using the NED
GSR distance of 69.6 Mpc. The deprojected axis ratio is 0.95. In
addition to the inner pseudoring, the $B-K_s$ color index map in
Figure~\ref{images3} shows a very well-defined, red nuclear ring.  In
the $B$-band, the ring is a clear dust feature with no recent star
formation evident.  A visual mapping of the ring in the color index map
gives an axis ratio of 0.82 and a diameter of
15\rlap{.}$^{\prime\prime}$4 or 5.2 kpc, unusually large for a nuclear
ring.  These are close to the face-on values since the galaxy is only
slightly inclined. There are no prominent bar dust lanes.

The fitted bulge model is nearly exponential, and corresponds to a Sersic
index of $n$=1.26. A
single gaussian component well-represents the $K_s$-band relative Fourier intensity
amplitudes, with only a slight departure of the peak $m$=2 amplitude from the 
gaussian fit. In spite of the conspicuousness of the bar, it is a relatively weak
feature and has a bar class of 1 and a ``$Q_b$ family" (Buta et al. 2005) 
of only SAB. The spiral class is 1.

{\it NGC 613} - A complicated object with at least 5 spiral arms, all of which are
still prominent in the $K_s$-band. After deprojection,
the inner pseudoring has an axis ratio of 0.67 and a diameter of 2\arcm 46 or 6.3 kpc,
and is aligned almost exactly parallel to the bar. The $B-K_s$ color index map
in Figure~\ref{images4} shows strong leading dust lanes in the bar. The $B$-band
image shows a small spiral in the center that in the $K_s$-band is either a nuclear
ansae bar or a nuclear ring highly elongated along the main bar (B\"oker
et al. 2008). This feature is
shown in the lower right panel of Figure~\ref{images4}. The projected diameter of 
the feature is 9\rlap{.}$^{\prime\prime}$8 (0.94 kpc). Peeples \& Martini (2006)
show a structure map of this same area, illustrating fine details of the
dust distribution.

The decomposition yielded a nearly exponential bulge with a flux contribution of
14.4\%. However, this is probably an overestimate due to the nuclear bar/ring.
The relative Fourier intensities show a bar that is not easily interpreted
in terms of single or double gaussians. The double-humped profiles in Figure~\ref{fourier}
are a symmetry-assumption mapping that was needed to get the maximum extent
of the bar but which fails to account for the significant asymmetry in that
feature that is apparent in the spiral plus disk image in Figure~\ref{separations}.
After separation, we find that NGC 613 is bar class 4 and spiral class 3.

{\it NGC 986} - This galaxy is characterized by apparently strong bar and spiral
patterns. The bar and spiral are closely connected, such that the bar smoothly changes
into the spiral. The $R-K_s$ color index map (Figure~\ref{images5}) 
shows very well-defined dust lanes on
the leading edges of the bar as well as strong red features in the inner parts of
the spiral arms. The nuclear region is complicated in both $R$ and $K_s$, and appears
to include a small nuclear dust ring 9\arcs 6 (1.2 kpc) in projected diameter. The short
black line in the lower right panel of Figure~\ref{images5} points to the object we have taken
to be the nucleus of the galaxy. From the color index map, this object has blue
colors.

The 2D decomposition gave a large value of $\beta$, 1.277, and a rather large
value of bulge-to-total luminosity ratio, 13.7\%. The bulge region of NGC 986 
is not very smooth and it is likely these values are unreliable.
Bar-spiral separation is also not clean in NGC 986 because the bar blends so smoothly
with the spiral. In our final analysis, we assumed what little bulge might
be present in this galaxy to be as flat as the disk.
The bar mapping in Figure~\ref{fourier} does a reasonable job
of separation.
Conversion of the separated images to a potential gives a bar class
of 4 and a spiral class of 5. By this measure, NGC 986 has the strongest spiral 
of the sample.

Kohno et al. (2008) recently observed NGC 986 in CO(3-2), and found the galaxy's
bar to be rich in dense molecular gas. These authors have suggested that the
complex central region of the galaxy is in a growing phase, being
fueled by the significant gas in the bar.

{\it NGC 1300} - The bar and spiral pattern in this galaxy are strong and
well-defined (Figure~\ref{images6}). The $B-K_s$ color index map reveals strong leading
dust lanes in the bar. Buta et al (2007) show a $B-I$ color index map that reveals
a small blue nuclear ring. In the $K_s$-band, this feature is a smooth, almost
circular lens-like feature of projected diameter 9\arcs 4 (0.9 kpc; see Figure~\ref{images6},
lower right panel). The 2D decomposition
gave a nearly exponential bulge including 9.9\% of the total $K_s$-band luminosity.

The bar representation in Figure~\ref{fourier} is a double-gaussian
mapping that is somewhat uncertain.
Figure~\ref{separations} shows that this mapping does a 
reasonable job of separating the bar from the spiral. With this separation, the galaxy is bar class 5 and spiral class 2.

{\it NGC 1566} - The bar lies inside the inner termination points of the bright
spiral (Figure~\ref{images7}), as first shown by Hackwell \& Schweizer (1983). 
The 2D decomposition (Figure~\ref{twodfits}) gave
a bulge having Sersic index $n$=2.4 and including 12.9\% of the total $K_s$ luminosity.
The relative Fourier amplitudes in Figure~\ref{fourier} show an asymmetric $m$=2 profile
inside $r$=40$^{\prime\prime}$  that is associated with the inner bar. The symmetry assumption
could be applied to all higher order terms but $m$=2. Although within the
broad definitions of the classes this galaxy is both bar and spiral class 2,
it is the second case in our sample where the spiral is stronger than the bar.

{\it NGC 4593} - Except for the obvious leading bar dust lanes and a nuclear dust ring,
the $B$ and $K_s$-band images of this galaxy are very similar. The nuclear dust ring
seen in the $B-K_s$ color index map
has a projected diameter of 11\arcs 0 (1.9 kpc). The 2D decomposition shown in Figure~\ref{twodfits}
has a Sersic index of 3.6 and a bulge-to-total $K_s$-band luminosity ratio of 34.5\%,
providing a possible classical bulge. The relative Fourier intensity amplitudes shown
in Figure~\ref{fourier} are well-fitted by double gaussians in all terms; this representation
does a reasonable job with the bar-spiral separation (Figure~\ref{separations}). With
these images, the bar class is 3 and the spiral class is 1.

{\it NGC 5101} - The $B$ and $K_s$-band images are similar, but the exceptional image
quality on the $K_s$-band image reveals short spiral arcs around the ends of the bar
(see lower left panel of Figure~\ref{images9}). The $B-K_s$ color index map reveals
only a weak trace of leading bar dust lanes, in addition to a tightly wrapped pattern
of spiral arms that wrap around the bar and which have slightly enhanced blue colors.

The 2D decomposition gave a Sersic index of 2.5 and a bulge contribution of 28\%.
The relative Fourier amplitudes of the bar are well-represented by double gaussians
to $m$=12, while for $m>$12, a single gaussian describes these amplitudes.
In the $K_s$-band, the spiral structure in NGC 5101 is very weak. The derived
bar class is 2 and the spiral class is 0.

{\it NGC 5335} - The morphology is characterized by a strong apparent bar and
a conspicuous inner ring that is weak near its projected major axis
(Figure~\ref{images1}). A visual mapping of the ring
in the deprojected image in Figure~\ref{separations} gives an axis ratio
of 0.89 and an alignment nearly perpendicular to the bar, very unusual for
an SB inner ring (Buta 1995). The ring also has an enormous linear diameter, 16.8 kpc (54\arcs 8),
also unusual for such features (de Vaucouleurs \& Buta 1980).

The 2D decomposition gave an exponential bulge and a bulge-to-total luminosity
ratio of 18.8\%.
The bar is the dominant feature in the relative Fourier amplitudes and is well-fitted by
a double gaussian in all even terms to $m$=20 at least. The spiral structure is
weak in the $K_s$-band, and we obtain a bar class of 4 and a spiral class of 0.

{\it NGC 5365} - The $K_s$-band image in Figure~\ref{images1} shows a well-defined
early-type barred S0 with an extended disk and a trace of an outer ring. A clear
secondary bar aligned nearly perpendicular to the primary bar is seen in this image,
a feature already noted by Mulchaey et al. (1997; see also Erwin 2004). The 2D
decomposition gave a Sersic index of $n$=1.8 and a bulge-to-total luminosity ratio
of 0.48, the latter a likely overestimate since the secondary bar was not fitted
separately.

A double-gaussian was needed to represent the even Fourier terms in the primary
bar of NGC 5365. Only the secondary bar provides any additional significant
amplitude. In spite of the apparent strength of the bar, the significant bulge
leads to a bar class of only 1. 

{\it NGC 6221} - Both the bar and spiral arms in this galaxy show considerable
dust content in the color index map in Figure~\ref{images10}. The outer spiral
pattern is disturbed and the galaxy is likely interacting, possibly with
neighbor NGC 6215 (Koribalski \& Dickey 2004). The azimuthally-averaged profile in Figure~\ref{twodfits}
smooths out much of this structure. The 2D decomposition gave a Sersic index
of $n$=1.6 and a B/T ratio of 9.1\%. A symmetry-assumption mapping of the 
relative Fourier intensity profiles of the bar was adopted in Figure~\ref{fourier}. 
The spiral plus disk image in Figure~\ref{separations} highlights some asymmetry
in the bar not accounted for by this mapping.

{\it NGC 6384} - The multi-armed nature of the spiral pattern in the $B$-band is
less evident in the $K_s$-band, where two arms seem to dominate (Figure~\ref{images11}).
In the $B$-band image, a weak inner ring surrounds a relatively weak-looking bar.
In the color index map, the inner ring surprisingly appears as a red feature. 
After visually mapping the feature, 
we find that the blue light inner ring has a projected diameter of 6.6 kpc, an axis
ratio of 0.69, and a major axis position angle of 29$^{\circ}$, while the red $B-K_s$
inner ring has a diameter of 5.2 kpc, an axis ratio of 0.65, and a major axis position 
angle of 39$^{\circ}$. The dust ring is largely confined to the inner edge of the inner 
ring and is enhanced on the near side, east of the center. The $B$-band inner ring is
not prominent in the color index map.

The 2D decomposition gave a bulge with Sersic index
$n$=3.1 and including 11.8\% of the total $K_s$-band luminosity.
The 2D decomposition left some decomposition pinch in the
deprojected image of NGC 6384. After removing this area, we find that the bar of
NGC 6384 is largely a single gaussian type. 

{\it NGC 6782} - NGC 6782 is an exceptional ringed
barred spiral. Although spiral structure is clearly evident in the $B$-band
image in Figure~\ref{images12}, the appearance of the galaxy in the $K_s$-band
is as a late S0, or type (R)SB(r)0$^+$. The object is well-known as a double-barred and triple-ringed
system, and was recently interpreted as type (R$_1$R$_2^{\prime}$)SB(r)a by
Buta et al. (2007). The $B-K_s$ color index map shows the strong leading dust lanes
and nearly circular blue star-forming nuclear ring. The 
galaxy was the subject of a dynamical study by Lin et al. (2008), who interpreted
the main features in terms of orbit resonances with the primary bar.

The 2D decomposition shown in Figure~\ref{twodfits} gave a Sersic index $n$=2.1 and
a B/T of 40.8\%. The latter is likely to be an overestimate because we have not
taken into account the secondary bar. The bar of NGC 6782 is well-represented
by a double-gaussian, although this does not include all of the light of an
extended oval which fills the deprojected minor axis of the outer ring. Bar-spiral
separation gives a bar class of 2 and a spiral class of 0. The $Q_b$ family of
NGC 6782 is SAB.

{\it NGC 6907} - The deprojected $K_s$-band image in Figure~\ref{images13} 
shows what could be interpreted as a bar in the inner regions, 
but the feature blends so smoothly with the spiral that the
relative Fourier amplitudes do not clearly distinguish it and the phase
of the $m$=2 component is only approximately constant in the apparent bar region.
The bar mapping in Figure~\ref{fourier} is imprecise since there is considerable
asymmetry in the apparent bar, but nevertheless it provides a reasonable separation.
The resulting bar and spiral classes are 3, with the spiral stronger than
the bar.

The color index map in Figure~\ref{images13} shows that red colors permeate the
apparent bar region. The red arm which breaks from the west end of the bar twists
sharply eastward and is unusual.

The 2D decomposition gave a nearly exponential bulge having a B/T of 12.7\%.

{\it NGC 7155} - The $K_s$-band image in Figure~\ref{images1} shows a well-defined
SB0 galaxy with a prominent bar. The image also shows no evidence for a
secondary bar. There is a faint trace of a diffuse inner ring.

The 2D decomposition gave a Sersic index of $n$=1.4 and a B/T of 35.6\%. Figure~\ref{fourier}
shows that a single gaussian well-represents the relative Fourier amplitudes of the
bar in NGC 7155. The bar class is 2 and there is no significant spiral.

{\it NGC 7329} - The images in Figure~\ref{images14} show a well-developed intermediate
type spiral where only the inner arms are prominent in the $K_s$-band. The color index
map shows mostly red colors in the bar region, but the map is uncertain because the seeing
on the $B$-band image is much poorer than on the $K_s$-band image. The 2D decomposition gave 
a Sersic index of $n$=1.4 and a B/T of 21.1\%. The deprojected $K_s$-band image shows
some decomposition pinch in the inner parts of the bar. This is smoothed over in the
bar plus disk image, but appears as a vertically-oriented oval in the spiral plus disk
image. We found that a single gaussian well-represents the relative Fourier profiles
of the bar, although the $m$=2 term is complicated by both the significant spiral
structure and the decomposition pinch. The separated images gave a bar class of 3
and a spiral class of 1.

\begin{figure}
\figurenum{2}
\plotone{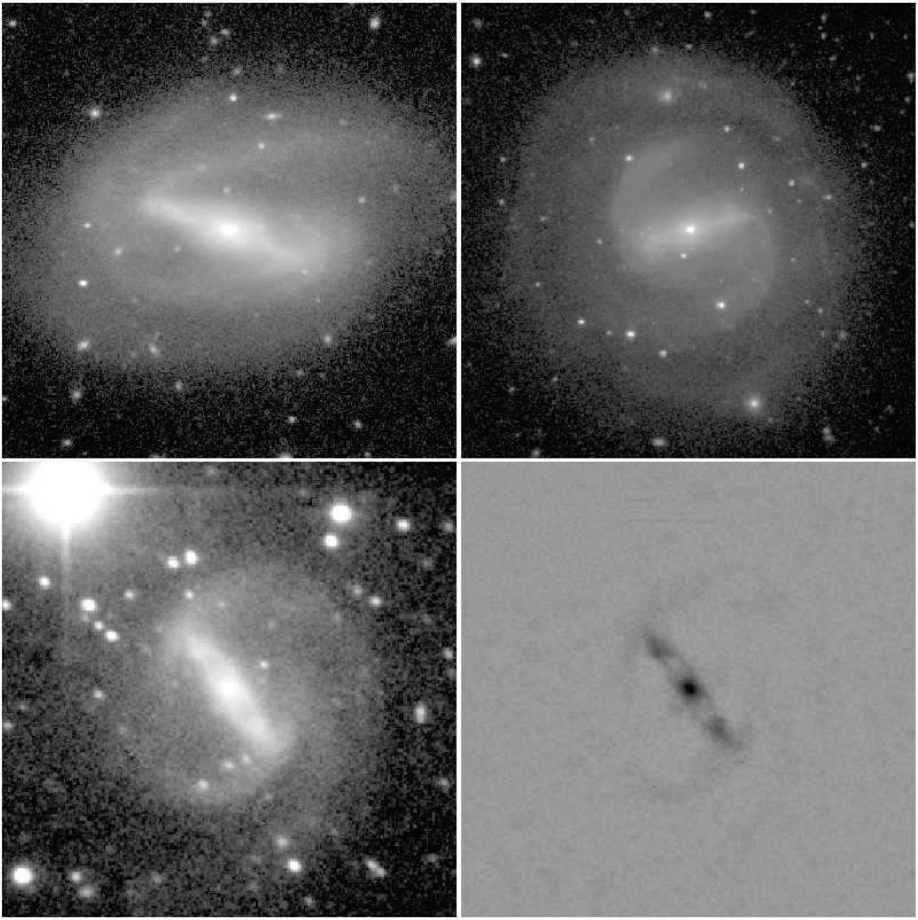}
\caption{ $K_s$-band images of NGC 7513 (upper left), IC 5092 (upper right),
and UGC 10862 (lower left). The images are logarithmic
in units of mag arcsec$^{-2}$ and the square fields have side lengths
of 2\arcm 98 for
NGC 7513 and IC 5092, 
and 1\arcm 91 for UGC 10862. The lower right image is also of UGC 10862,
but is an unsharp-masked image after subtracting a 31x31 pixel median
smoothed version of the cleaned galaxy image.
North is at the top and east 
is to the left in each frame.}
\label{images2}
\end{figure}

\begin{figure}
\figurenum{3}
\plotone{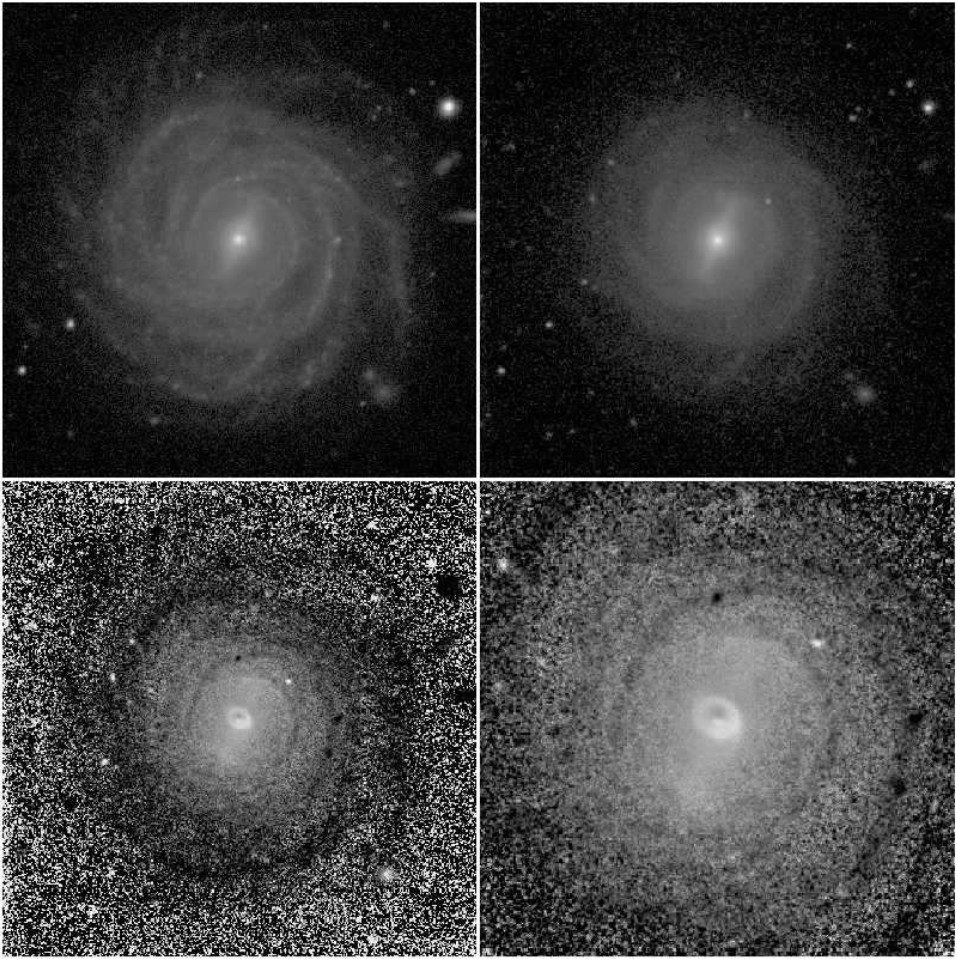}
\caption{Images of NGC 521. (upper left): $B$-band (Galaz et al. 2006); (upper right): $K_s$-band;
(lower left): $B-K_s$ color index map; (lower right) same as at lower left, at twice
the scale.  The images are logarithmic
in units of mag arcsec$^{-2}$ and the main fields have side lengths of
5\arcm 81. The color index map for this galaxy and all the others is coded
such that blue features are dark and red features are light.
North is at the top and east is to the left.}
\label{images3}
\end{figure}

\begin{figure}
\figurenum{4}
\plotone{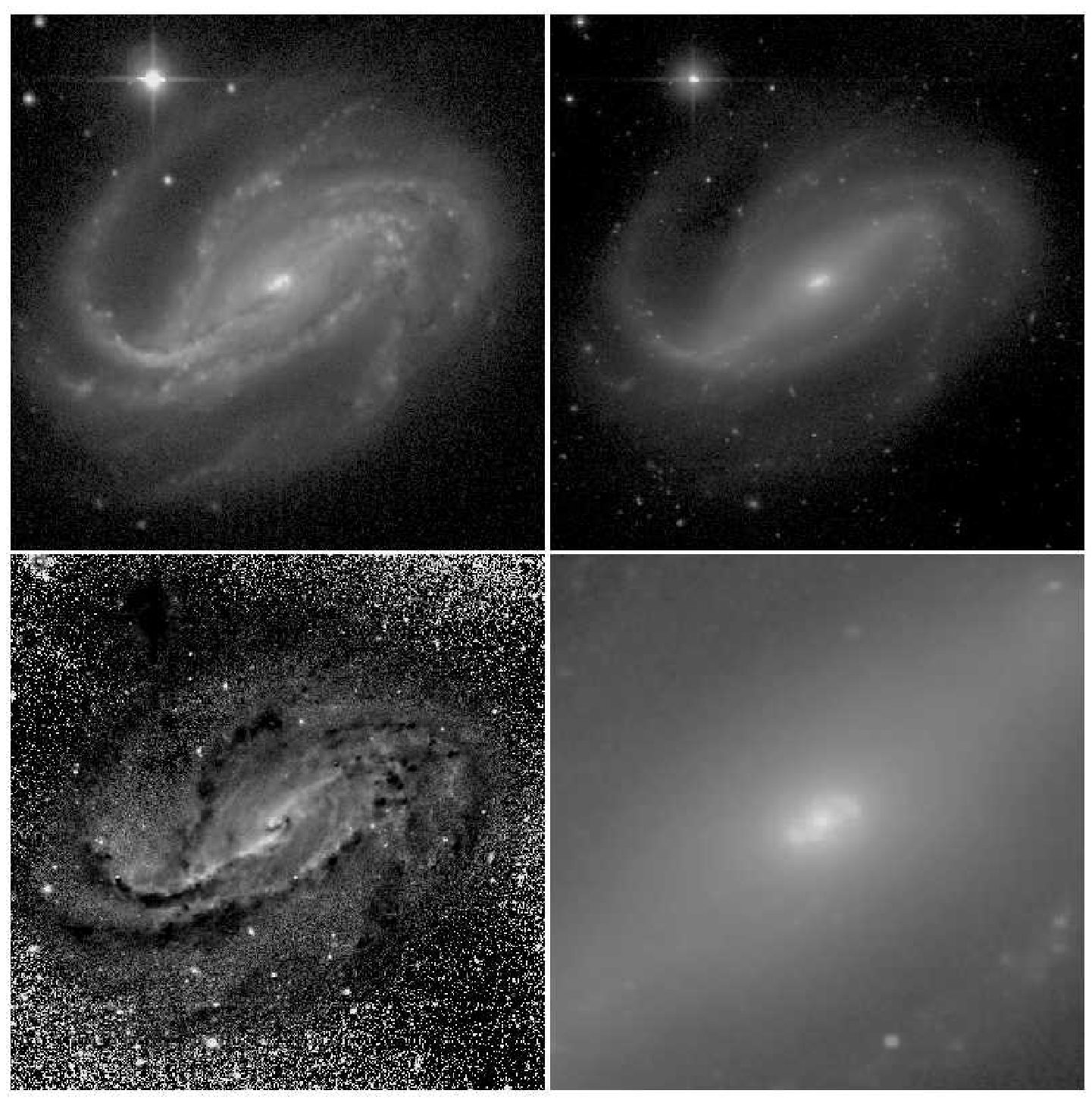}
\caption{Images of NGC 613. (upper left): $B$-band (Eskridge et al. 2002); (upper right): $K_s$-band;
(lower left): $B-K_s$ color index map; (lower right) same as at upper right, at four
times the scale. The images are logarithmic
in units of mag arcsec$^{-2}$ and the main fields have side lengths of
5\arcm 30. North is at the top and east is to the left.}
\label{images4}
\end{figure}

\begin{figure}
\figurenum{5}
\plotone{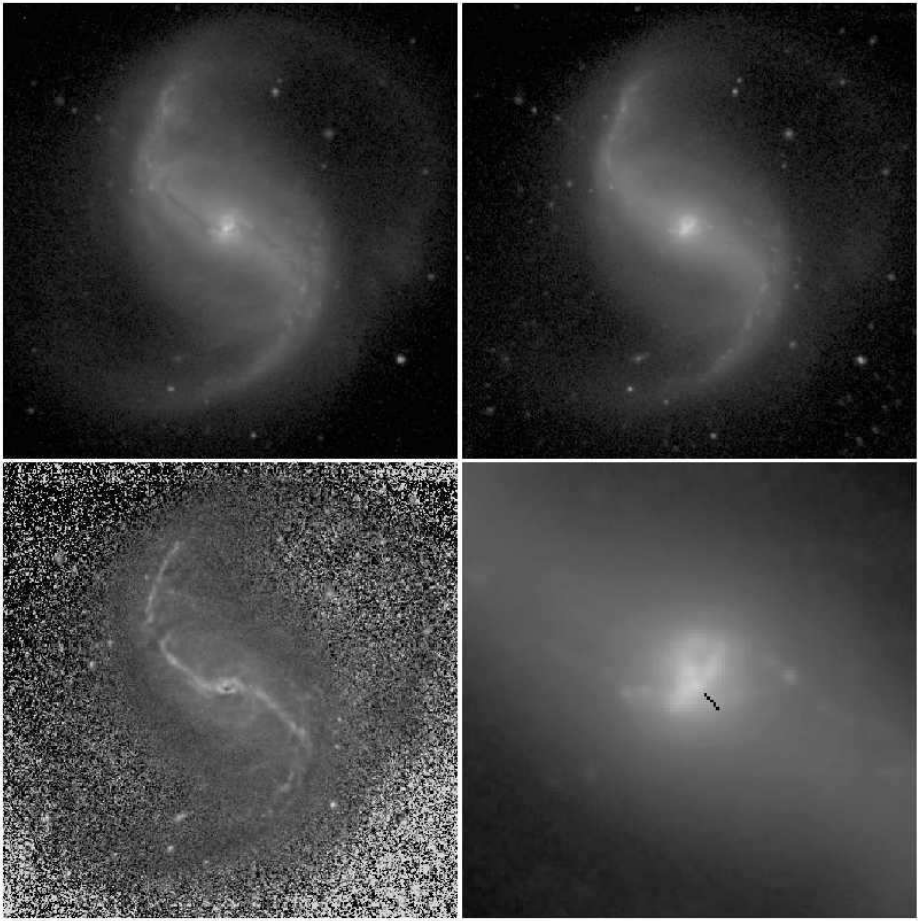}
\caption{Images of NGC 986. (upper left): $R$-band (Hameed \& Devereaux 1999); (upper right): $K_s$-band;
(lower left): $R-K_s$ color index map; (lower right) same as at upper right, at 
four times the scale. The images are logarithmic
in units of mag arcsec$^{-2}$ and the main fields have side lengths of
3\arcm 73. North is at the top and east is to the left. In the lower right
frame, the short dark line points to the feature recognized as the nucleus
in our analysis.}
\label{images5}
\end{figure}

\begin{figure}
\figurenum{6}
\plotone{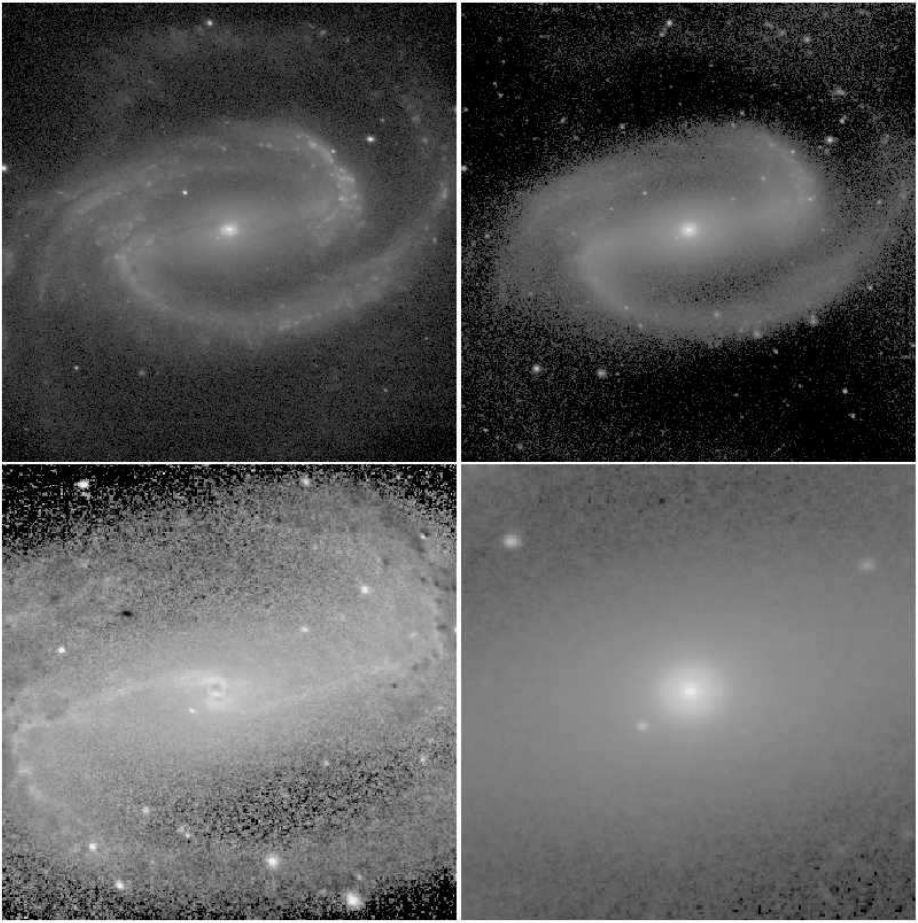}
\caption{Images of NGC 1300. (upper left): $B$-band (R. B. Tully, deVA); (upper right): $K_s$-band;
(lower left): $B-K_s$ color index map at twice the scale of the upper panels; 
(lower right) same as at upper right, at 
four times the scale. The images are logarithmic
in units of mag arcsec$^{-2}$ and the main fields have side lengths of
3\arcm 73. North is at the top and east is to the left.}
\label{images6}
\end{figure}

\begin{figure}
\figurenum{7}
\plotone{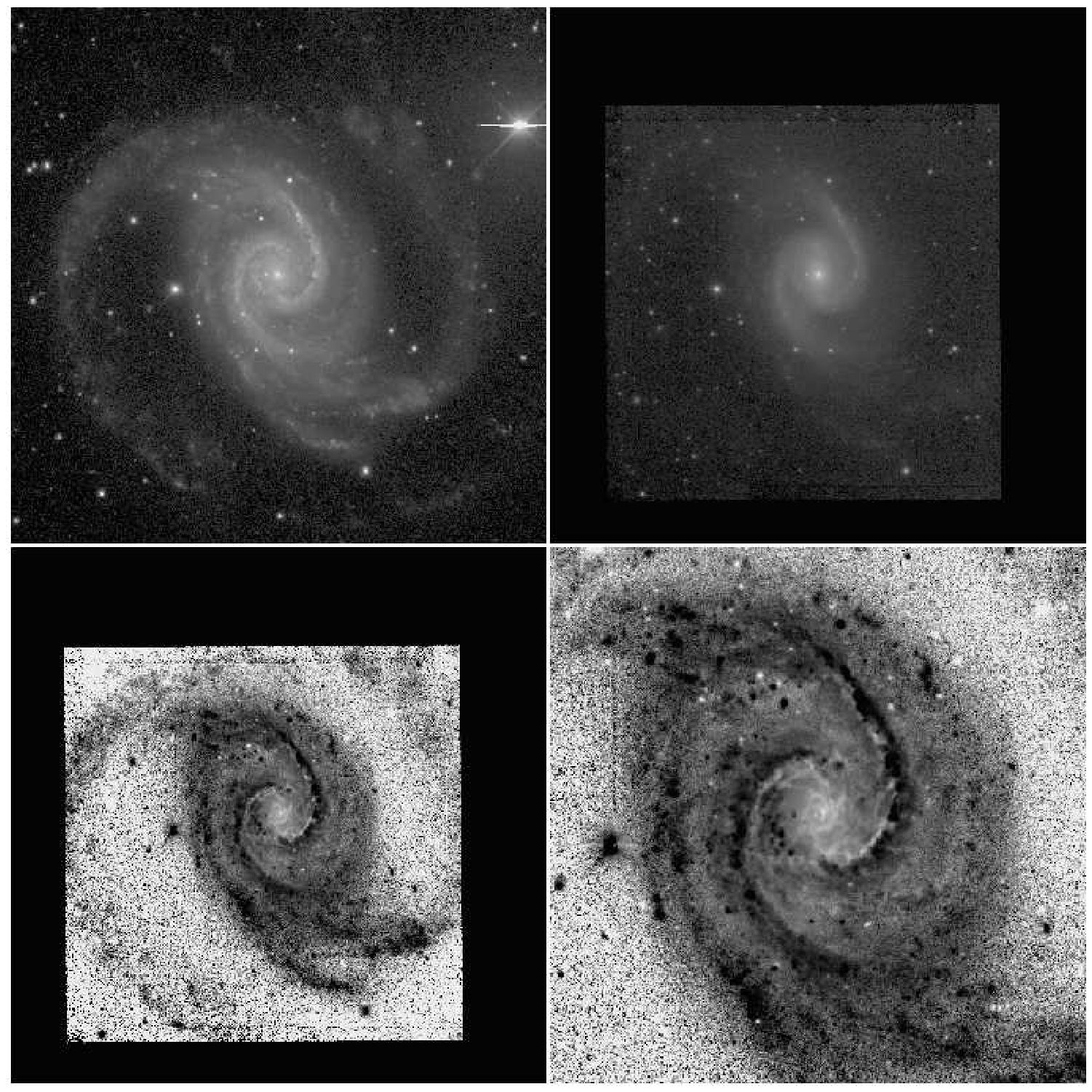}
\caption{Images of NGC 1566. (upper left): $B$-band (Kennicutt et al. 2003); (upper right): $K_s$-band;
(lower left): $B-K_s$ color index map; (lower right) same as at lower left, at twice
the scale. The images are logarithmic
in units of mag arcsec$^{-2}$ and on the same scale. The $B$-band field has a side length of
10\arcm 15. North is at the top and east is to the left.}
\label{images7}
\end{figure}

\begin{figure}
\figurenum{8}
\plotone{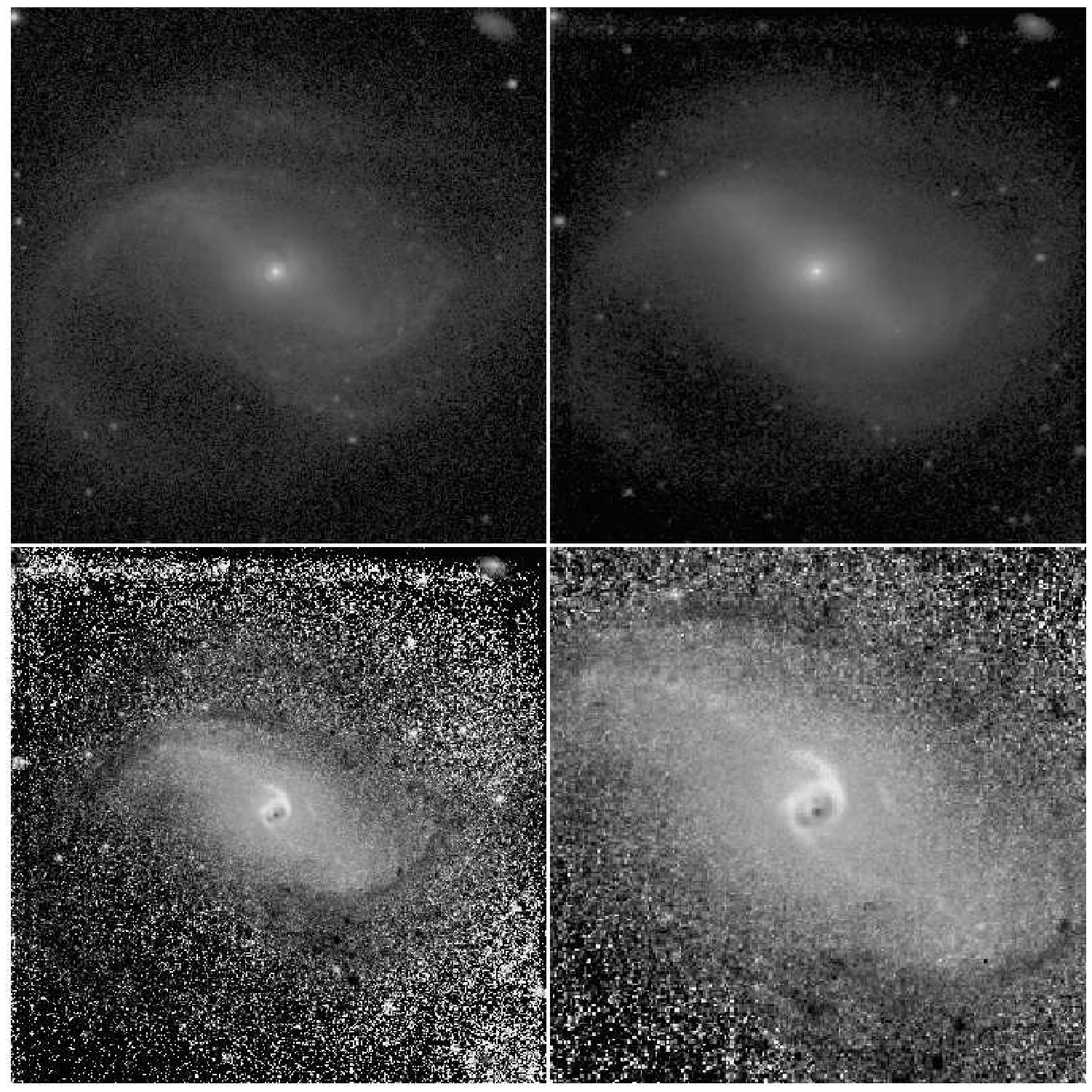}
\caption{Images of NGC 4593. (upper left): $B$-band (Eskridge et al.
2002); (upper right): $K_s$-band;
(lower left): $B-K_s$ color index map; (lower right) same as at lower left, at twice
the scale. The images are logarithmic
in units of mag arcsec$^{-2}$ and the main fields have side lengths of
3\arcm 73. North is at the top and east is to the left.}
\label{images8}
\end{figure}

\begin{figure}
\figurenum{9}
\plotone{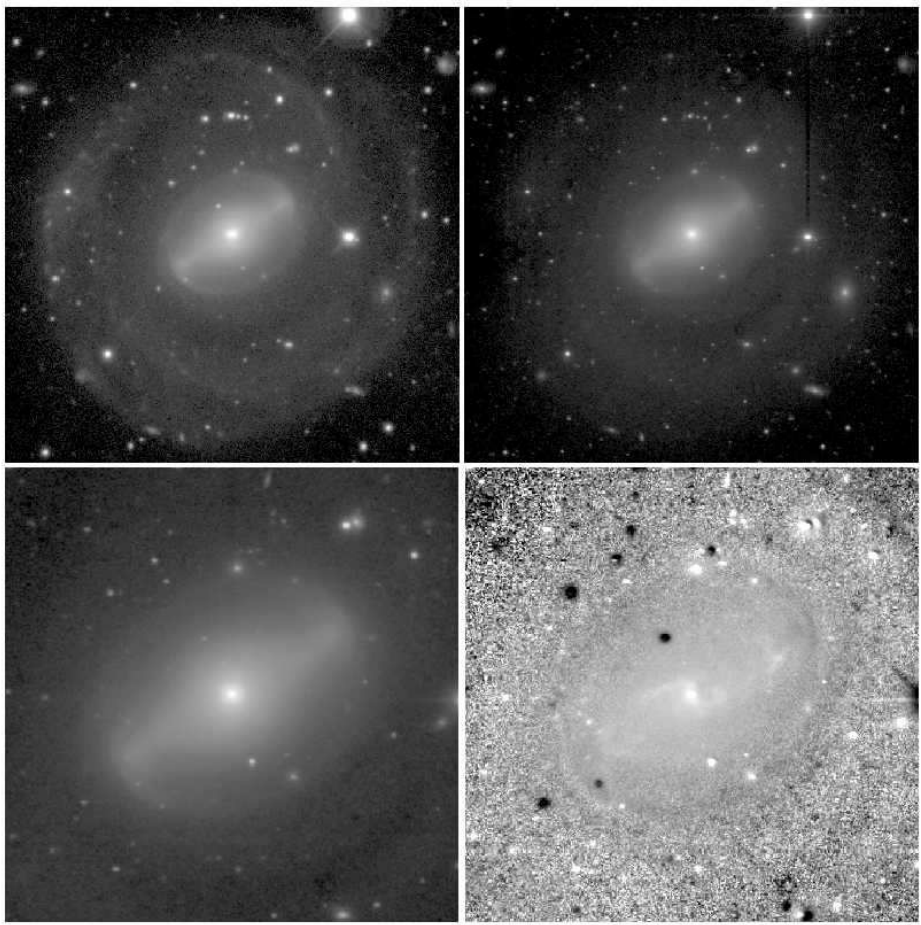}
\caption{Images of NGC 5101. (upper left): $B$-band  (Eskridge et al.
2002); (upper right): $K_s$-band;
(lower left): same as at upper right, at twice the scale;
(lower right): $B-K_s$ color index map, also at twice the scale
of the upper panels. The images are logarithmic
in units of mag arcsec$^{-2}$ and the main fields have side lengths of
5\arcm 96. North is at the top and east is to the left.}
\label{images9}
\end{figure}

\begin{figure}
\figurenum{10}
\plotone{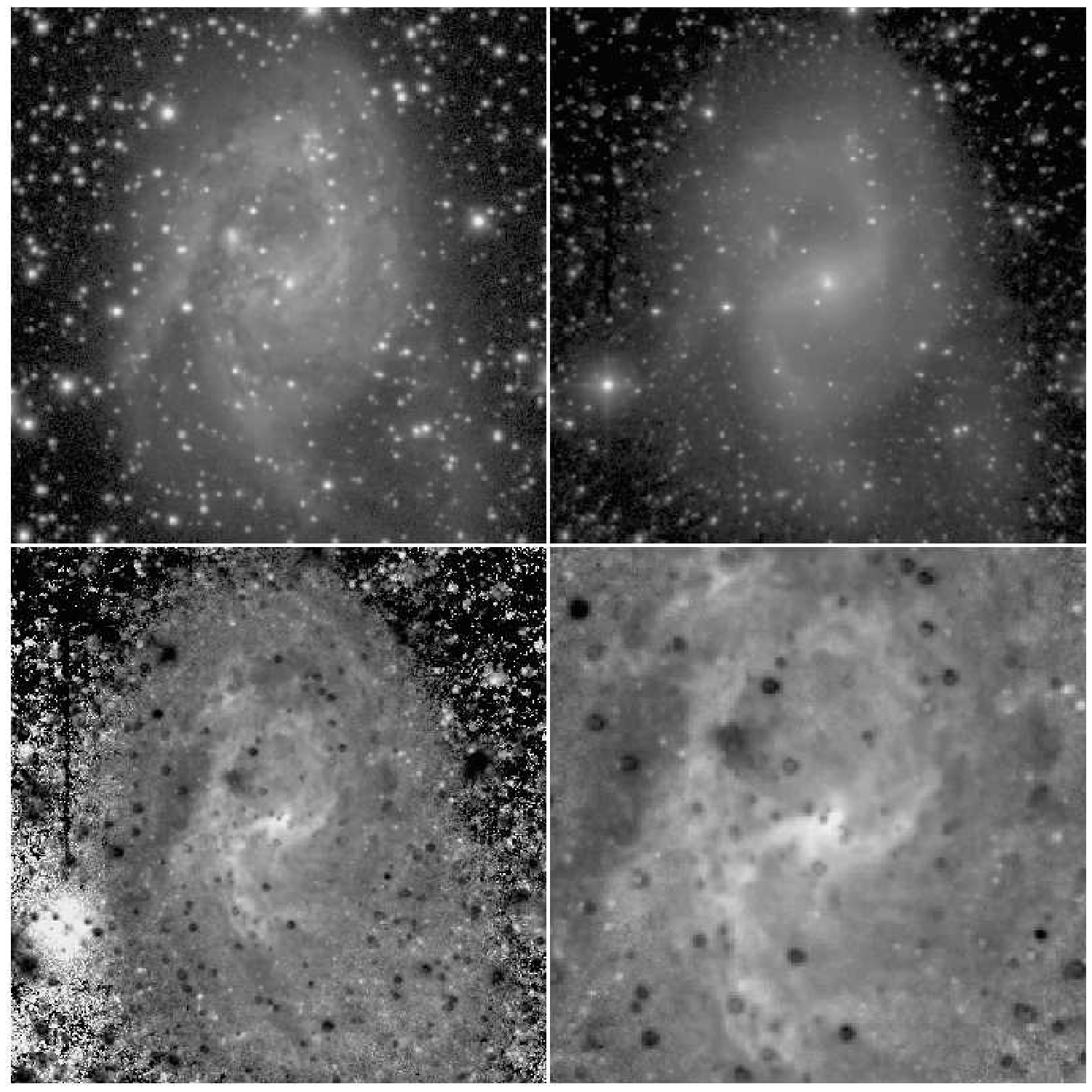}
\caption{Images of NGC 6221. (upper left): $B$-band (Eskridge et al.
2002); (upper right): $K_s$-band;
(lower left): $B-K_s$ color index map; (lower right) same as at lower left, at twice
the scale. The images are logarithmic
in units of mag arcsec$^{-2}$ and the main fields have side lengths of
3\arcm 73. North is at the top and east is to the left. In the color index
maps, central colors are uncertain owing to a seeing mis-match between the
$B$ and $K_s$-band images.}
\label{images10}
\end{figure}

\begin{figure}
\figurenum{11}
\plotone{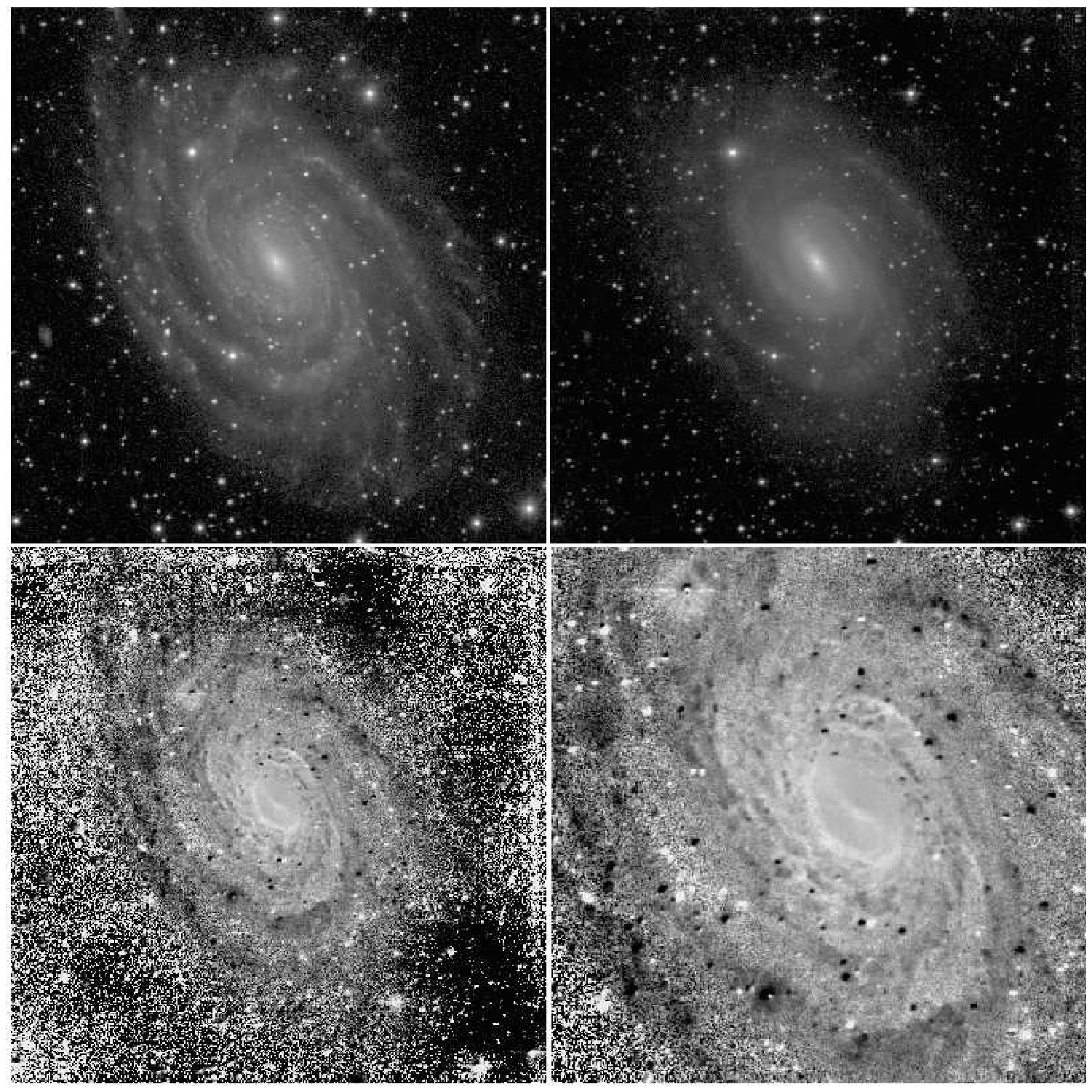}
\caption{Images of NGC 6384. (upper left): $B$-band (S. C. Odewahn,
deVA); (upper right): $K_s$-band;
(lower left): $B-K_s$ color index map; (lower right) same as at lower left, at twice
the scale. The images are logarithmic
in units of mag arcsec$^{-2}$ and the main fields have side lengths of
6\arcm 71. North is at the top and east is to the left.}
\label{images11}
\end{figure}

\begin{figure}
\figurenum{12}
\plotone{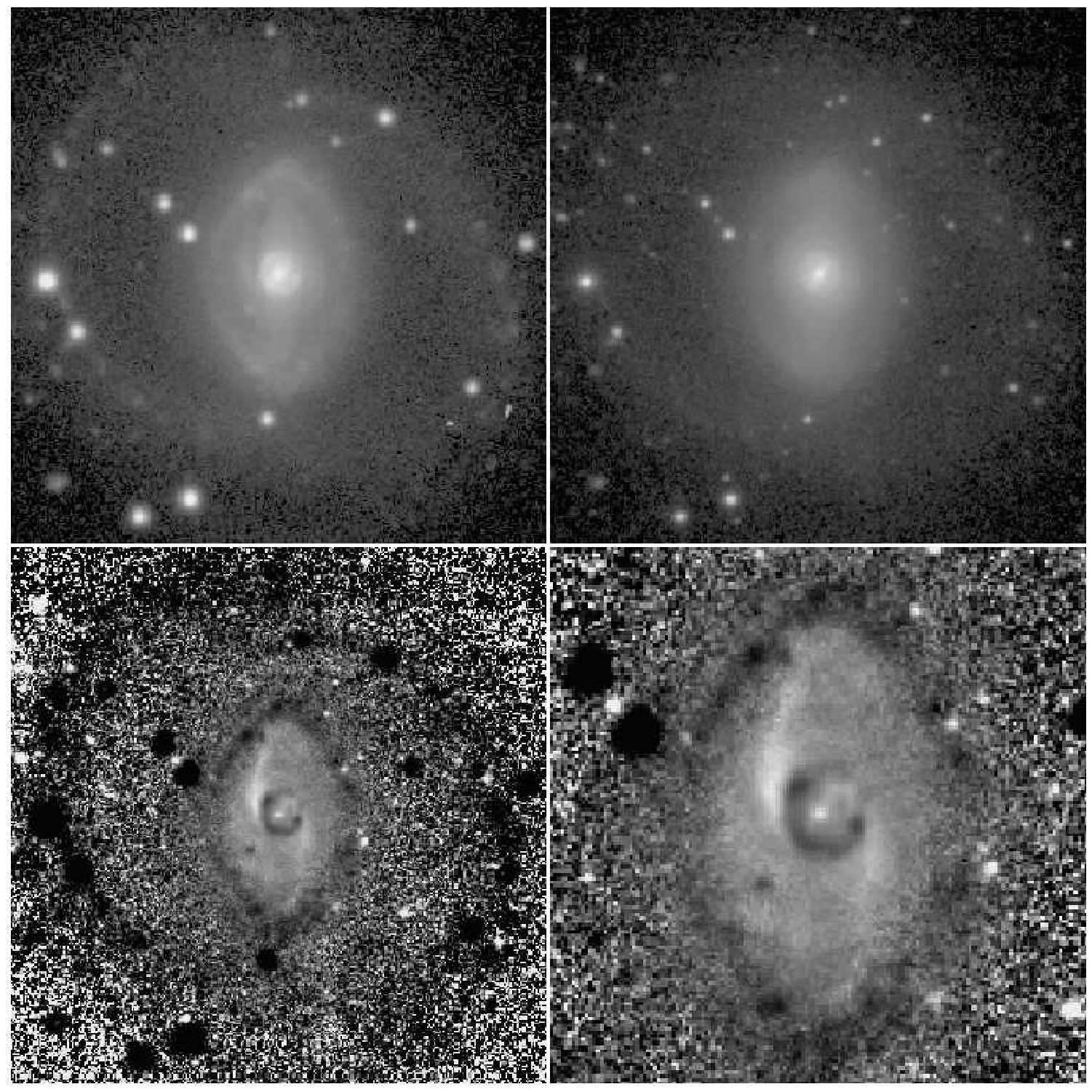}
\caption{Images of NGC 6782. (upper left): $B$-band (R. Buta, deVA);
(upper right): $K_s$-band; (lower left): $B-K_s$ color index map;
(lower right) same as at lower left, at twice the scale. The images are
logarithmic in units of mag arcsec$^{-2}$ and the main fields have side
lengths of 1\arcm 49. North is at the top and east is to the left.}
\label{images12}
\end{figure}

\begin{figure}
\figurenum{13}
\plotone{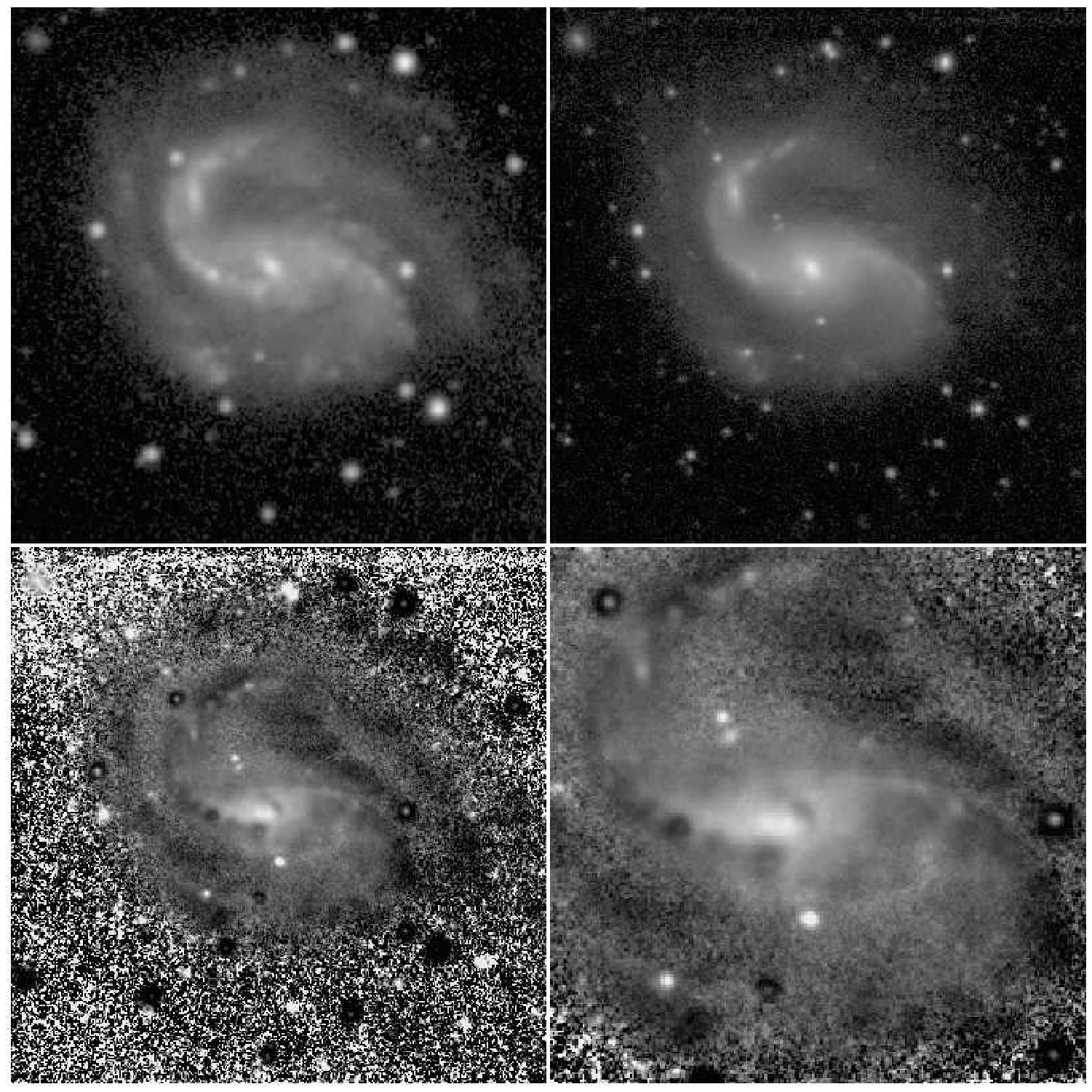}
\caption{Images of NGC 6907. (upper left): $B$-band (B. Canzian, deVA);
(upper right): $K_s$-band; (lower left): $B-K_s$ color index map;
(lower right) same as at lower left, at twice the scale. The images are
logarithmic in units of mag arcsec$^{-2}$ and the main fields have side
lengths of 3\arcm 62. North is at the top and east is to the left. In
the color index maps, central colors are uncertain owing to a seeing
mis-match between the $B$ and $K_s$-band images.}
\label{images13}
\end{figure}

\begin{figure}
\figurenum{14}
\plotone{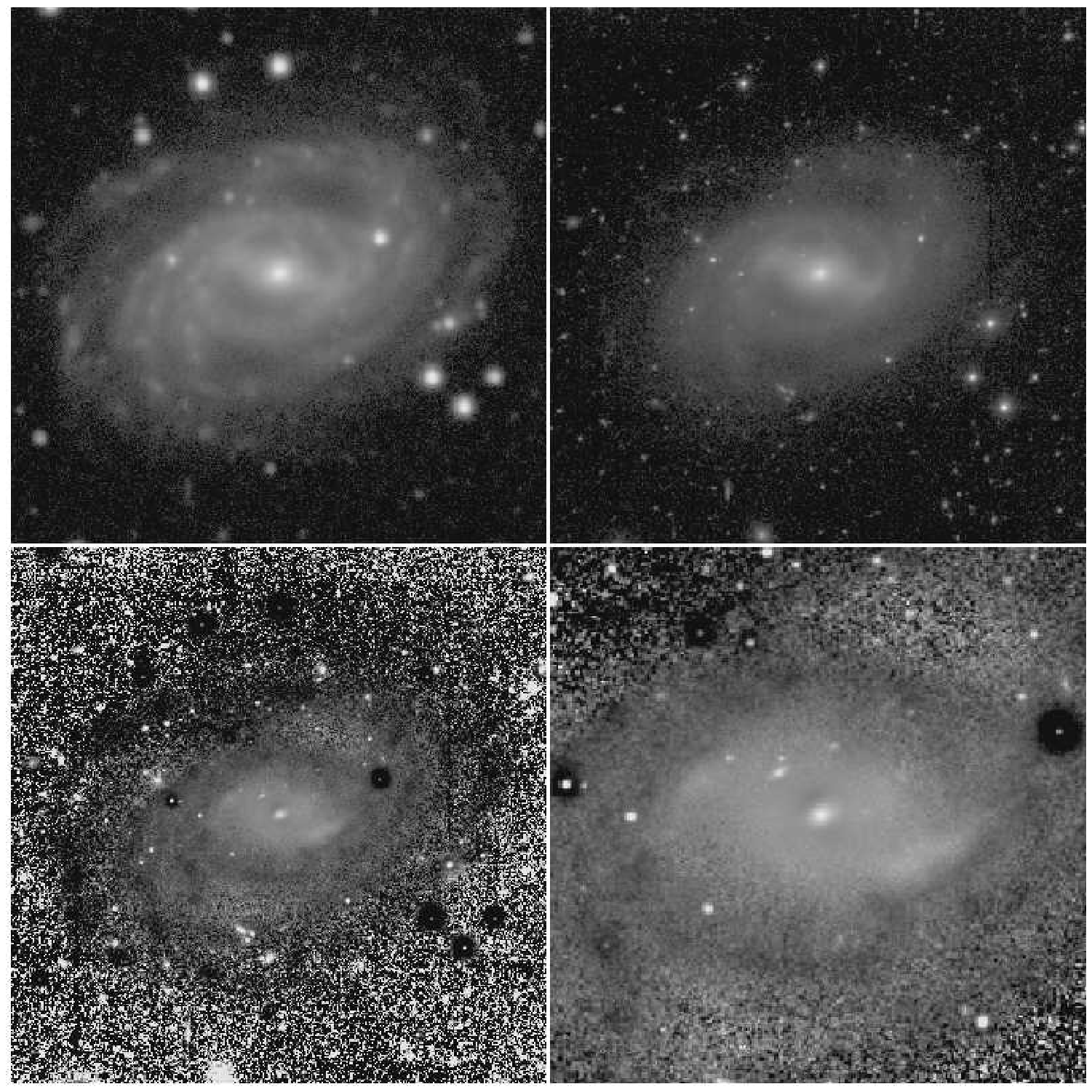}
\caption{Images of NGC 7329. (upper left): $B$-band (G. B. Purcell,
deVA); (upper right): $K_s$-band;
(lower left): $B-K_s$ color index map; (lower right) same as at lower left, at twice
the scale. The images are logarithmic
in units of mag arcsec$^{-2}$ and the main fields have side lengths of
4\arcm 41. North is at the top and east is to the left. In the color index
maps, central colors are uncertain owing to a seeing mis-match between the
$B$ and $K_s$-band images.}
\label{images14}
\end{figure}

\begin{figure}
\figurenum{15}
\plotone{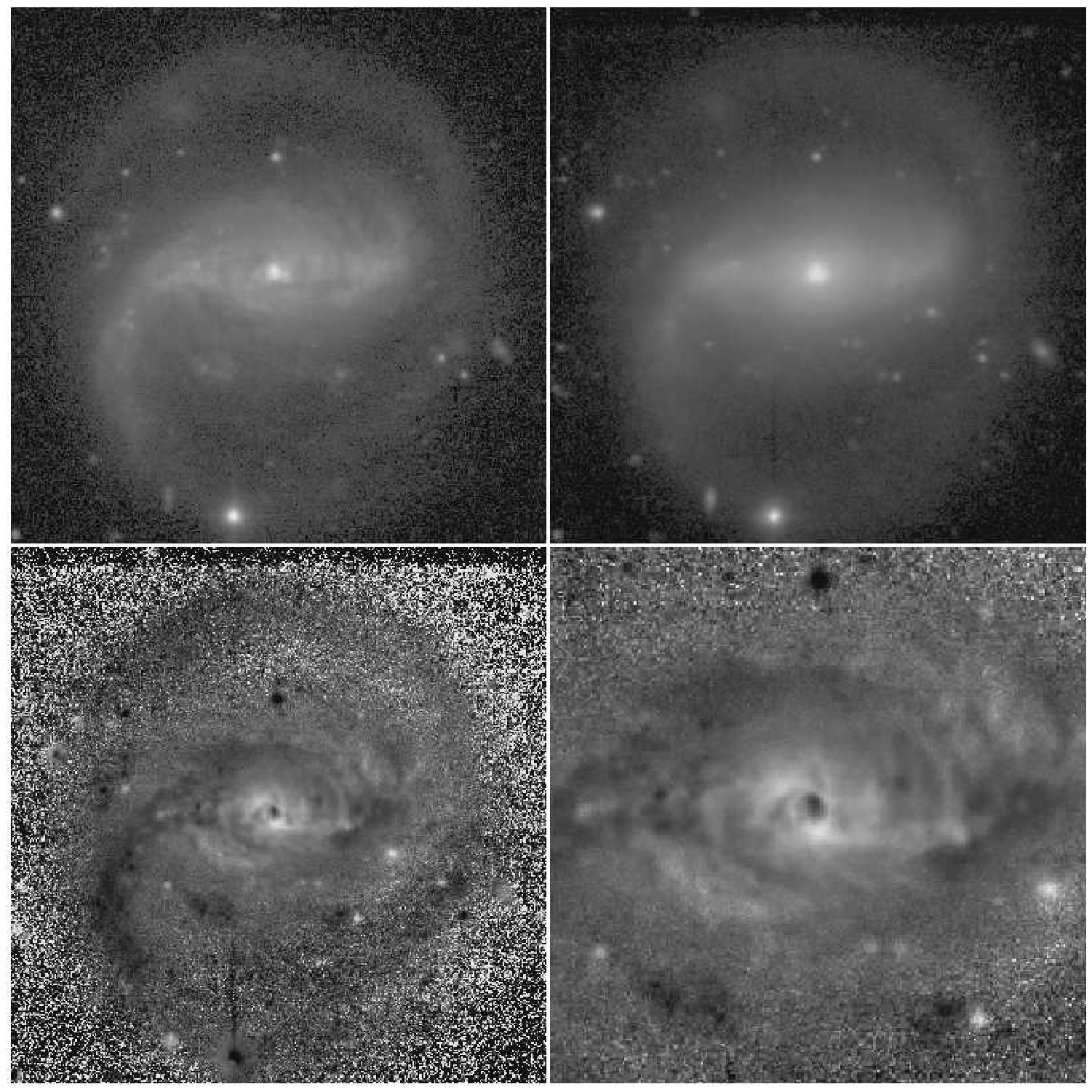}
\caption{Images of NGC 7552. (upper left): $B$-band (Eskridge et al. 2002); 
(upper right): $K_s$-band;
(lower left): $B-K_s$ color index map; (lower right) same as at lower left, at twice
the scale. The images are logarithmic
in units of mag arcsec$^{-2}$ and the main fields have side lengths of
3\arcm 73. North is at the top and east is to the left.}
\label{images15}
\end{figure}

\begin{figure}
\figurenum{16}
\plotone{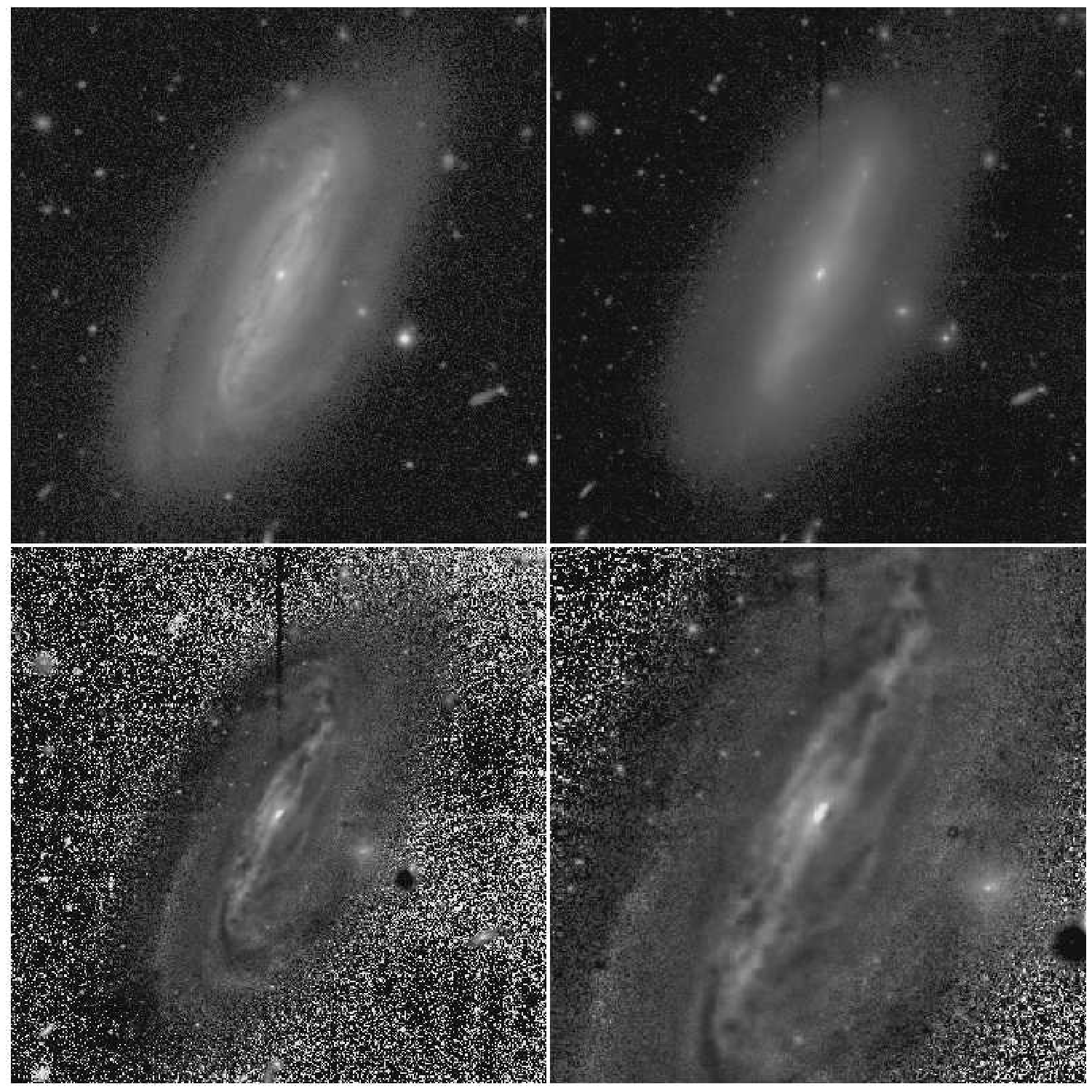}
\caption{Images of NGC 7582. (upper left): $B$-band (Eskridge et al.
2002); (upper right): $K_s$-band;
(lower left): $B-K_s$ color index map; (lower right) same as at lower left, at twice
the scale. The images are logarithmic
in units of mag arcsec$^{-2}$ and the main fields have side lengths of
5\arcm 30. North is at the top and east is to the left.}
\label{images16}
\end{figure}

{\it NGC 7513} - The $K_s$-band image in Figure~\ref{images2} shows a 
well-defined bar and faint spiral arms. The 2D decomposition 
(Figure~\ref{twodfits}) gave an approximately exponential bulge with
a $B/T$ of only 3.3\%. The bar was found to be well-represented by a
double-gaussian for all even Fourier terms to $m$=20 (Figure~\ref{fourier}).
However, we found that including only even Fourier terms provided a
poor mapping of the bar. Our analysis in this case includes extrapolations
of odd Fourier terms in the same manner as the even terms.
Figure~\ref{separations} shows that there is little residual asymmetry
in the bar when we account for such terms. From the separated
images, NGC 7513 is found to have a very strong bar with a bar class of 7
and a spiral class of 1. It has the weakest known spiral in the presence
of one of the strongest known bars.

{\it NGC 7552} - The images in Figure~\ref{images15} show a strong bar
and conspicuous spiral pattern in a relatively face-on disk. The bar
shows strong dust absorption, and there is a blue nucleus. The 2D
decomposition gave a Sersic index of $n$=2.2 and a $B/T$ of 38.4\%.
The relative Fourier amplitudes (Figure~\ref{fourier}) show a dominant
bar that is best represented by the symmetry assumption for $m$=2
and a double gaussian by all even terms having $m$$>$2. The residual spiral plus disk
image in Figure~\ref{separations} shows some asymmetry in the bar region.
From the separated images, the bar class is 4 while the spiral class 
is 1.

{\it NGC 7582} - Except for a higher inclination, this galaxy is very
similar to NGC 7552. The bar is very dusty and the spiral is fairly
conspicuous in the $K_s$-band (Figure~\ref{images16}). The 2D
decomposition gave a Sersic index of $n$=2.7 and a $B/T$ of 17.8\%.
The relative Fourier amplitudes (Figure~\ref{fourier}) are well-fitted
by double-gaussians in all even terms to $m$=20. After deprojection,
the images show strong decomposition pinch in the inner regions, and
some asymmetry in the bar region is highlighted in the spiral plus
disk image (Figure~\ref{separations}). The separated images gave a 
bar class of 4 and a spiral class of 1.

\begin{figure}
\figurenum{17}
\plotone{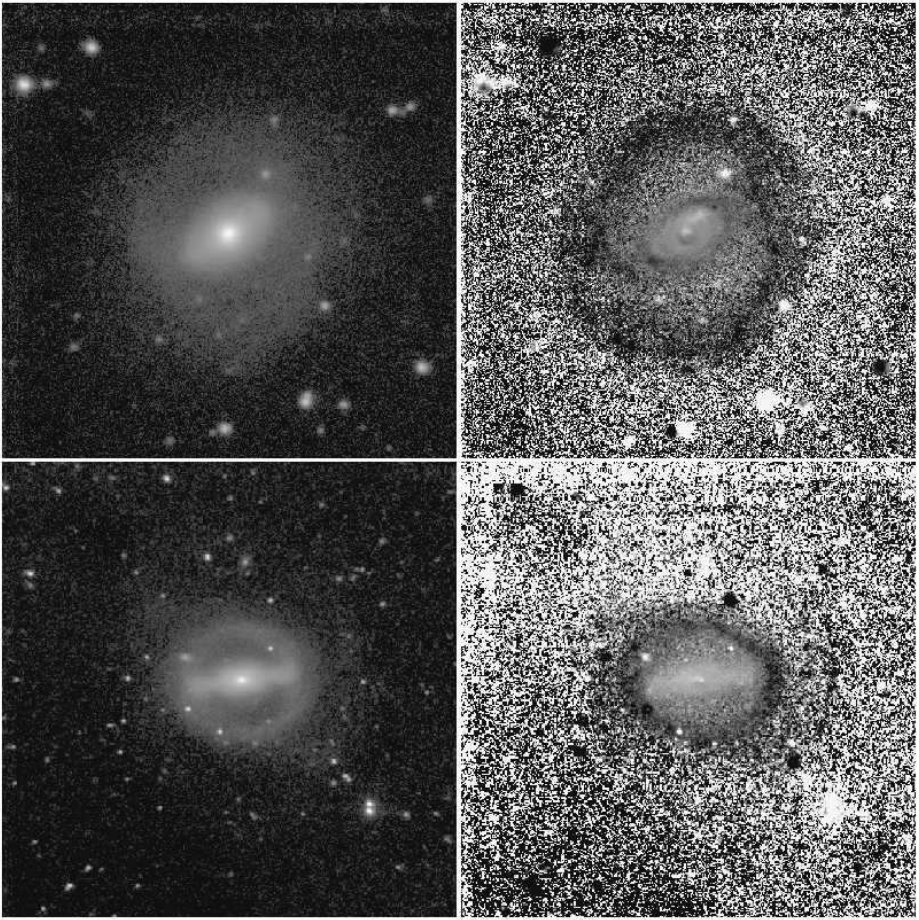}
\caption{Images of IC 1438 (upper left): $K_s$-band; (upper right):
$B-K_s$; and IC 4290 (lower left): $K_s$-band; (lower right) $B-K_s$.
Both $B$-band images are from the deVA (R. Buta).  The images are
logarithmic in units of mag arcsec$^{-2}$ and the fields have side
lengths of 3\arcm 55 for IC 1438 and 3\arcm 76 for IC 4290.  North is
at the top and east is to the left.}
\label{images17}
\end{figure}

\begin{figure}
\figurenum{18}
\plotone{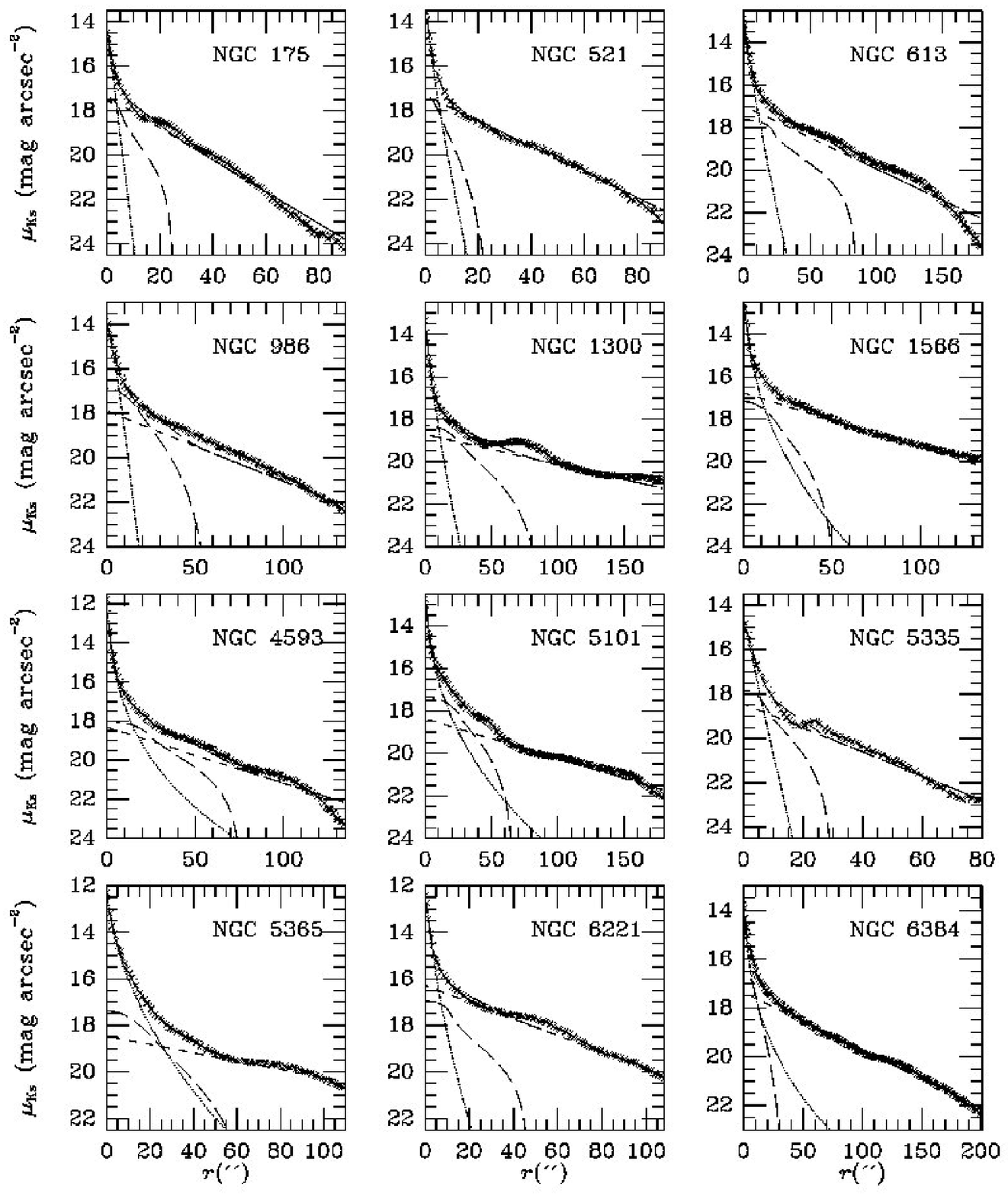}
\caption{}
\label{twodfits}
\end{figure}

{\it IC 1438} - The $K_s$-band image in Figure~\ref{images17} shows
a nearly face-on, weakly-barred galaxy with a faint outer ring.
The color index map shows enhanced star formation in nuclear, inner,
and outer rings/pseudorings. The 2D decomposition (Figure~\ref{twodfits})
gave a nearly exponential bulge and a $B/T$ of 32.5\%. The mapping of
the bar in Figure~\ref{fourier} required no special treatment (either 
gaussian or symmetry assumption) since
the Fourier amplitudes decline to zero near the ends of the bar
for each term. The $m$=2 term 
is more complicated than $m$=4 and 6 because the primary bar is 
imbedded within a clear oval that contributes mainly to $m$=2.
From the separated images in Figure~\ref{separations}, the galaxy
is bar class 1 and spiral class 1 with the bar stronger than the
spiral.

{\it IC 4290} - The bar and inner ring are the most conspicuous
features seen in the $K_s$-band image (Figure~\ref{images17}).
Both features are also seen in the color index map. The galaxy
is relatively face-on and a visual mapping of the deprojected 
inner ring gave a diameter of 18.4 kpc, an axis ratio of 0.89,
and (within 8$^{\circ}$) 
an alignment nearly parallel to the bar. The bar itself has
a strong inner boxy zone noted in previous studies by Buta \& Crocker
(1991) and Buta et al. (1998). The 2D decomposition gave a nearly
exponential bulge and a $B/T$ of 14.1\%. The relative Fourier amplitudes
in the bar are well-represented by double gaussians for all even terms
to $m$=18 (Figure~\ref{fourier}). The separated images 
(Figure~\ref{separations}) gave a bar class of 5 and a spiral class of 1.

{\it IC 5092} - The $K_s$-band morphology of this galaxy is in the
unusual form of an s-shaped barred spiral encompassed by a conspicuous
outer ring (Figure~\ref{images2}). The 2D decomposition gave a Sersic 
index of $n$=3.1 and a $B/T$ of 5.0\% (Figure~\ref{twodfits}). A single
gaussian was adopted for the bar mapping in Figure~\ref{fourier}.
The separated images (Figure~\ref{separations}) gave a bar class of 
5 and a spiral class of 2.

{\it UGC 10862} - The bar in this late-type spiral is unusual: it appears
in the form of a highly elongated ring with ansae. This character is
shown best in the unsharp mask image in Figure~\ref{images2}. We have
no optical images of this galaxy that can be used to determine the colors
of the bar features. The 2D decomposition gave a bulge with a Sersic
index of $n$=0.68 and a $B/T$ of 0.9\%. The bulge is weak and the
solution may not be reliable. The bar mapping in Figure~\ref{fourier}
uses a single gaussian representation and the separated images in 
Figure~\ref{separations} gave a bar class of 8 and a spiral class of 2, 
the strongest bar in the sample.

\section{Discussion: Do the Stronger Bars Have Stronger Spirals?}

We investigate this issue not only using $Q_s$ as a measure of spiral
strength, but also the $m$=2 and 4 spiral contrasts. The reason for
including the spiral contrasts is because these can be considered
the response to the spiral driver, the bar torque. The spiral torque 
$Q_s$ is not the same as the response amplitude,
but is diluted by the inner bulge and the bar radial force.

Figure~\ref{newqbqs} shows graphs of spiral contrast, spiral strength,
and total maximum relative non-axisymmetric torque strength versus bar
strength $Q_b$, parameters all listed in Table 7.  The $Q_g$ versus
$Q_b$ plot is shown only to highlight that for a sample like ours,
$Q_g$ is also a good indicator of bar strength.  The spiral contrasts
$A_{2s}$ and $A_{4s}$ were estimated from the spiral plus disk images
after separation of the bar. Note that the radii of these maxima (also
listed in Table 7) are often comparable but can differ considerably.

Figure~\ref{newqbqs} shows that the spiral parameters $A_{2s}$,
$A_{4s}$, and $Q_s$ all have little correlation with $Q_b$. (We
note that the strongest spiral occurs for intermediate values of
$Q_b$.) 
We can nevertheless quantify the correlations with a few statistical
tests using programs from Press et al. (1986). For the 23 galaxies, the
linear correlation coefficient for the $Q_s,Q_b$ plot is $r_{\ell}$=0.26 and
the null hypothesis of zero correlation is disproved only at the 22.5\%
significance level. The Spearman rank-order correlation coefficient is
$r_{sp}$=0.41 with a significance level of 5\%, while the Kendall Tau rank
coefficient is $\tau$=0.27 with a significance level of 7\%.  Similar
parameters for the $A_{2s},Q_b$ plot are $r_{\ell}$=0.16 ($P_{r_{\ell}}$ $<$ 47\%),
$r_{sp}$=0.24 ($P_{r_{sp}}$ $<$ 27\%), and $\tau$=0.16 ($P_{\tau}$ $<$29\%).
Similar parameters for the $A_{4s},Q_b$ plot are $r_{\ell}$=0.27 ($P_{r_{\ell}}$ $<$
22\%), $r_{sp}$=0.39 ($P_{r_{sp}}$ $<$ 7\%), and $\tau$=0.25 ($P_{\tau}$
$<$10\%). For $Q_s$ versus $Q_b$, the non-parametric rank-order
coefficients don't completely rule out some correlation, but any
correlation is weak. This is certainly partly due to the small number
of galaxies in our present sample.

On the other hand, $Q_s$ (and to some extent also $A_{2s}$ and
$A_{4s}$) will have an inverse correlation with $Q_b$ due to the nature
of bar-spiral separation. In a reliable separation, the radial profiles
of the torques for the bar and spiral must lie wholly within the curve
for the total torque. This leads to correlated uncertainties in $Q_s$
and $Q_b$: if the bar is overestimated, then the spiral is
underestimated and vice versa. In principle, this could weaken a real
positive correlation between $Q_s$ and $Q_b$ if the radial torque
curves for the bar and the spiral significantly overlap. However, for
19 of the 23 galaxies in our sample, the curves for the bar and the
spiral (or other outer components) do not greatly overlap. The worst
cases are NGC 613, 986, 1300, and 6907 (see Figure~\ref{separations}).
Also, Buta, Block, \& Knapen (2003) showed that a $\pm$10\% uncertainty
in $r_{2b}$ for the bar Fourier mapping would move a galaxy like NGC
6951 ($Q_b$=0.28, $Q_s$=0.21) along a line having $\Delta Q_s$/$\Delta
Q_b$ =0.044/0.021 =$-$2.1. Thus, the correlated uncertainties may
spread the points out more in $Q_s$ than in $Q_b$. Since NGC 6951 is a
very typical case of significant overlap between the bar and the spiral
(see Figure 2 of Buta, Block, \& Knapen 2003), and the effect is still
small, we conclude that correlated uncertainties between $Q_s$ and
$Q_b$ are not causing a significant false correlation, nor masking
completely a real one, between $Q_s$ and $Q_b$ over all the data
points in our sample.

In Figure~\ref{newqbmeanqs}c, we have combined our 23 galaxies with
previous studies of bar and spiral strengths. This graph shows $Q_s$
versus $Q_b$ for 177 galaxies including the samples of Buta et al.
(2005), Block et al. (2004), and six early-type spirals from Buta
(2004), in addition to the AAT sample galaxies. For 15 galaxies having
two sources of parameters, the values were averaged so that only a
single point is plotted. The agreement between duplicate values, which
involves calculations either from OSUBGS $H$-band images (the $Q_{b1}$
and $Q_{s1}$ values) or from $K_s$ images (the $Q_{b2}$ and $Q_{s2}$
values), is shown in Figure~\ref{newqbmeanqs}a,b. No error bars are
indicated on the individual points in Figure~\ref{newqbmeanqs}c, but
the errors would be similar to those shown in Figure~\ref{newqbqs}c.
Figure~\ref{newqbmeanqs}d shows means of $Q_s$ in bins of $Q_b$
indicated by the dotted horizontal lines. The correlation analysis for
this sample gives $r_{\ell}$=0.35, $r_{sp}$=0.31, and $\tau$=0.22, all fairly low
values but with probabilities indicating a significant correlation.  We
see in panel (b) that the means of $Q_s$ do increase slightly with
increasing $Q_b$.  However, there is little correlation for $Q_b$ $<$
0.3. The most interesting case added by the AAT sample is NGC 7513
which, as we have noted, is a class 7 bar accompanied by only a weak
class 1 spiral.

Figures~\ref{revabsmag} and ~\ref{revrc3t} show plots of $Q_s$ versus
$Q_b$ for the combined sample of 177 galaxies, but subdivided according
to absolute blue magnitude and RC3 numerical stage (type) index.
Several individual galaxies are labeled for reference. These plots
reveal in a more convincing way that some correlation between $Q_s$ and
$Q_b$ is indeed present. Among the more luminous galaxies, the
strongest bars have the strongest spirals (NGC 1530, 7479). Among
intermediate luminosity galaxies, NGC 1042 and 7412 stand out as spiral
outliers at low $Q_b$, although they appear to be relatively normal
late-types. NGC 1042 is classified as type SAB and NGC 7412 as type SB
in RC3, yet only weak bars were detected in the near-infrared.
These galaxies are also outliers in the rightmost panels in
Figure~\ref{revrc3t}. Only weak trends are evident in most of the panels, in
the sense of a slow increase in spiral strength with increasing bar
strength. Two of the strongest bars with the weakest spirals that we
added, NGC 7513 and UGC 10862, are of lower luminosity than cases like
NGC 986, 1530, and 7479 which have very strong spirals. The most
significant-looking correlations appear for later types.

On the basis of 17 spirals, Block et al. (2004) found a correlation
between $Q_s$ and $Q_b$ and suggested that this implies that bars and
spirals grow together and have the same pattern speed.  Our
comparably-sized AAT sample also supports this finding but shows in
addition that some very strong bars can have rather weak near-infrared
spirals.

These results suggest that some spirals probably are driven responses
to a strong bar, although we may need more information to decide which
ones. We suggest that cases like NGC 986, 1530, and 7479 are in this
category. All three of these objects have been analyzed using the
potential-density phase-shift method (Zhang \& Buta 2007), and all
three show phase-shift distributions consistent with a single pattern
speed of the main bar and spiral. Zhang \& Buta (2007) discuss NGC
1530, while Buta \& Zhang (2008) will provide the information on NGC
7479. We will show the phase-shift results for NGC 986 in a separate
paper, where we will also show that NGC 175 has a phase-shift
distribution consistent with a decoupling between the spiral and the
bar.

We suggested in section 1 that bars may drive spirals only when the bar
is growing or if there is gaseous dissipation. We suspect that there
could be a saturated state where there is a bar but it cannot do much
to drive a spiral. An example of such saturation is an SB0 galaxy. All
the stars are in steady orbits, and nothing is growing fast enough or
dissipating fast enough to be a significant source of direction.  A
large galaxy with a small inner bar is also not likely to have a
bar-driven spiral. A variety of factors undoubtedly conspire to make
the scatter significant in a plot of $Q_s$ versus $Q_b$.

\section{Conclusions}

We have analyzed near-infrared images of 23 barred galaxies covering
a wide range of types and apparent bar strengths. Using Fourier
techniques, we have separated the bars from the spirals and have
derived maximum relative bar and spiral torque strengths. Our results
are as follows:

\noindent
1. The sample is morphologically diverse and includes strong two-armed
barred spirals as well as multi-armed barred spirals.

\noindent
2. As in previous studies (e.g, Buta et al. 2005, 2006), the relative
Fourier intensity amplitudes of some of the bars in this sample can
be mapped with single and double gaussian representations. Others
can be mapped with the symmetry assumption (Buta, Block, \& Knapen
2003). 

\noindent
3. We showed that Fourier amplitude fractions from $m$=2 and 4
bar Fourier profiles could be a useful way to define bar radii.

\noindent
4. We showed that stronger bars have relatively sharper ends. This
could mean that the orbits crowd an outer resonance, as if the volume
in phase space that contains bar-reinforcing orbits is nearly filled.

\noindent
5. In answer to our main question, we find weak but definite
indications that stronger spirals are associated with stronger bars.
This is consistent with our previous findings, but two of the galaxies
in our present sample, NGC 7513 and UGC 10862, show that exceptionally
strong bars can have weak near-infrared spirals. Some galaxies with
strong bars, like NGC 986, do have strong spirals.  Because spirals
having bars with $Q_b$ $>$ 0.4 are very rare, our study is still
affected by small number statistics at the strong bar end.

Thus, our main conclusion of this study is similar to that of Block et
al. (2004): some bars and spirals probably grow together in a global
disk instability, leading to the average increase of $Q_s$ with $Q_b$
for $Q_b$ $>$ 0.3. For $Q_b$ $<$ 0.3, bars and spirals may be more
independent features in general. Cases where the strength of the bar
and the spiral are comparably large, as in NGC 986, could be genuine
``bar-driven" spirals. Nevertheless, the existence of cases like NGC
7513 and UGC 10862, which lie in what was previously an empty region in
the $Q_s$, $Q_b$ plot, shows that other factors probably complicate the
relationship between bar and spiral torque strengths.

Although our analysis did not provide a definitive answer to the
question posed in the title of this paper, owing in part to the
limitations of our samples as well as the depth and quality of some of
the near-IR images used, this situation will change soon with the {\it
Spitzer Survey of Stellar Structure in Galaxies} (S$^4$G) (Sheth et al.
2009). This survey will provide a nearly complete sample of 2300
galaxies of all types within 40Mpc to a depth that would be very
difficult to achieve in the $K_s$-band from the ground. With such a
large sample, we can improve the statistics in all regions of the
$Q_s$,$Q_b$ diagram and further examine the questions we have raised
here.

We thank Emma Allard for help during the observations and with the data
reduction, Stuart Ryder for his support during the observing run at the
AAT, and the anonymous referee for helpful comments. RB acknowledges
the support of NSF Grant AST 05-07140 to the University of Alabama. JHK
acknowledges support through IAC project 3I/2407. DME acknowledges
publication support from the Vassar College Research Committee. HS and
EL acknowledge the support from the Academy of Finland. IP acknowledges
support from the Mexican foundation CONACyT under project 35947-E. DLB
warmly thanks the Board of Trustees of the Anglo-American Chairman's
Fund for their financial support.  This research has made use of the
NASA/IPAC Extragalactic Database (NED), which is operated by the Jet
Propulsion Laboratory, California Institute of Technology, under
contract with NASA.

REFERENCES 

\noindent
Athanassoula, E. \& Misiriotis, A. 2002, \mnras, 330, 35

\noindent
Block, D. L. \& Wainscoat, R. J. 1991, Nature, 353, 48

\noindent 
Block, D. L., Puerari, I., Frogel, J. A., Eskridge, P. B., Stockton, A., Fuchs, B. 1999, Ap\&SS, 269, 5

\noindent 
Block, D. L., Buta, R., Knapen, J. H., Elmegreen, D. M., Elmegreen, B.
 G., \& Puerari, I. 2004, \aj, 128, 183

\noindent
B{\"o}ker, T., Falc{\'on}-Barroso, J., Schinnerer, E., 
Knapen, J. H., \& Ryder, S. 2008, \aj, 135, 479

\noindent
Buta, R. 1995, \apjs, 96, 39

\noindent
Buta, R. 2004, Astrophysics and Space Science Library, 319, 101

\noindent 
Buta, R., Block, D. L., and Knapen, J. H. 2003, \aj, 126, 1148

\noindent
Buta, R. J., Corwin, H. G., \& Odewahn, S. C. 2007, The de Vaucouleurs Atlas of
Galaxies, Cambridge: Cambridge U. Press (deVA)

\noindent Buta, R. \& Crocker, D. A. 1991, \aj, 102, 1715

\noindent Buta, R., Alpert, A., Cobb, M. L., et al. 1998, AJ, 116, 1142

\noindent Buta, R., Vasylyev, S., Salo, H., and Laurikainen, E. 2005, \aj, 130,
506

\noindent
Buta, R. \& Zhang, X. 2008, in preparation

\noindent 
Combes, F. \& Sanders, R. H. 1981, \aap, 96, 164

\noindent
de Grijs, R. 1998, \mnras, 299, 595

\noindent 
de Vaucouleurs, G. and Buta, R. 1980, \apjs, 44, 451

\noindent
de Vaucouleurs, G., de Vaucouleurs, A., Corwin, H. G., Buta, R. J., Paturel, G., 
\& Fouque, P. 1991, Third Reference Catalog of Bright Galaxies (New York: Springer) (RC3)

\noindent
Elmegreen, B. G. \& Elmegreen, D. M. 1985, \apj, 288, 438

\noindent
Elmegreen, B. G., Elmegreen, D. M., \& Hirst, A. C. 2004, \apj, 612, 191

\noindent
Elmegreen, B. G., Elmegreen, D. M., Knapen, J. H., Buta, R. J., Block, D. L.,
\& Puerari, I. 2007, \apj, 670, L97

\noindent
Elmegreen, D. M. \& Elmegreen, B. G. 1982, \mnras, 201, 1021

\noindent 
Erwin, P. 2004, \aap, 415, 941

\noindent
Erwin, P. 2005, \mnras, 364, 283

\noindent 
Eskridge, P. B., Frogel, J. A., Pogge, R. W., et al. 2002, ApJS, 143, 73

\noindent
Gadotti, D, Athanassoula, E., Carrasco, L., Bosma, A., de Souza, R. E.,
Recillas, E. 2007, \mnras, 381, 943

\noindent
Galaz, G., Villalobos, A., Infante, L., \& Donzelli, C. 2006, \aj, 131, 2035

\noindent 
Hackwell, J. A. and Schweizer, F. 1983, \apj, 265, 643

\noindent
Hameed, S. \& Devereaux, N. 1999, \aj, 118, 730

\noindent
Jogee, S. et al. 2004, \apj, 615, 105

\noindent
Kennicutt, R. C. et al. 2003, \pasp, 115, 928

\noindent
Knapen, J. H., Shlosman, I., \& Peletier, R. F. 2000, \apj, 529, 93

\noindent
Kohno, K. et al. 2008, PASJ, 60, 457

\noindent
Koribalski, B. \& Dickey, J. M. 2004, \mnras, 348, 1255

\noindent
Kormendy, J. \& Norman, C. A. 1979, \apj, 233, 539

\noindent 
Kormendy, J. and Kennicutt, R. 2004, ARAA, 42, 603

\noindent 
Laurikainen, E. \& Salo, H. 2002, \mnras, 337, 1118

\noindent
Laurikainen, E., Salo, H., \& Buta, R. 2004, \apj, 607, 103

\noindent
Laurikainen, E., Salo, H., \& Buta, R. 2005, \mnras, 362, 1319

\noindent
Laurikainen, E., Salo, H., Buta, R., Knapen, J., Speltincx, T.,
\& Block, D. L.  2006, \mnras, 132, 2634

\noindent
Laurikainen, E., Salo, H., Buta, R., \& Knapen, J. H. 2007,
\mnras, 381, 401

\noindent
Lin, L.-H., Yuan, C., \& Buta, R. 2008, arXiv0805.3613

\noindent
Lynden-Bell, D. \& Ostriker, J. P. 1967, \mnras, 136, 293

\noindent
Marinova, I. \& Jogee, S. 2007, \apj, 659, 1176

\noindent
Menendez-Delmestre, K., Sheth, K., Schinnerer, E., Jarrett, T., \& Scoville,
N. 2007, \apj, 657, 790

\noindent 
Mulchaey, J. S., Regan, M. W., \& Kundu, A. 1997, \apjs, 110, 299

\noindent
Peeples, M. S. \& Martini, P. 2006, \apj, 652, 1097

\noindent
Press, W. H. et al. 1986, Numerical Recipes

\noindent
Regan, M. \& Elmegreen, D. 1997, AJ, 114, 965

\noindent 
Quillen, A. C., Frogel, J. A., \& Gonz\'alez, R. A. 1994, \apj, 437, 162

\noindent
Sersic, J. L. 1968, Atlas de galaxias Australes, Cordoba, Argentina: Observatorio
Astronomico

\noindent
Sheth, K. et al. 2008, \apj, 675, 1141

\noindent
Sheth, K. et al. 2009, Spitzer Proposal ID \#60007

\noindent
Toomre, A., 1969, \apj, 158, 899

\noindent
Toomre, A., 1981, in The Structure and Evolution of Normal Galaxies, S. M. Fall,
ed., Cambridge, Cambridge University press, p. 111

\noindent
Verley, S., Combes, F., Verdes-Montenegro, L., Bergond, G., \& 
Leon, S. 2007, \aap, 474, 43

\noindent
Yuan, C. \& Kuo, C.-L. 1997, \apj, 486, 750

\noindent
Zhang, X. \& Buta, R. 2007, \aj, 133, 2584

\clearpage
\include{t01}
\clearpage
\include{t02}
\clearpage
\include{t03}
\clearpage
\include{t04}
\clearpage
\include{t05}
\clearpage
\include{t06}
\clearpage
\include{t07}
\clearpage
\include{t08}

\clearpage

\clearpage

\clearpage

\begin{figure}
\vspace{-1.0truecm}
\figurenum{18 (cont.)}
\plotone{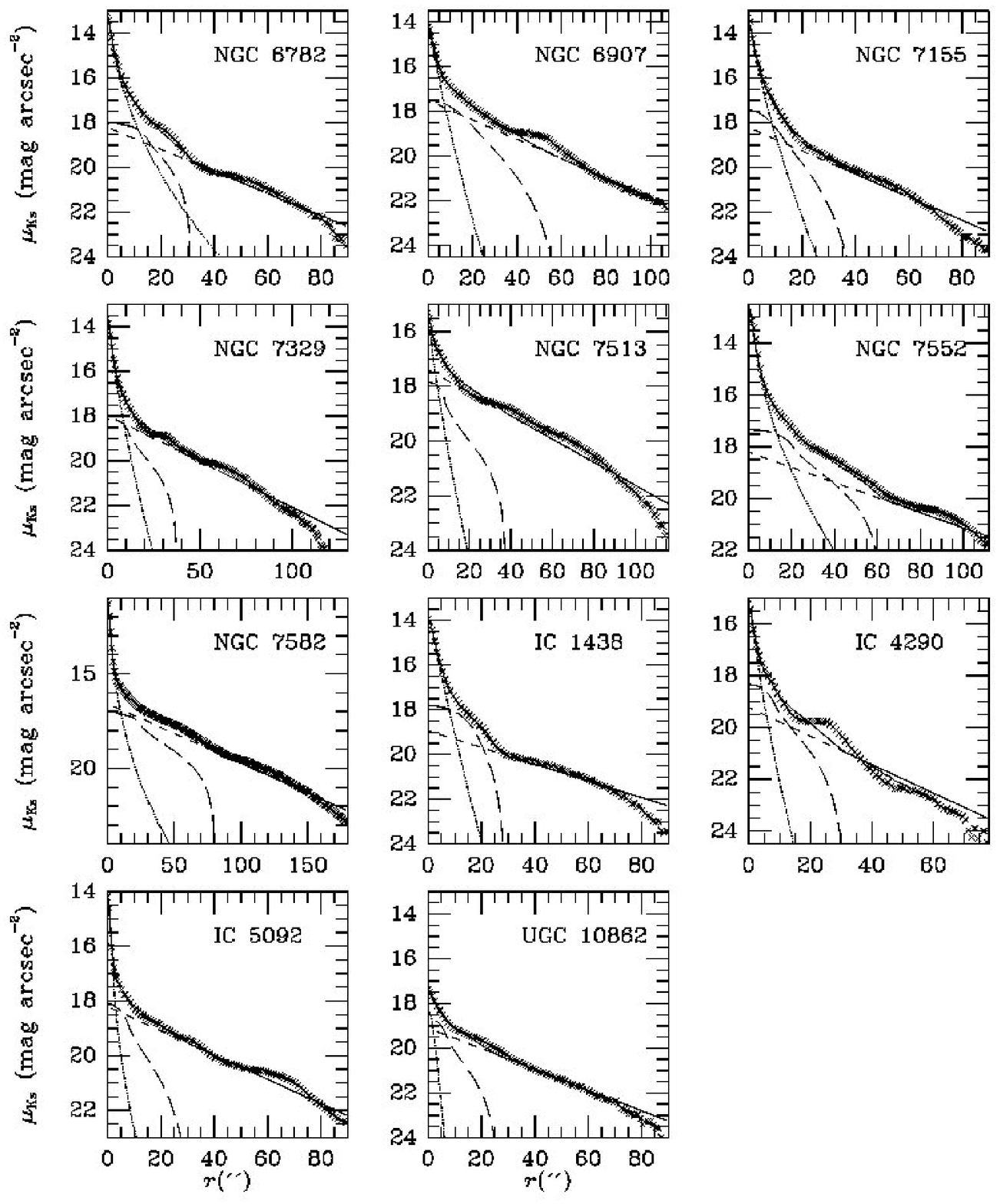}
\end{figure}
\begin{figure}
\figurenum{18 (cont.)}
\caption{Azimuthally-averaged $K_s$-band surface brightness profiles showing
the results of the 2D bulge/disk/bar decomposition fits. The luminosity
distributions have been averaged within fixed ellipses having the orientation
parameters listed in Table 1. Crosses refer to the observed profiles, dotted
curves to the bulge models, short dashed lines to the disk model, long dashed
curves to the bar model, and solid curves to the total model. These models
assume a spherical bulge.}
\end{figure}

\clearpage

\begin{figure}
\figurenum{19}
\plotone{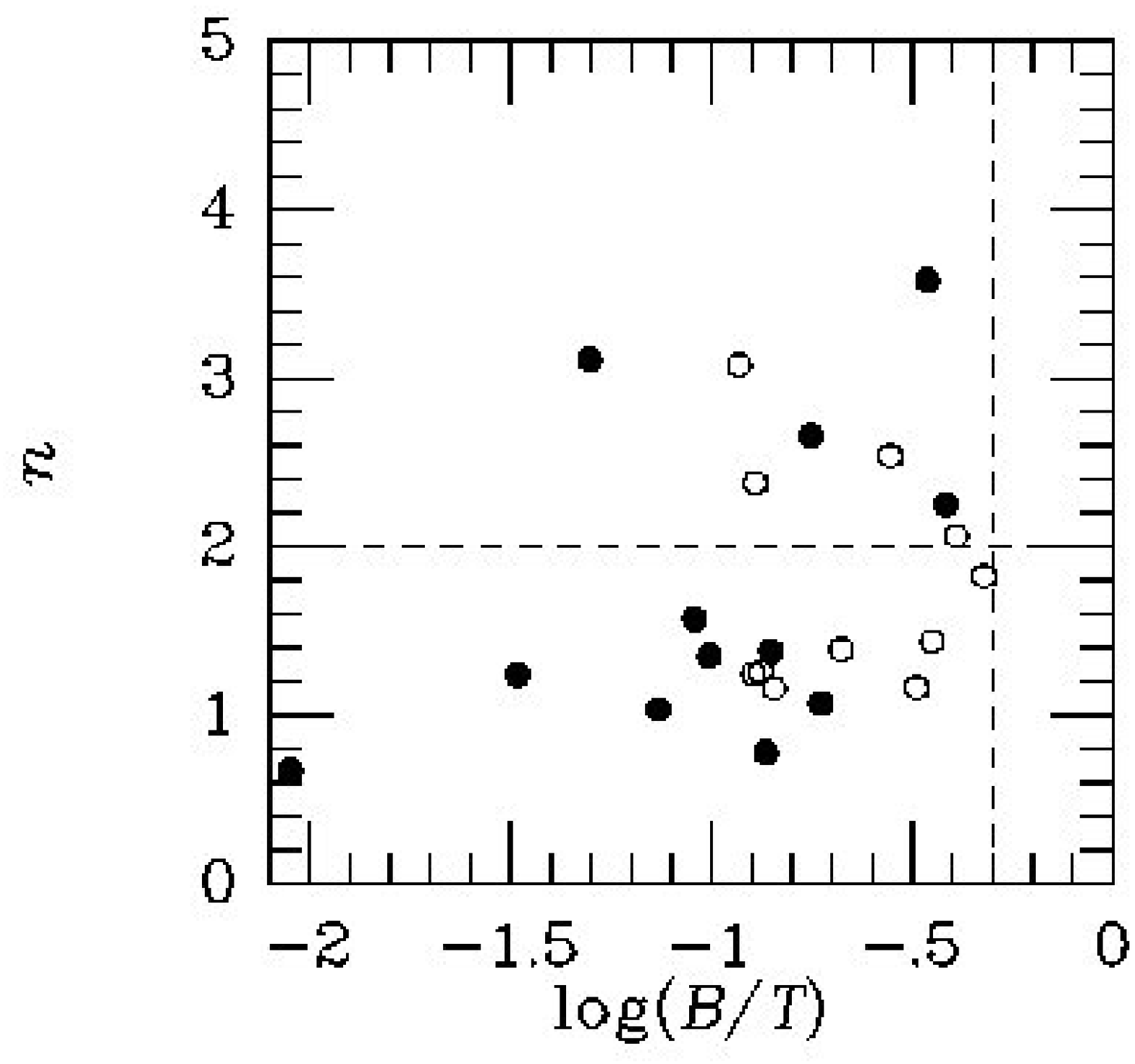}
\vspace{1.0truecm}
\caption{Graph of Sersic index $n$ and bulge-to-total luminosity ratio
$B/T$ for 23 barred galaxies. Filled circles are galaxies having $Q_b$ $\geq$ 0.28,
and open circles are galaxies having $Q_b$ $<$ 0.28.
The dashed lines indicate limits for
classical and pseudobulges from Kormendy \& Kennicutt (2004). Points
having $n$ $<$ 2 and $log (B/T) <$ $-$0.3 are considered pseuodbulges.}
\label{pbulges}
\end{figure}

\clearpage 

\begin{figure}
\figurenum{20}
\plotone{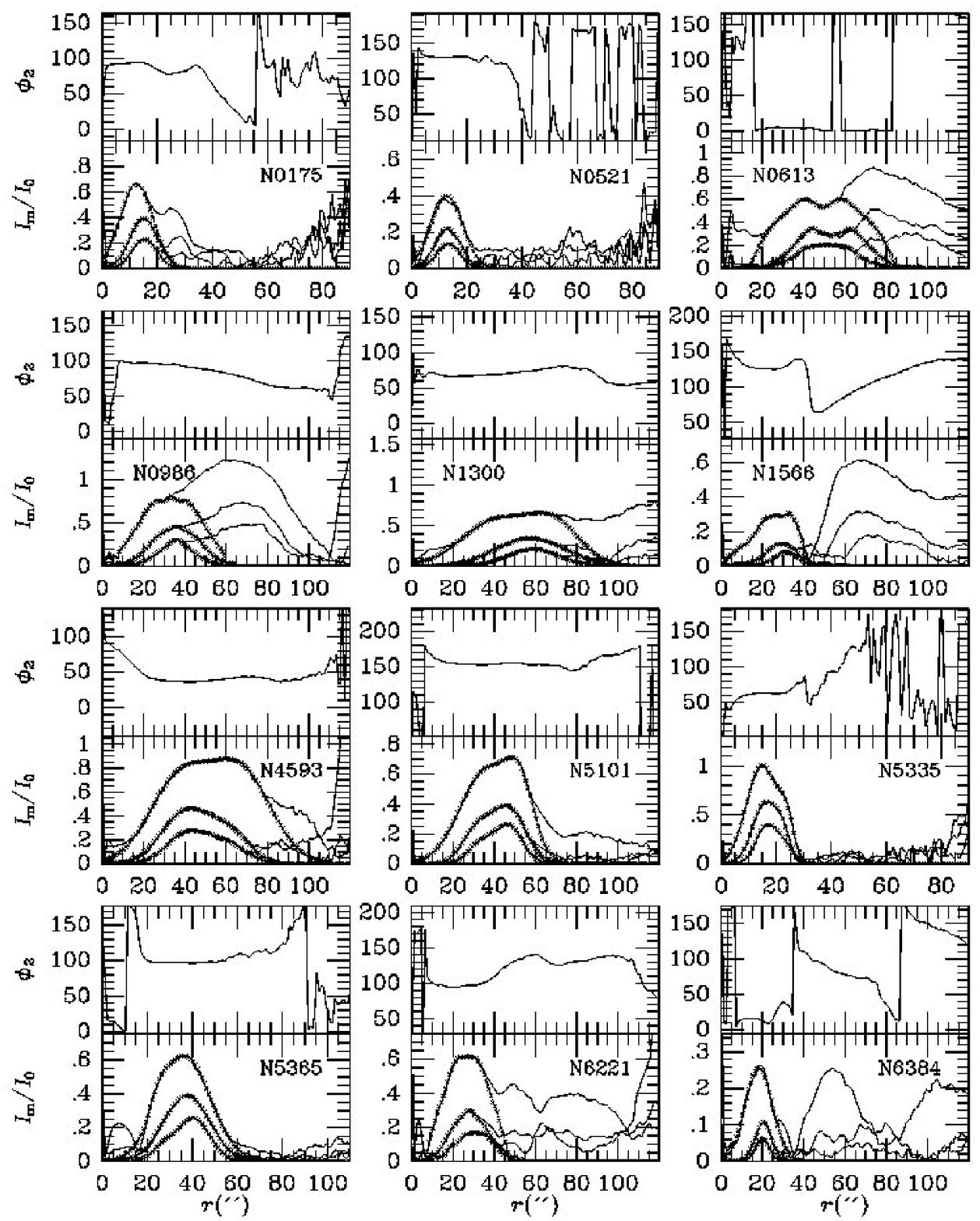}
\caption{}
\label{fourier}
\end{figure}
\begin{figure}
\vspace{-1.0truecm}
\figurenum{20 (cont.)}
\plotone{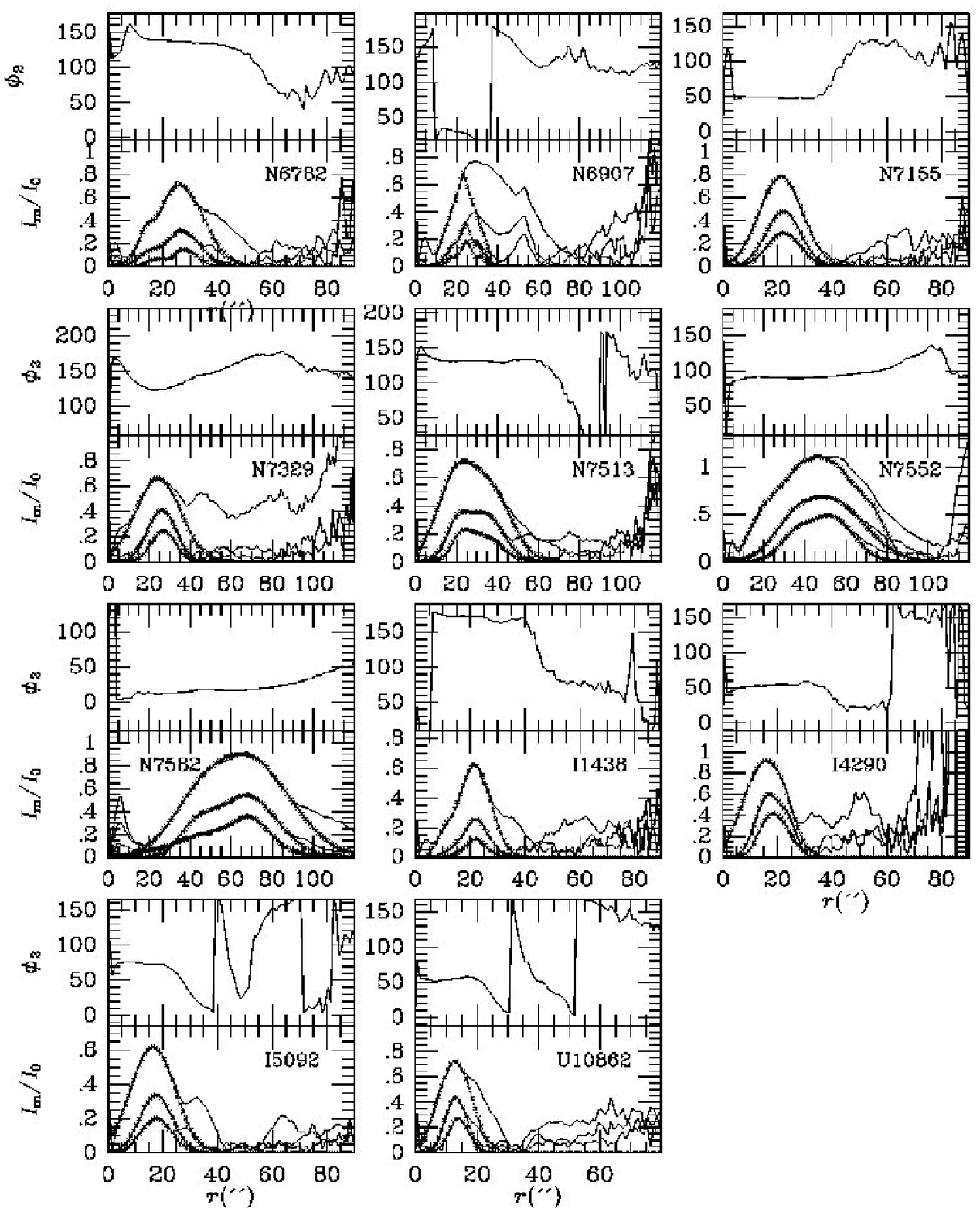}
\figurenum{20 (cont.)}
\caption{Plots of relative Fourier amplitudes $I_m/I_0$ ($m$=2,4,6) and
$m$=2 phase $\phi_2$ (degrees) for 23 galaxies. The crossings show mappings of
the bar used for bar-spiral separation and estimation of bar and spiral
strengths. Some of these mappings are based on the symmetry assumption,
while others are based on single or double gaussian fits.}
\end{figure}

\clearpage

\begin{figure}
\figurenum{21}
\plotone{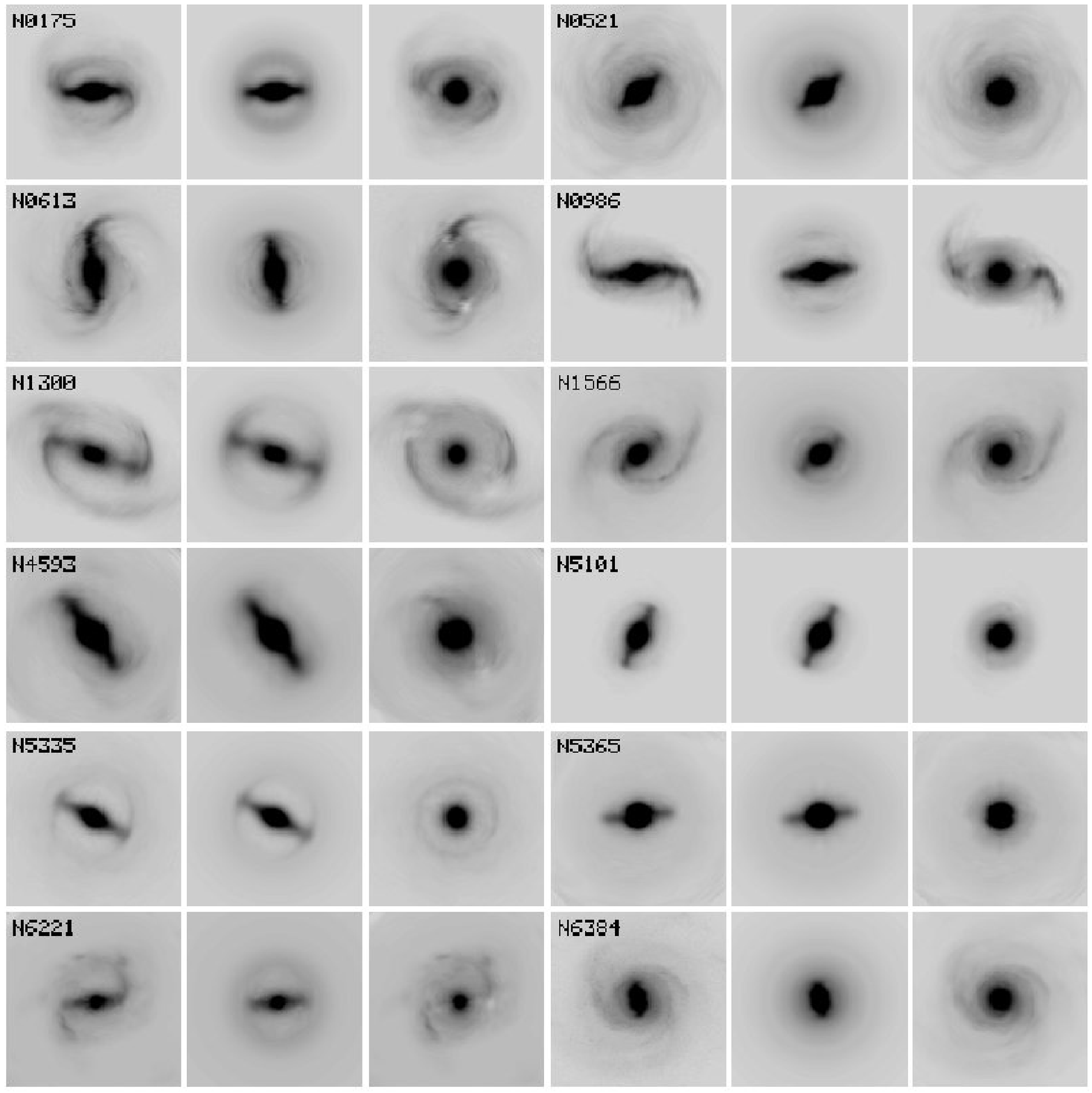}
\caption{}
\label{separations}
\end{figure}
\begin{figure}
\vspace{-1.0truecm}
\figurenum{21 (cont.)}
\plotone{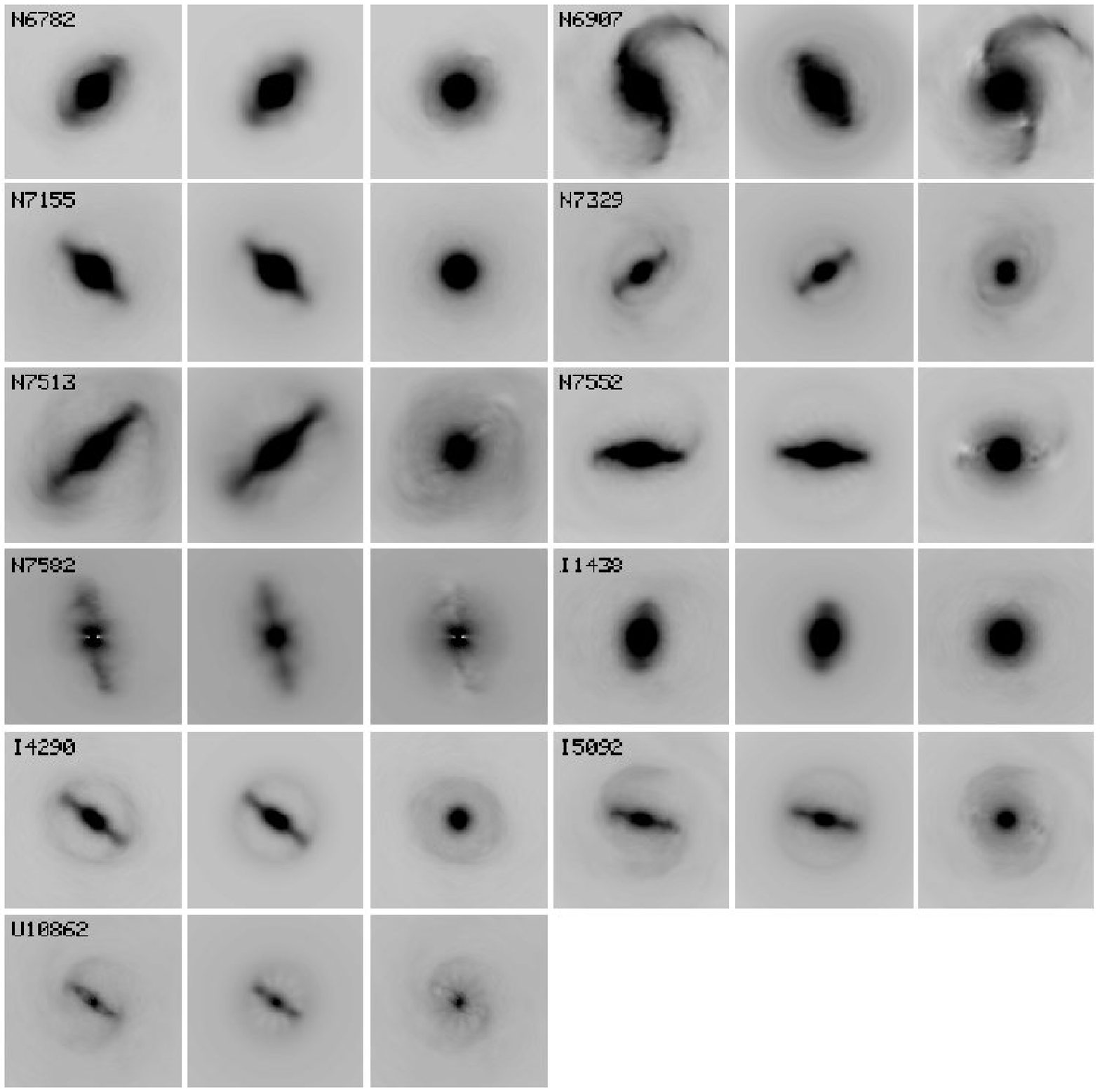}
\figurenum{21 (cont.)}
\caption{Bar-spiral separated images of the sample galaxies. Three images are
shown for each galaxy (left to right): the deprojected $K_s$-band image cleaned
of foreground and background objects (with the galaxy name at upper left), 
the bar plus disk image based on the
mapping given in Figure~\ref{fourier}, and the spiral plus disk image after
bar removal. The images focus mainly on the inner regions and are in linear
intensity units.}
\end{figure}

\begin{figure}
\vspace{2.0truecm}
\figurenum{22}
\plotone{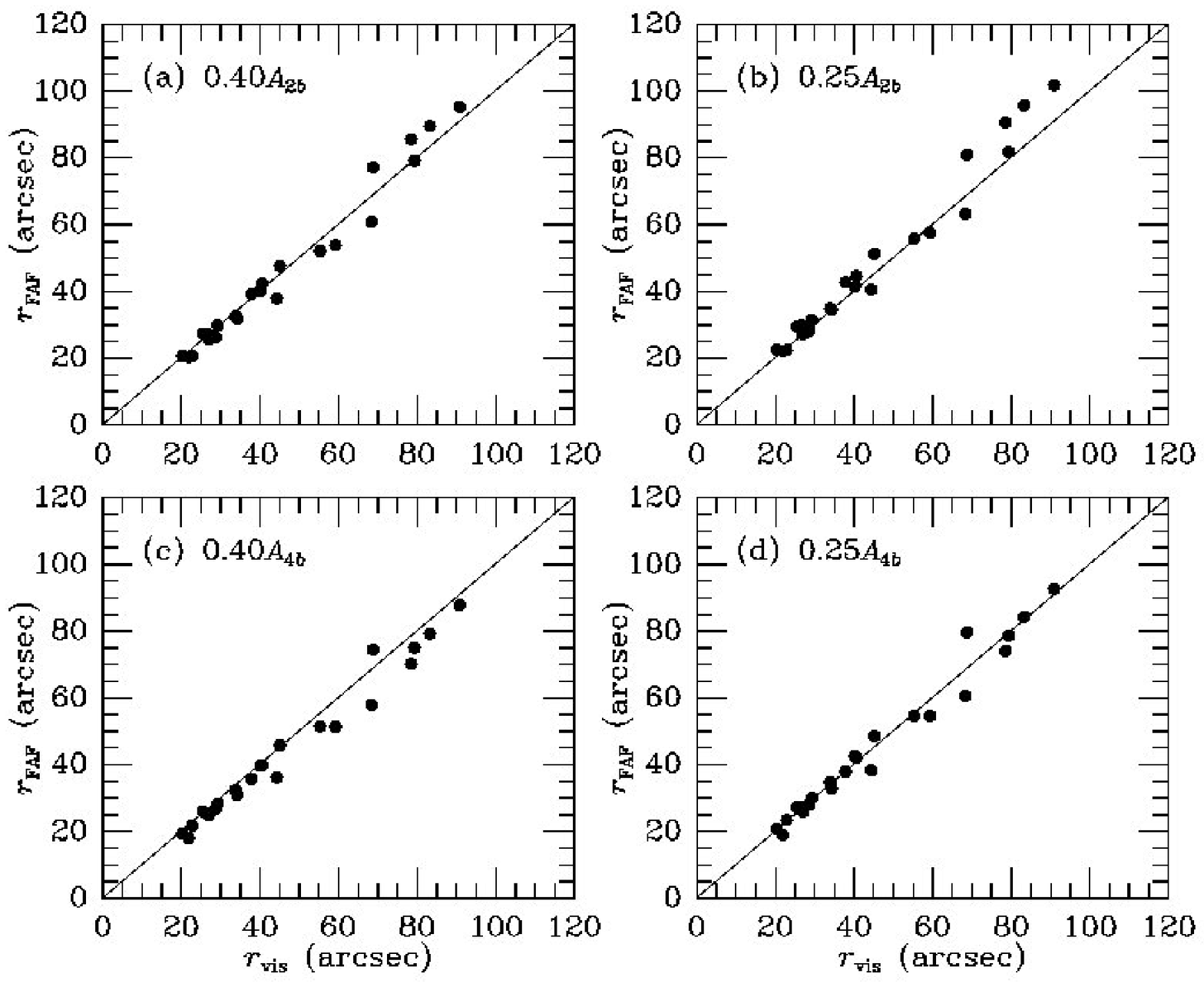}
\caption{Graphs of Fourier-estimated bar radii versus visual
bar radii for 23 barred galaxies. The radii $r_{FAF}$ are ``Fourier
amplitude fraction" radii for the indicated fractions of the bar contrast
parameters $A_{2b}$ and $A_{4b}$.}
\label{rbar}
\end{figure}

\begin{figure}
\vspace{2.0truecm}
\figurenum{23}
\plotone{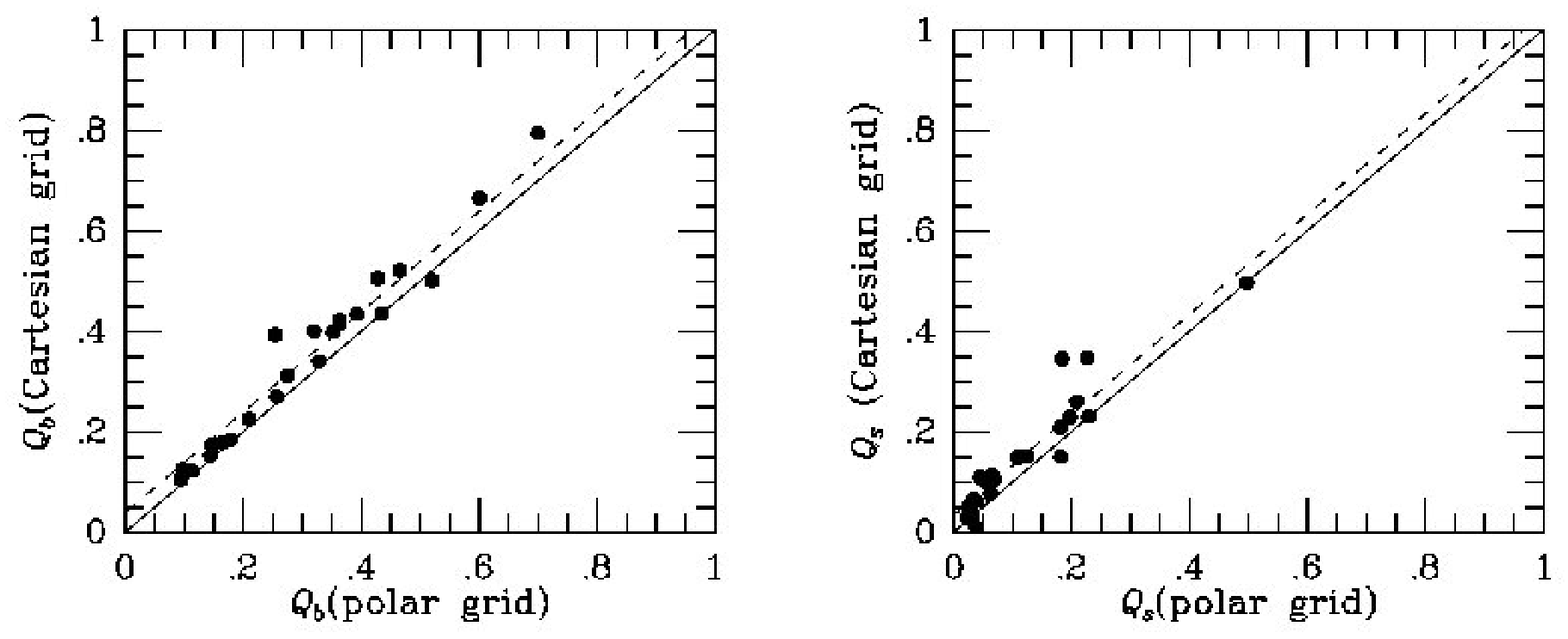}
\caption{A comparison of bar and spiral strengths based on potentials
derived from the Cartesian method (Quillen et al. 1994) and the polar
grid method (Laurikainen \& Salo 2002). The solid lines are for perfect
correlation while the dashed lines allow for average offsets. The differences
are not due to the integration method used but are most likely related to
small differences in the treatment of the bulge and the vertical scale-height,
as well as the mappings of the maxima in different quadrants which are
independent of the potential calculations.}
\label{compare}
\end{figure}

\begin{figure}
\vspace{2.0truecm}
\figurenum{24}
\plotone{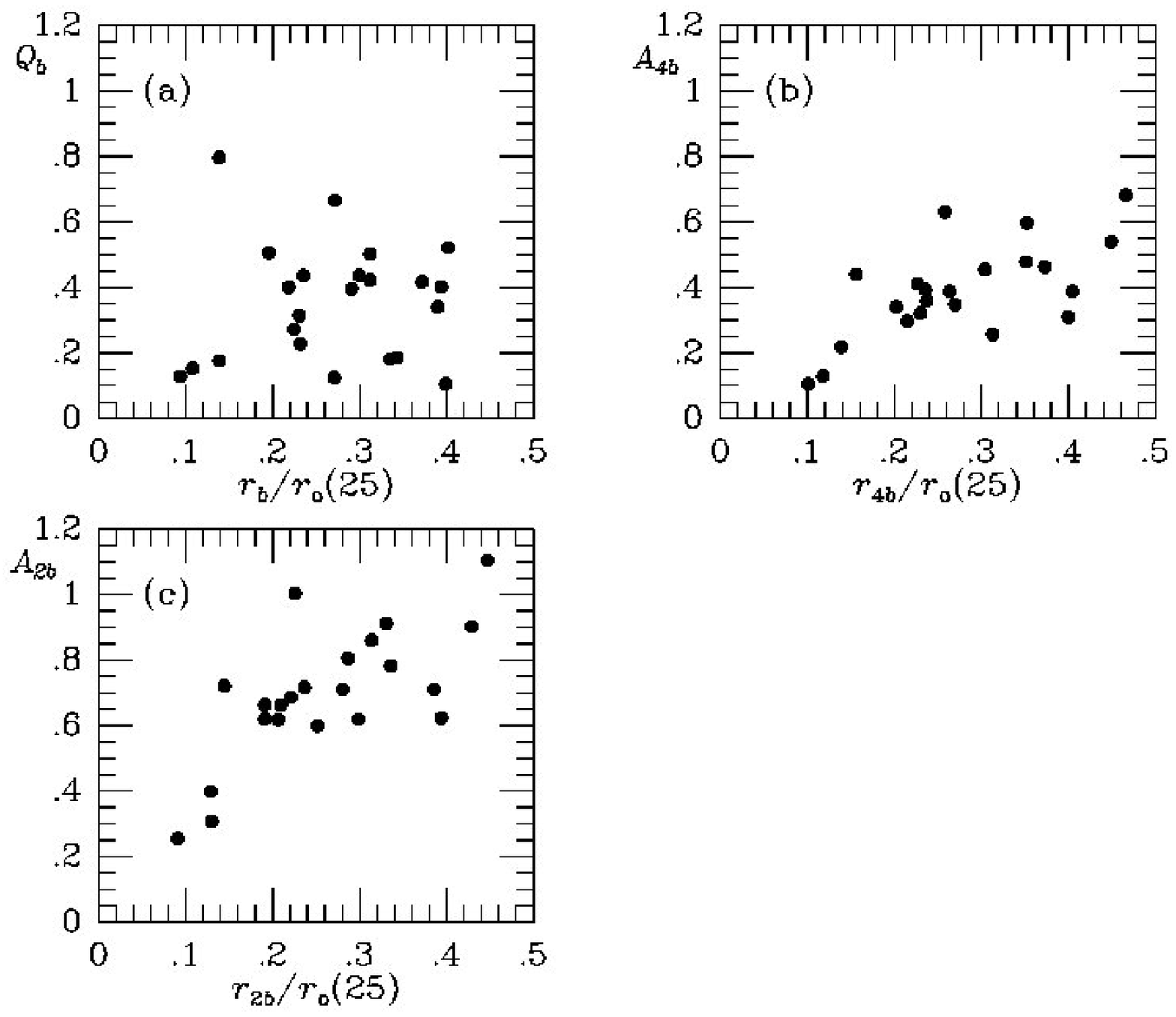}
\caption{Graphs of bar strength and contrast with normalized radii for 23 barred galaxies.}
\label{strengths}
\end{figure}

\begin{figure}
\vspace{2.0truecm}
\figurenum{25}
\plotone{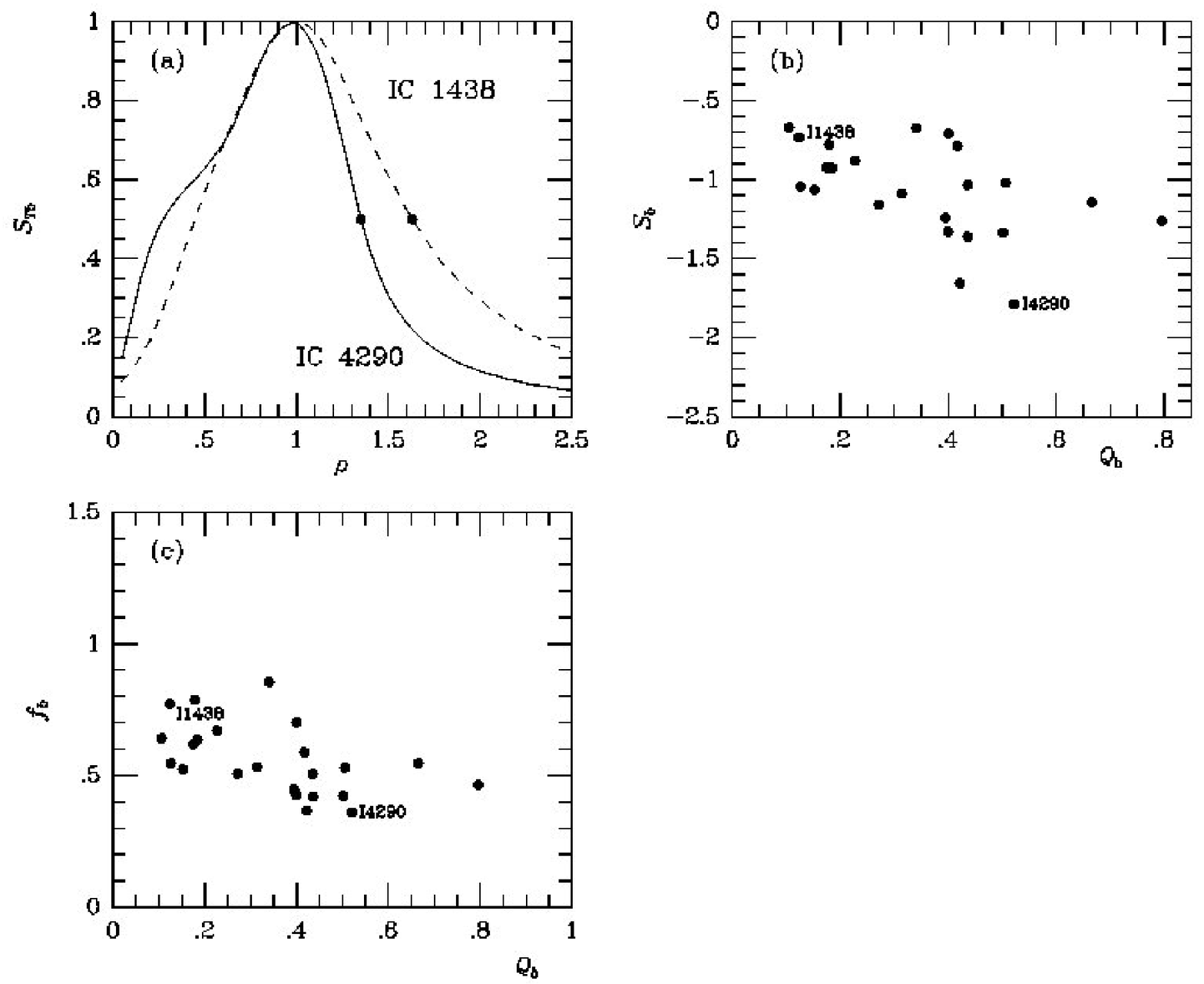}
\caption{(a) Normalized bar torque profiles for the weakly-barred
galaxy IC 1438 and the strongly-barred galaxy IC 4290. The parameters
are $S_{Tb}=Q_{Tb}(r)/Q_b$ and $\rho=r/r_b$. (b) Plot of slope
$dS_{Tb}/d\rho$ at $S_{Tb}$=0.5 (filled circles in (a)). (c) A graph of
the relative bar-end drop-off fraction defined as
$f_b={(r(0.25Q_b)-r(0.75Q_b))\over r_b}$, versus the bar strength
$Q_b$. In (b) and (c), IC 1438 and IC 4290 are indicated.} 
\label{writeqb} 
\end{figure}

\begin{figure}
\vspace{2.0truecm}
\figurenum{26}
\plotone{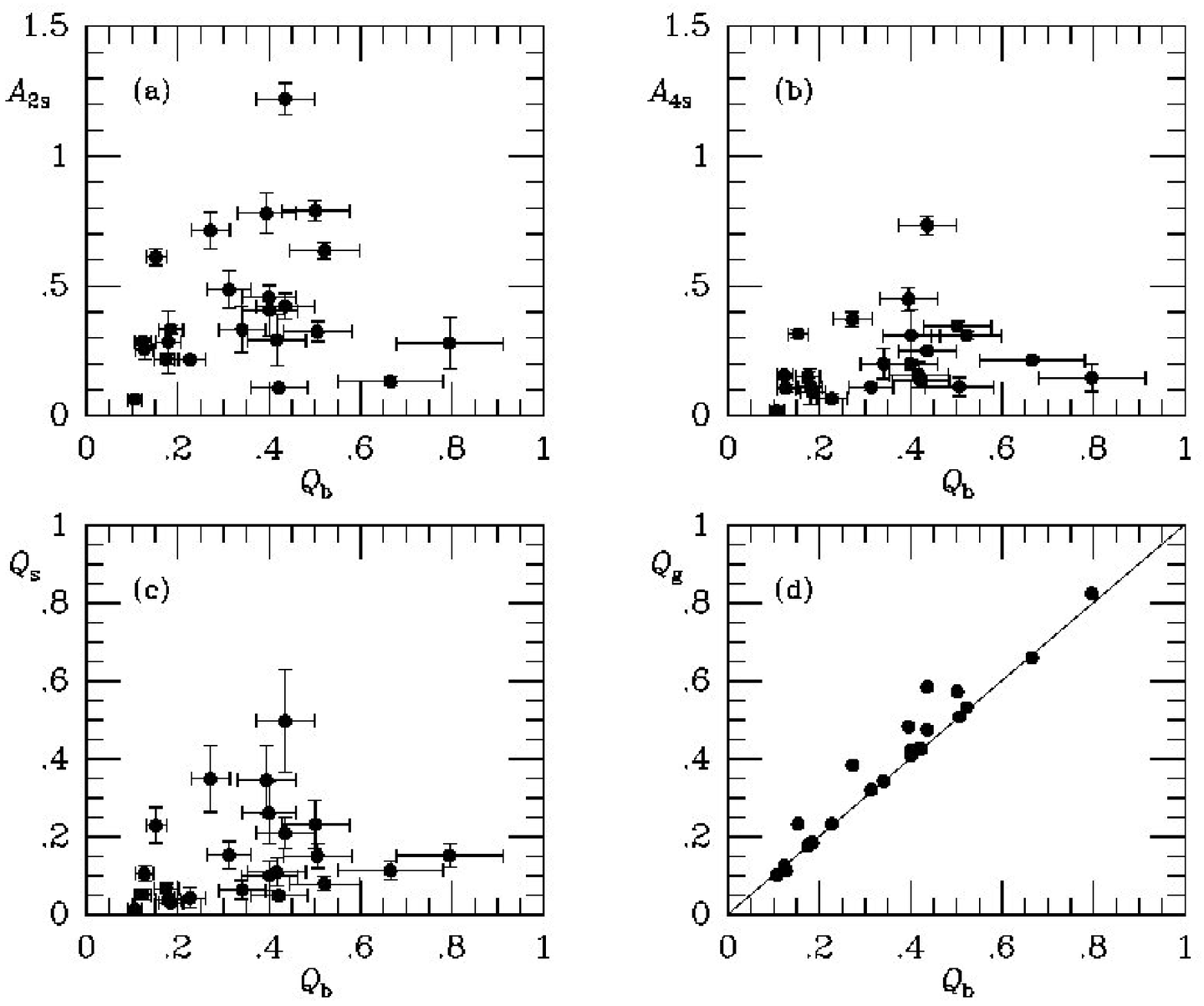}
\caption{Graphs of (a) $m$=2 spiral contrast $A_{2s}$, (b)
$m$=4 spiral contrast $A_{4s}$, (c) spiral strength $Q_s$,
and (d) maximum total relative gravitational torque strength $Q_g$,
versus bar strength $Q_b$.}
\label{newqbqs}
\end{figure}

\begin{figure}
\vspace{2.0truecm}
\figurenum{27}
\plotone{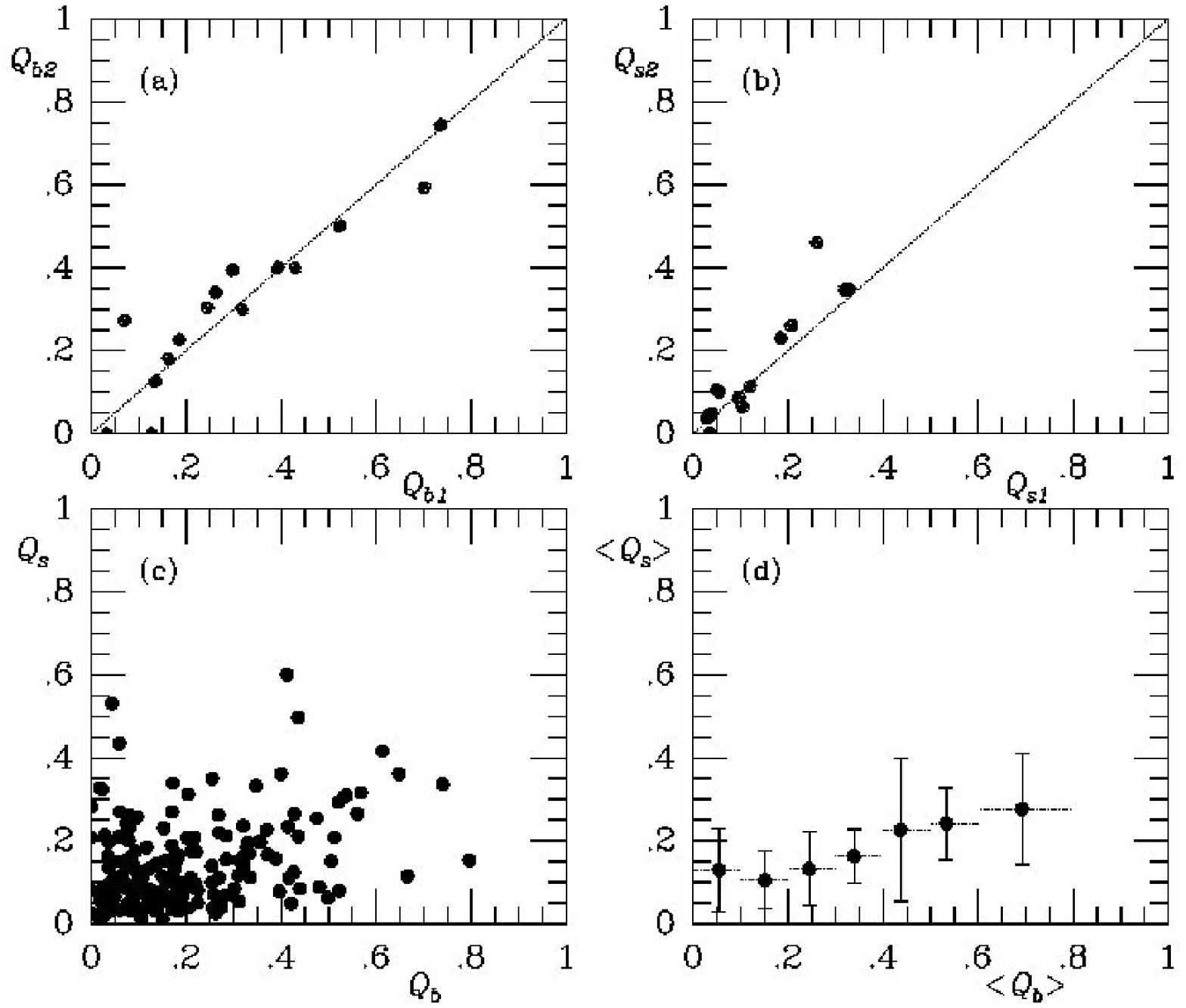}
\caption{
(a) Comparison between $Q_b$ values from different studies.  The
$Q_{b1}$ are values from Buta et al. (2005), while the $Q_{b2}$ are
values from this paper, Block et al. (2004), and Buta (2004). The solid
line is for unit slope.
(b) Comparison between $Q_s$ values from different studies.  The
$Q_{s1}$ are values from Buta et al. (2005), while the $Q_{s2}$ are
values from this paper, Block et al. (2004), and Buta (2004). The solid
line is for unit slope.
(c) Graph of spiral strength $Q_s$ versus bar strength $Q_b$ for 
a combined sample of 177 galaxies including:
the present AAT sample,
the Block et al. (2004) sample, the Buta et al. (2005) sample, and a small
sample of early-type spirals from Buta (2004).
(c) Means and standard deviations of $Q_s$ in bins of 0.1 in $Q_b$.
The horizontal dashed lines indicate the $Q_b$ bins within which $Q_s$ is averaged.
$<Q_b>$ is the average of $Q_b$ only within these bins and not at a given $Q_s$.}
\label{newqbmeanqs}
\end{figure}

\begin{figure}
\vspace{2.0truecm}
\figurenum{28}
\plotone{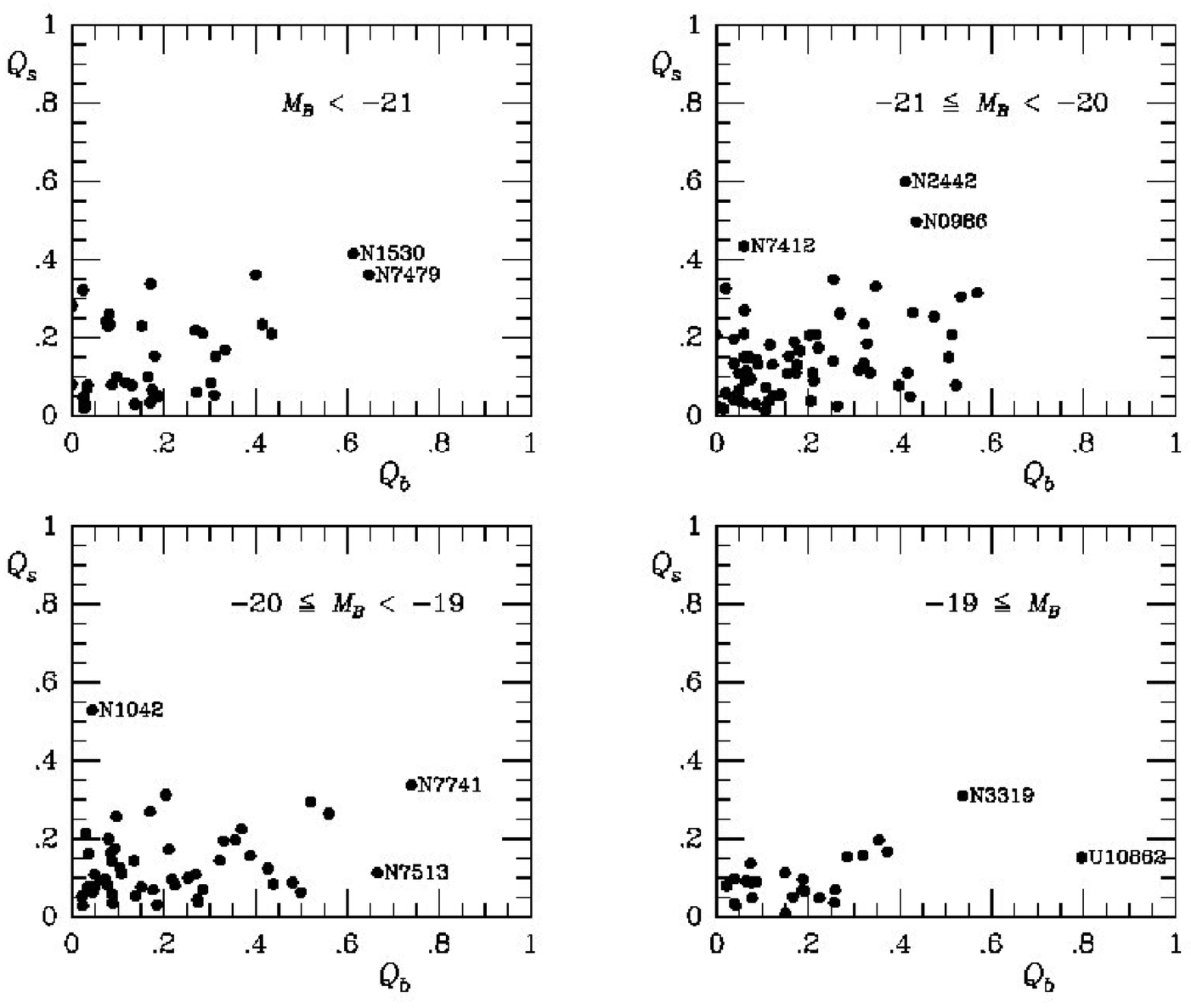}
\caption{Graphs of spiral strength $Q_s$ versus bar strength $Q_b$ for 
a combined sample of 177 galaxies subdivided according to absolute blue
magnitude. Several outliers are labeled.}
\label{revabsmag}
\end{figure}

\begin{figure}
\vspace{2.0truecm}
\figurenum{29}
\plotone{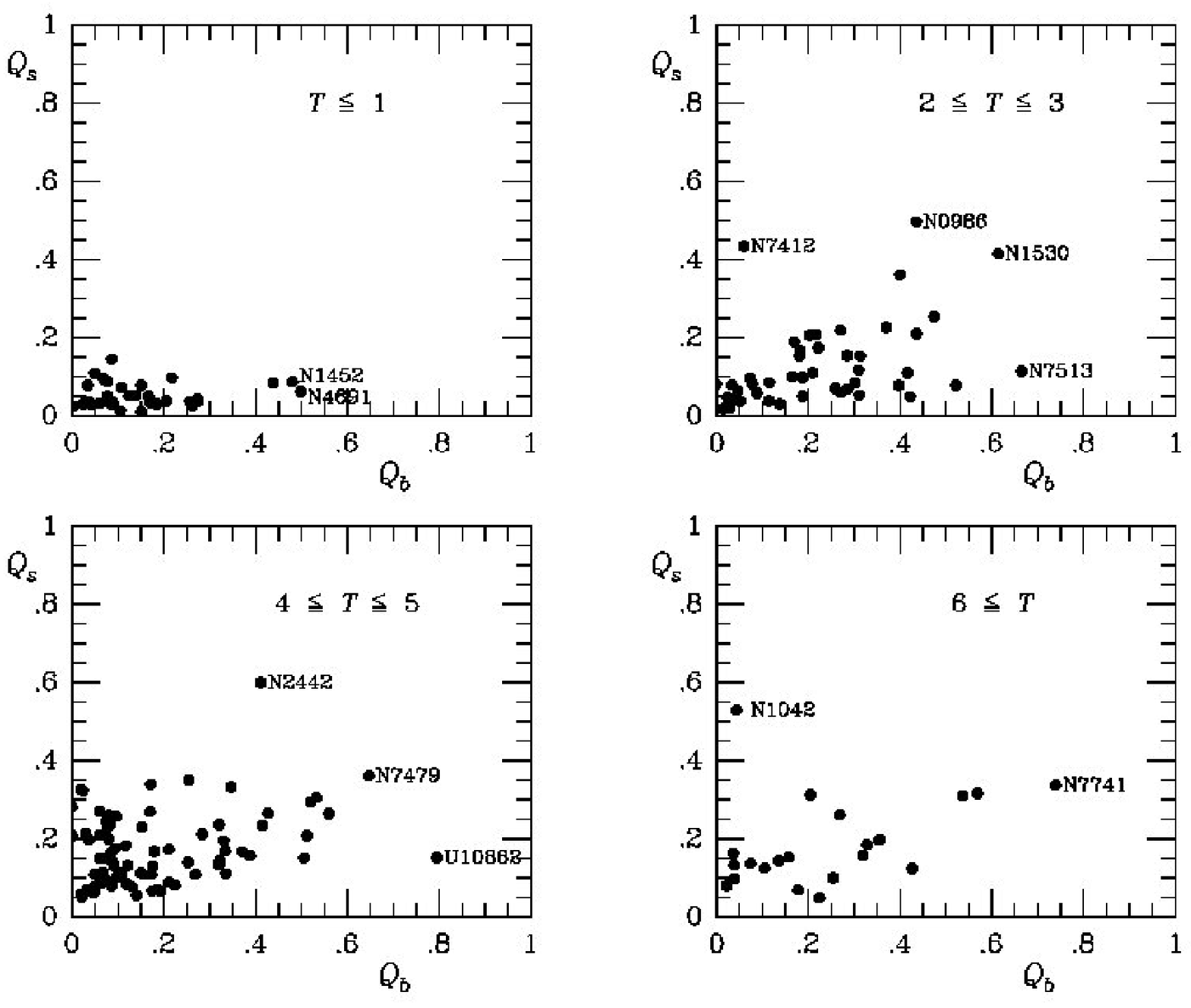}
\caption{Graphs of spiral strength $Q_s$ versus bar strength $Q_b$ for 
a combined sample of 177 galaxies subdivided according to RC3 stage
index. Several outliers are labeled.}
\label{revrc3t}
\end{figure}

\end{document}

%% file: t01.tex
\makeatletter
\def\jnl@aj{AJ}
\ifx\revtex@jnl\jnl@aj\let\tablebreak=\nl\fi
\makeatother
 
\begin{deluxetable}{llcrrrcl}
\tablenum{1}
\tablewidth{40pc}
\tablecaption{Revised Classifications and Orientation Parameters\tablenotemark{a}}
\tablehead{
\colhead{Galaxy} &
\colhead{Type} &
\colhead{$<q>$}&
\colhead{$<\phi>$} &
\colhead{$<\phi>$} &
\colhead{range} &
\colhead{FWHM} &
\colhead{Ori.} 
\\
\colhead{} &
\colhead{} &
\colhead{disk}&
\colhead{disk} &
\colhead{bar} &
\colhead{($^{\prime\prime}$)} &
\colhead{(pix)} &
\colhead{pars.} 
\\
\colhead{1} &
\colhead{2} &
\colhead{3} &
\colhead{4} &
\colhead{5} &
\colhead{6} &
\colhead{7} &
\colhead{8} 
}
\startdata
NGC  175 & SB($\underline{\rm r}$s)ab                     &  0.965$\pm$0.002  &  32.5$\pm$1.5 &125.1 &  54- 74   &3.06  &  $K_s$  \\
NGC  521 & SB($\underline{\rm r}$s)bc                     &  0.980$\pm$0.002  &  25.8$\pm$7.3 &157.3 &  94-111   &2.47  &  $B$   \\
NGC  613 & SB(rs)bc                                       &  0.749$\pm$0.003  & 121.5$\pm$0.4 &122.8 & 150-205   &2.59  &  $B$   \\
NGC  986 & (R$_1^{\prime}$)SB(rs)b                        &  0.822$\pm$0.001  & 141.6$\pm$2.0 & 54.8 & 111-123   &2.98  &  $R$   \\
NGC 1300 & SB(s)b                                         &  0.849$\pm$0.019  & 117.2$\pm$1.2 &106.6 & 185-195   &2.85  &  $B$   \\
NGC 1566 & (R$_1^{\prime}$)SAB(s)bc                    &  0.887$\pm$0.004  & 49.2$\pm$0.8 & 17.2, 2.7 & 117-153   &3.29  &  $B$   \\
NGC 4593 & (R$^{\prime}$)SB(rs)ab                         &  0.737$\pm$0.004  &  99.5$\pm$0.5 & 54.2 & 117-127   &3.21  &  $B$   \\
NGC 5101 & (R$_1$R$_2^{\prime}$)SB($\underline{\rm r}$s)a &  0.929$\pm$0.003  & 145.0$\pm$0.5 &121.4 & 164-184   &2.95  &  $B$   \\
NGC 5335 & SB(r)b                                         &  0.844$\pm$0.003  &  95.4$\pm$0.6 &152.7 &  51- 67   &3.76  &  $K_s$ \\
NGC 5365 & (R)SB0$^-$                                     &  0.583$\pm$0.002  &   6.8$\pm$0.4 &112.0 &  85-105   &2.60  &  $K_s$ \\
NGC 6221 & SB(s)bc pec                                    &  0.665$\pm$0.009  &  12.4$\pm$0.3 &113.9 & 110-164   &2.64  &  $B$   \\
NGC 6384 & SAB(r)bc                                       &  0.605$\pm$0.003  &  30.6$\pm$0.3 & 35.9 & 230-261   &2.72  &  $B$   \\
NGC 6782 & (R$_1$R$_2^{\prime}$)SB(r)a                    &  0.894$\pm$0.002  &  34.3$\pm$0.5 &177.9 &  70 -89   &2.63  &  $B$   \\
NGC 6907 & SAB(s)bc                                       &  0.837$\pm$0.003  &  69.5$\pm$0.6 & 93.8 &  87-106   &4.10  &  $B$   \\
NGC 7155 & SB(r)0$^o$                                     &  0.950$\pm$0.006  &  49.9$\pm$5.3 & 95.9 &  77- 88   &2.78  &  $K_s$ \\
NGC 7329 & SB(r)b                                         &  0.775$\pm$0.001  & 119.0$\pm$0.1 & 76.0 & 132-140   &2.51  &  $B$   \\
NGC 7513 & SB(s)b                                         &  0.675$\pm$0.020  & 104.6$\pm$0.3 & 70.8 &  74-104   &2.80  &  $K_s$ \\
NGC 7552 & (R$_1^{\prime}$)SB(s)ab                        &  0.910$\pm$0.008  & 184.7$\pm$3.5 & 92.9 & 102-124   &4.31  &  $B$   \\
NGC 7582 & (R$_1^{\prime}$)SB(s)ab                        &  0.446$\pm$0.002  & 150.4$\pm$0.1 &156.1 & 189-219   &3.18  &  $B$   \\
IC  1438 & (R$_1$R$_2^{\prime}$)SAB(r)a                   &  0.862$\pm$0.002  & 128.6$\pm$0.9 &123.0  &  90-100  &4.83  &  opt   \\
IC  4290\tablenotemark{b} & (R$^{\prime}$)SB(r)a          &  0.906            &  48.4         & 97.6  &          &2.45  &  opt, kin\\
IC  5092 & (R)SB(s)c                                      &  0.906$\pm$0.004  &  32.3$\pm$0.7 &106.3  &  73 -88  &2.62  &  $K_s$  \\
UGC 10862& SB(rs)c                                        &  0.920$\pm$0.003  & 164.9$\pm$2.0 & 35.8  &  82 -92  &3.28  &  $K_s$  \\
\enddata
\tablenotetext{a}{Col. 1: galaxy name; 2: classification either from the de Vaucouleurs Atlas of Galaxies (Buta et al.
2007) or in the same system by R. Buta;
3: mean disk axis ratio and mean error of ellipse
fits to isophotes, based on $B$ or near-IR images as indicated in col. 8; 4: mean disk position angle (degrees) based on same
ellipse fits, in frame of the $K_s$-band image.
J2000 position angles may be derived
as $\phi$(disk,J2000)=$<\phi>$(disk)+0\rlap{.}$^{\circ}$58;
5: bar position angle (degrees) based on ellipse fits to bar isophotes on $K_s$-band 
image, in the frame of the same image. J2000 position angles may be derived as $\phi$(bar,J2000)=$<\phi>$(bar)+0\rlap{.}$^{\circ}$58; 6: mean FWHM of
stellar profile in pixels (1 pix=0\rlap{.}$^{\prime\prime}$447) on image; 
7: range
in arcsecs used to get $<q>$ and $<\phi>$; 8: bandpass used for orientation parameters, often a
deeper $B$-band image, available mainly from the OSUBGS, NED, or our unpublished image library. "Opt" means based on
several optical filters, "kin" means based partly on kinematic parameters as well.}
\tablenotetext{b}{Orientation parameters from Buta et al. 1998, AJ, 116 1142.}
\end{deluxetable}

%% file: t02.tex
\makeatletter
\def\jnl@aj{AJ}
\ifx\revtex@jnl\jnl@aj\let\tablebreak=\nl\fi
\makeatother
 
\begin{deluxetable}{lcrccr}
\tablenum{2}
\tablewidth{25pc}
\tablecaption{Absolute Magnitudes and Angular Sizes\tablenotemark{a}}
\tablehead{
\colhead{Galaxy} &
\colhead{$\mu_o$} &
\colhead{$(K_s)_T^o$} &
\colhead{$M_{K_s}^o$}&
\colhead{$M_B^o$} &
\colhead{$r_o(25)$} 
\\
\colhead{} &
\colhead{} &
\colhead{} &
\colhead{} &
\colhead{} &
\colhead{($^{\prime\prime}$)} 
\\
\colhead{1} &
\colhead{2} &
\colhead{3} &
\colhead{4} &
\colhead{5} &
\colhead{6} 
}
\startdata
NGC  175  & 33.66 &  9.21 &  $-$24.5 &  $-$21.0 &    65.6  \\
NGC  521  & 34.21 &  8.57 &  $-$25.6 &  $-$21.9 &    97.1  \\
NGC  613  & 31.48 &  7.02 &  $-$24.5 &  $-$21.0 &   164.9  \\
NGC  986  & 32.05 &  7.77 &  $-$24.3 &  $-$20.6 &   116.7  \\
NGC 1300  & 31.57 &  7.55 &  $-$24.0 &  $-$20.8 &   189.3  \\
NGC 1566  & 31.30 &  6.88 &  $-$24.4 &  $-$21.1 &   249.5  \\
NGC 4593  & 32.75 &  7.98 &  $-$24.8 &  $-$21.3 &   116.7  \\
NGC 5101  & 31.88 &  7.13 &  $-$24.8 &  $-$20.7 &   172.6  \\
NGC 5335  & 34.00 & 10.08 &  $-$23.9 &  $-$20.7 &    64.1  \\
NGC 5365  & 32.50 &  7.89 &  $-$24.6 &  $-$20.5 &    92.7  \\
NGC 6221  & 31.40 &  7.06 &  $-$24.3 &  $-$21.6 &   128.0  \\
NGC 6384  & 31.94 &  7.48 &  $-$24.5 &  $-$21.3 &   202.8  \\
NGC 6782  & 33.61 &  8.85 &  $-$24.8 &  $-$21.4 &    68.7  \\
NGC 6907  & 33.25 &  8.28 &  $-$25.0 &  $-$21.9 &   106.4  \\
NGC 7155  & 32.13 &  8.96 &  $-$23.2 &  $-$19.2 &    64.1  \\
NGC 7329  & 33.17 &  8.86 &  $-$24.3 &  $-$21.1 &   116.7  \\
NGC 7513  & 31.70 &  8.94 &  $-$22.8 &  $-$19.4 &    99.3  \\
NGC 7552  & 31.68 &  7.53 &  $-$24.1 &  $-$20.5 &   104.0  \\
NGC 7582  & 31.64 &  7.31 &  $-$24.3 &  $-$20.8 &   150.4  \\
IC  1438  & 32.82 &  9.28 &  $-$23.5 &  $-$20.4 &    72.0  \\
IC  4290  & 34.04 & 10.31 &  $-$23.7 &  $-$20.2 &    49.8  \\
IC  5092  & 33.18 &  9.40 &  $-$23.8 &  $-$20.6 &    86.5  \\
UGC 10862 & 31.97 & 11.66 &  $-$20.3 &  $-$18.9 &    86.5  \\
\enddata
\tablenotetext{a}{Col. 1: galaxy name; 2: Galactic standard of rest (GSR) distance modulus
from NED; 3: total $K_s$$-$band magnitude corrected for Galactic extinction (NED);
4: absolute $K_s$-band magnitude; 5: absolute $B$-band magnitude based on
$B_T^o$ from RC3 and the NED distance modulus in col. 2; 6: $r_o(25)=D_o/2$ is the
radius of the $\mu_B$=25.0 mag arcsec$^{-2}$ isophote, corrected for Galactic
extinction (from RC3).}
\end{deluxetable}

%% file: t03.tex
\makeatletter
\def\jnl@aj{AJ}
\ifx\revtex@jnl\jnl@aj\let\tablebreak=\nl\fi
\makeatother
 
\begin{deluxetable}{lcccccccccccc}
\tablenum{3}
\tablewidth{45pc}
\tablecaption{2D Decomposition Parameters\tablenotemark{a}}
\tablehead{
\colhead{Galaxy} &
\colhead{$\mu(0)$} &
\colhead{$r_b$} &
\colhead{$\beta$} &
\colhead{$\mu$(0)} &
\colhead{$h_r$}&
\colhead{$a_{bar}$} &
\colhead{$b_{bar}$} &
\colhead{$\mu$(0)} &
\colhead{$\phi_{bar}$} &
\colhead{$n_{bar}$} &
\colhead{$B/T$} &
\colhead{$bar/T$} 
\\
\colhead{} &
\colhead{bulge} &
\colhead{$^{\prime\prime}$} &
\colhead{} &
\colhead{disk} &
\colhead{$^{\prime\prime}$} &
\colhead{$^{\prime\prime}$} &
\colhead{$^{\prime\prime}$} &
\colhead{bar}    &
\colhead{deg} &
\colhead{}    &
\colhead{}    &
\colhead{}    
\\
\colhead{1} &
\colhead{2} &
\colhead{3} &
\colhead{4} &
\colhead{5} &
\colhead{6} &
\colhead{7} & 
\colhead{8} & 
\colhead{9} & 
\colhead{10} &
\colhead{11} &
\colhead{12} &
\colhead{13} 
}
\startdata
NGC 175  &  14.10 &   0.99 &  0.966 &  17.35 &   15.2 &   25.0 &    5.9 &  17.62 & 182.4 &   0.25 &  0.074 &  0.118  \\
NGC 521  &  13.06 &   0.81 &  0.796 &  17.32 &   18.8 &   26.8 &    8.8 &  17.30 &  40.5 &   3.14 &  0.133 &  0.062  \\
NGC 613  &  12.85 &   1.82 &  0.861 &  17.30 &   36.8 &   84.9 &   22.0 &  17.96 &  92.5 &   0.00 &  0.144 &  0.178  \\
NGC 986  &  13.93 &   2.78 &  1.277 &  18.12 &   34.5 &   59.9 &   23.9 &  17.11 &   4.8 &   2.00 &  0.137 &  0.282  \\
NGC1300  &  12.97 &   1.03 &  0.736 &  18.89 &   76.8 &   89.4 &   30.8 &  18.47 &  71.3 &   1.00 &  0.099 &  0.153  \\
NGC1566  &  10.83 &   0.15 &  0.421 &  16.89 &   42.5 &   53.6 &   21.3 &  17.30 &  38.4 &   1.25 &  0.129 &  0.076  \\
NGC4593  &   8.06 &   0.00 &  0.279 &  18.65 &   38.1 &   76.0 &   24.5 &  18.35 &  38.1 &   0.30 &  0.345 &  0.232  \\
NGC5101  &  11.10 &   0.16 &  0.394 &  18.46 &   58.2 &   67.1 &   28.1 &  17.41 &  64.2 &   0.71 &  0.280 &  0.210  \\
NGC5335  &  14.33 &   1.34 &  0.930 &  18.64 &   20.1 &   29.9 &    8.0 &  18.02 & 152.2 &   0.56 &  0.188 &  0.177  \\
NGC5365  &  11.61 &   0.59 &  0.547 &  19.07 &   61.3 &   78.0 &   28.7 &  17.99 & 186.8 &   2.98 &  0.480 &  0.131  \\
NGC6221  &  11.40 &   0.43 &  0.636 &  16.75 &   30.5 &   46.9 &   11.7 &  17.39 &   7.2 &   0.10 &  0.091 &  0.096  \\
NGC6384  &  11.37 &   0.04 &  0.325 &  17.91 &   45.9 &   39.8 &   17.3 &  17.02 & 100.5 &   4.00 &  0.118 &  0.059  \\
NGC6782  &  11.86 &   0.28 &  0.485 &  18.35 &   22.0 &   31.3 &   14.6 &  18.16 &  47.2 &   0.00 &  0.408 &  0.171  \\
NGC6907  &  13.57 &   1.25 &  0.803 &  17.71 &   25.1 &   63.7 &   25.1 &  17.71 & 102.7 &   2.19 &  0.127 &  0.236  \\
NGC7155  &  12.70 &   0.85 &  0.695 &  18.29 &   21.0 &   44.7 &   15.9 &  17.51 & 137.8 &   2.56 &  0.356 &  0.196  \\
NGC7329  &  12.92 &   0.82 &  0.718 &  18.23 &   26.2 &   38.0 &   11.0 &  18.40 &  38.0 &   0.00 &  0.211 &  0.121  \\
NGC7513  &  15.19 &   1.12 &  0.806 &  17.81 &   25.6 &   38.0 &    7.5 &  18.28 &  43.4 &   0.00 &  0.033 &  0.088  \\
NGC7552  &  10.51 &   0.19 &  0.445 &  18.32 &   37.3 &   62.6 &   20.4 &  17.45 &   0.8 &   0.12 &  0.384 &  0.272  \\
NGC7582  &   7.81 &   0.02 &  0.376 &  17.46 &   34.4 &   80.5 &   24.0 &  17.96 & 105.1 &   0.00 &  0.178 &  0.215  \\
IC 1438   &  13.21 &   1.26 &  0.857 &  19.11 &   29.0 &   29.1 &   19.7 &  17.98 &  81.5 &   0.72 &  0.325 & 0.235  \\
IC 4290   &  14.43 &   0.64 &  0.724 &  19.34 &   19.7 &   31.3 &    8.7 &  18.46 & 142.9 &   0.45 &  0.141 & 0.264  \\
IC 5092   &   9.35 &   0.00 &  0.321 &  18.32 &   24.5 &   31.3 &    8.5 &  18.22 & 164.4 &   0.68 &  0.050 & 0.107  \\
UGC10862 &  17.98 &   1.70 &  1.468 &  19.20 &   23.5 &   26.8 &    7.0 &  18.76 & -36.2 &   0.49 &  0.009 & 0.125  \\
\enddata
\tablenotetext{a}{Col. 1: galaxy name; 2: bulge central surface brightness (mag arcsec$^{-2}$); 3: bulge characteristic
radius; (4) Sersic $\beta$ parameter; (5) disk central surface brightness (mag arcsec$^{-2}$); (6) disk radial
scale length; (7,8): maximum bar major and minor axis radii; (9) bar central surface brightness (mag arcsec$^{-2}$);
(10) bar position angle in galaxy plane relative to line of nodes; (11) bar exponent; 
(12) bulge-to-total luminosity ratio; (13) bar-to-total luminosity ratio}
\end{deluxetable}

%% file: t04.tex
\makeatletter
\def\jnl@aj{AJ}
\ifx\revtex@jnl\jnl@aj\let\tablebreak=\nl\fi
\makeatother
 
\begin{deluxetable}{lcrcccc}
\tablenum{4}
\tablewidth{30pc}
\tablecaption{Relative Bar Parameters and Fourier Component Radii\tablenotemark{a}}
\tablehead{
\colhead{Galaxy} &
\colhead{$A_{2b}$} &
\colhead{$r_{2b}$} &
\colhead{$r_{2b}/r_o(25)$} &
\colhead{$A_{4b}$} &
\colhead{$r_{4b}$} &
\colhead{$r_{4b}/r_o(25)$} 
\\
\colhead{} &
\colhead{} &
\colhead{($^{\prime\prime}$)} &
\colhead{} &
\colhead{} &
\colhead{($^{\prime\prime}$)} &
\colhead{} 
\\
\colhead{1} &
\colhead{2} &
\colhead{3} &
\colhead{4} &
\colhead{5} &
\colhead{6} & 
\colhead{7} 
}
\startdata
NGC 175 & 0.664 &  12.5 & 0.191 & 0.394 &  15.5 & 0.236 \\
NGC 521 & 0.398 &  12.5 & 0.129 & 0.218 &  13.5 & 0.139 \\
NGC 613 & 0.598 &  57.5 & 0.349 & 0.347 &  62.5 & 0.379 \\
NGC 986 & 0.805 &  33.5 & 0.287 & 0.454 &  35.5 & 0.304 \\
NGC1300 & 0.861 &  59.5 & 0.314 & 0.455 &  57.5 & 0.304 \\
NGC1566 & 0.308 &  32.5 & 0.130 & 0.129 &  29.5 & 0.118 \\
NGC4593 & 0.876 &  60.5 & 0.518 & 0.462 &  43.5 & 0.373 \\
NGC5101 & 0.711 &  48.5 & 0.281 & 0.388 &  45.5 & 0.264 \\
NGC5335 & 1.003 &  14.5 & 0.226 & 0.630 &  16.5 & 0.258 \\
NGC5365 & 0.623 &  36.5 & 0.394 & 0.389 &  37.5 & 0.405 \\
NGC6221 & 0.618 &  26.5 & 0.207 & 0.297 &  27.5 & 0.215 \\
NGC6384 & 0.257 &  18.5 & 0.091 & 0.106 &  20.5 & 0.101 \\
NGC6782 & 0.711 &  26.5 & 0.386 & 0.308 &  27.5 & 0.400 \\
NGC6907 & 0.687 &  23.5 & 0.221 & 0.322 &  24.5 & 0.230 \\
NGC7155 & 0.783 &  21.5 & 0.336 & 0.478 &  22.5 & 0.351 \\
NGC7329 & 0.663 &  24.5 & 0.210 & 0.411 &  26.5 & 0.227 \\
NGC7513 & 0.717 &  23.5 & 0.237 & 0.360 &  23.5 & 0.237 \\
NGC7552 & 1.104 &  46.5 & 0.447 & 0.682 &  48.5 & 0.466 \\
NGC7582 & 0.902 &  64.5 & 0.429 & 0.539 &  67.5 & 0.449 \\
IC 1438  & 0.620 &  21.5 & 0.299 & 0.256 &  22.5 & 0.313 \\
IC 4290  & 0.912 &  16.5 & 0.331 & 0.596 &  17.5 & 0.352 \\
IC 5092  & 0.621 &  16.5 & 0.191 & 0.340 &  17.5 & 0.202 \\
UGC 10862 & 0.722 &  12.5 & 0.145 & 0.440 &  13.5 & 0.156 \\
\enddata
\tablenotetext{a}{Col. 1: galaxy name; 2: maximum relative $m$=2 Fourier intensity amplitude
$A_{2b} = (I_2/I_0)_{max}$ of the bar, in the $K_s$ band;
3: radius of $A_{2b}$; 4: ratio of $r_{2b}$ to the radius of the corrected isophotal diameter
$D_o$(25) from RC3 (Table 2); 5-7: same parameters for $m$=4}.
\end{deluxetable}

%% file: t05.tex
\makeatletter
\def\jnl@aj{AJ}
\ifx\revtex@jnl\jnl@aj\let\tablebreak=\nl\fi
\makeatother
 
\begin{deluxetable}{lccccrrrrrrrrc}
\tablenum{5}
\tablewidth{0pc}
\tablecaption{Gaussian Fourier Components for 23 Galaxies\tablenotemark{a}}
\tablehead{
\colhead{Galaxy} &
\colhead{$A_{2i}$} &
\colhead{$A_{4i}$} &
\colhead{$A_{6i}$} &
\colhead{$A_{8i}$} &
\colhead{$r_{2i}$} &
\colhead{$r_{4i}$} &
\colhead{$r_{6i}$} &
\colhead{$r_{8i}$} &
\colhead{$\sigma_{2i}$} &
\colhead{$\sigma_{4i}$} &
\colhead{$\sigma_{6i}$} &
\colhead{$\sigma_{8i}$} 
\\
\colhead{1} &
\colhead{2} &
\colhead{3} &
\colhead{4} &
\colhead{5} &
\colhead{6} &
\colhead{7} &
\colhead{8} &
\colhead{9} &
\colhead{10} &
\colhead{11} &
\colhead{12} &
\colhead{13} 
} 
\startdata
NGC  175   &  0.66 &  0.40 &  0.23 &  0.14 &  12.5 &  15.1 &  15.4 &
15.3 &   5.9 &   5.0 &   4.5 &   4.0 \\
NGC  521   &  0.40 &  0.22 &  0.14 &  0.09 &  12.8 &  13.1 &  13.4 &
13.5 &   5.7 &   4.6 &   4.2 &   3.4 \\
NGC 1300-1 &  0.34 &  0.46 &  0.28 &  0.16 &  34.3 &  57.3 &  58.8 &
59.1 &  11.2 &  16.2 &  12.2 &  12.8 \\
NGC 1300-2 &  0.84 &  0.00 &  0.00 &  0.00 &  62.4 &   0.0 &   0.0 &
0.0 &  20.3 &   0.0 &   0.0 &   0.0 \\
NGC 4593-1 &  0.69 &  0.44 &  0.24 &  0.14 &  37.3 &  41.0 &  40.4 &
40.1 &  14.4 &  11.9 &   9.5 &   8.1 \\
NGC 4593-2 &  0.76 &  0.23 &  0.17 &  0.13 &  67.1 &  62.4 &  59.1 &
58.3 &  14.7 &   9.5 &  10.1 &   8.6 \\
NGC 5101-1 &  0.62 &  0.22 &  0.13 &  0.17 &  36.3 &  31.7 &  33.4 &
43.7 &  12.6 &   7.0 &   6.0 &   8.7 \\
NGC 5101-2 &  0.39 &  0.37 &  0.25 &  0.00 &  52.7 &  47.2 &  47.2 &
0.0 &   6.7 &   8.2 &   6.7 &   0.0 \\
NGC 5335-1 &  1.00 &  0.61 &  0.36 &  0.25 &  14.9 &  16.0 &  16.2 &
18.5 &   5.7 &   4.5 &   3.7 &   4.2 \\
NGC 5335-2 &  0.35 &  0.25 &  0.19 &  0.06 &  23.7 &  22.3 &  22.8 &
25.7 &   2.6 &   2.6 &   3.5 &   1.2 \\
NGC 5365-1 &  0.11 &  0.01 &  0.08 &  0.04 &  25.3 &  25.1 &  27.0 &
27.0 &   3.9 &   3.8 &   4.6 &   4.7 \\
NGC 5365-2 &  0.62 &  0.39 &  0.25 &  0.16 &  36.8 &  37.8 &  40.5 &
41.2 &  11.3 &  10.1 &   7.2 &   8.0 \\
NGC 6384   &  0.26 &  0.11 &  0.06 &  0.03 &  18.4 &  20.6 &  20.1 &
19.9 &   6.6 &   4.0 &   3.3 &   3.3 \\
NGC 6782-1 &  0.12 &  0.09 &  0.06 &  0.03 &  13.5 &  14.1 &  17.4 &
17.5 &   2.2 &   3.5 &   3.0 &   2.9 \\
NGC 6782-2 &  0.71 &  0.31 &  0.15 &  0.07 &  26.5 &  27.3 &  28.3 &
28.4 &   8.5 &   6.6 &   4.4 &   4.2 \\
NGC 7155   &  0.78 &  0.48 &  0.29 &  0.18 &  21.4 &  22.2 &  22.3 &
22.9 &   7.9 &   6.4 &   6.1 &   6.1 \\
NGC 7329-1 &  0.66 &  0.41 &  0.25 &  0.13 &  24.6 &  26.3 &  27.1 &
28.1 &  10.9 &   6.9 &   5.9 &   5.4 \\
NGC 7329-2 &  0.00 &  0.00 &  0.00 &  0.00 &   0.0 &   0.0 &   0.0 &
0.0 &   0.0 &   0.0 &   0.0 &   0.0 \\
NGC 7513-1 &  0.57 &  0.24 &  0.11 &  0.12 &  19.5 &  20.3 &  22.2 &
24.0 &   9.2 &   5.6 &   4.2 &   5.4 \\
NGC 7513-2 &  0.49 &  0.34 &  0.19 &  0.11 &  37.0 &  34.3 &  32.1 &
35.3 &  10.2 &   8.7 &   9.5 &   7.3 \\
NGC 7552-1 &  0.00 &  0.18 &  0.23 &  0.18 &   0.0 &  33.2 &  34.9 &
36.5 &   0.0 &   9.0 &   7.8 &   7.4 \\
NGC 7552-2 &  0.00 &  0.66 &  0.47 &  0.31 &   0.0 &  52.2 &  53.2 &
53.2 &   0.0 &  16.7 &  10.7 &   9.2 \\
NGC 7582-1 &  0.21 &  0.21 &  0.22 &  0.08 &  41.5 &  42.5 &  54.8 &
46.3 &  12.4 &   7.9 &  21.1 &  10.6 \\
NGC 7582-2 &  0.87 &  0.54 &  0.18 &  0.23 &  67.6 &  67.4 &  70.3 &
69.0 &  20.9 &  15.1 &   8.0 &   9.8 \\
IC  1438-1 &  0.62 &  0.26 &  0.12 &  0.05 &  21.7 &  22.2 &  22.3 &
21.9 &   5.9 &   4.6 &   3.9 &   3.4 \\
IC  1438-2 &  0.11 &  0.03 &  0.01 &  0.01 &  10.2 &  12.9 &  16.6 &
12.0 &   2.8 &   3.1 &   5.7 &   3.8 \\
IC  4290-1 &  0.82 &  0.58 &  0.29 &  0.21 &  13.8 &  16.5 &  16.2 &
16.5 &   6.8 &   4.5 &   3.7 &   3.4 \\
IC  4290-2 &  0.33 &  0.27 &  0.26 &  0.16 &  22.3 &  24.2 &  22.2 &
22.7 &   4.8 &   3.1 &   4.1 &   3.7 \\
IC  5092   &  0.62 &  0.34 &  0.20 &  0.11 &  16.5 &  17.6 &  17.8 &
18.1 &   8.0 &   5.9 &   5.6 &   6.0 \\
UGC 10862  &  0.72 &  0.44 &  0.27 &  0.17 &  12.9 &  13.2 &  14.0 &
14.4 &   5.5 &   3.5 &   3.2 &   3.0 \\
\enddata
\tablenotetext{a}{Col. (1) Galaxy name. If a double gaussian was fitted to the $I_m/I_0$
profiles, the first gaussian is listed as "-1" while the second is "-2". (2-5): gaussian relative amplitudes
$A_{mi}$ for $m$ = 2, 4, 6, and 8. The index $i$=1 for a single gaussian
fit, and 1 and 2 for a double gaussian fit. For a double gaussian fit, 
$A_{m1}$ is listed on the first line and
$A_{m2}$ is listed on the second line for a given galaxy. (6-9): mean radii $r_{mi}$
in arcseconds. (10-13): gaussian width $\sigma_{mi}$ in arcseconds.}
\end{deluxetable}

%% file: t06.tex
\makeatletter
\def\jnl@aj{AJ}
\ifx\revtex@jnl\jnl@aj\let\tablebreak=\nl\fi
\makeatother
 
\begin{deluxetable}{lrrrrrr}
\tablenum{6}
\tablewidth{40pc}
\tablecaption{Bar Radii\tablenotemark{a}}
\tablehead{
\colhead{Galaxy} &
\colhead{$r(0.4A_{2b})$} &
\colhead{$r(0.25A_{2b})$} &
\colhead{$r(0.4A_{4b})$} &
\colhead{$r(0.25A_{4b})$} &
\colhead{$r(vis)$} &
\colhead{$r(0.25A_{4b})$}
\\
\colhead{} &
\colhead{($^{\prime\prime}$)} &
\colhead{($^{\prime\prime}$)} &
\colhead{($^{\prime\prime}$)} &
\colhead{($^{\prime\prime}$)} &
\colhead{($^{\prime\prime}$)} &
\colhead{(kpc)} 
\\
\colhead{1} &
\colhead{2} &
\colhead{3} &
\colhead{4} &
\colhead{5} & 
\colhead{6} &
\colhead{7} 
}
\startdata
NGC 175 &       20.5 &       22.4 &       21.9 &       23.4 &       22.9 &     6.1 \\
NGC 521 &       20.6 &       22.4 &       19.4 &       20.8 &       20.4 &     7.0 \\
NGC 613 &       79.1 &       81.7 &       75.1 &       78.7 &       79.4 &     7.6 \\
NGC 986 &       53.9 &       57.6 &       51.3 &       54.6 &       59.3 &     6.8 \\
NGC 1300 &       89.5 &       95.8 &       79.3 &       84.3 &       83.3 &     8.4 \\
NGC 1566 &       40.2 &       41.3 &       39.6 &       42.6 &       40.2 &     3.8 \\
NGC 4593 &       85.6 &       90.5 &       70.2 &       74.2 &       78.6 &    12.8 \\
NGC 5101 &       60.8 &       63.3 &       58.0 &       60.6 &       68.4 &     7.0 \\
NGC 5335 &       25.8 &       27.1 &       24.8 &       25.9 &       27.1 &     7.9 \\
NGC 5365 &       52.1 &       55.7 &       51.5 &       54.6 &       55.3 &     8.4 \\
NGC 6221 &       42.3 &       44.6 &       39.8 &       42.1 &       40.6 &     3.9 \\
NGC 6384 &       27.3 &       29.4 &       26.0 &       27.3 &       25.6 &     3.2 \\
NGC 6782 &       38.0 &       40.6 &       36.2 &       38.2 &       44.4 &     9.7 \\
NGC 6907 &       32.5 &       34.9 &       32.3 &       34.7 &       33.9 &     7.5 \\
NGC 7155 &       32.0 &       34.5 &       30.8 &       32.8 &       34.3 &     4.2 \\
NGC 7329 &       39.3 &       42.7 &       35.7 &       37.9 &       37.9 &     7.9 \\
NGC 7513 &       47.6 &       51.4 &       45.7 &       48.5 &       45.2 &     5.1 \\
NGC 7552 &       77.1 &       81.0 &       74.4 &       79.7 &       68.8 &     8.4 \\
NGC 7582 &       95.3 &      101.9 &       87.8 &       92.6 &       90.9 &     9.6 \\
IC 1438  &       29.7 &       31.5 &       28.4 &       29.9 &       29.3 &     5.3 \\
IC 4290  &       26.4 &       28.3 &       26.7 &       28.0 &       28.7 &     8.7 \\
IC 5092  &       27.3 &       29.8 &       25.5 &       27.3 &       26.6 &     5.7 \\
UGC 10862 &       20.3 &       22.0 &       18.0 &       19.1 &       21.9 &     2.3 \\
\enddata
\tablenotetext{a}{Col. 1: galaxy name; 
2: radius at which the bar $m$=2 amplitude is 0.4$A_{2b}$;
3: radius at which the bar $m$=2 amplitude is 0.25$A_{2b}$;
4: radius at which the bar $m$=4 amplitude is 0.4$A_{4b}$;
5: radius at which the bar $m$=4 amplitude is 0.25$A_{4b}$;
6: visual bar radius (arcsec);
7: col. 5 bar radius (kpc)}
\end{deluxetable}

%% file: t07.tex
\makeatletter
\def\jnl@aj{AJ}
\ifx\revtex@jnl\jnl@aj\let\tablebreak=\nl\fi
\makeatother
 
\begin{deluxetable}{lcccccccccc}
\tablenum{7}
\tablewidth{40pc}
\tablecaption{Maximum Relative Gravitational Torques and Spiral Contrasts\tablenotemark{a}}
\tablehead{
\colhead{Galaxy} &
\colhead{$Q_g$} &
\colhead{$Q_b$} &
\colhead{$Q_s$} &
\colhead{$A_{2s}$} &
\colhead{$A_{4s}$} &
\colhead{$r_g$} &
\colhead{$r_b$} &
\colhead{$r_s$} &
\colhead{$r_{2s}$} &
\colhead{$r_{4s}$} 
\\
\colhead{1} &
\colhead{2} &
\colhead{3} &
\colhead{4} &
\colhead{5} &
\colhead{6} &
\colhead{7} & 
\colhead{8} & 
\colhead{9} & 
\colhead{10} & 
\colhead{11}  
}
\startdata
NGC   175 & 0.475 & 0.436 & 0.210 & 0.422 & 0.249 &  16.0 &  15.5 &  26.0 &  26.5 &  28.5 \\
          & \llap{$\pm$}0.073 & \llap{$\pm$}0.064 & \llap{$\pm$}0.040 & \llap{$\pm$}0.050 & \llap{$\pm$}0.010 &       &       &       &       &       \\
NGC   521 & 0.176 & 0.175 & 0.067 & 0.218 & 0.149 &  13.0 &  13.5 &  54.0 &  57.5 &  66.5 \\
          & \llap{$\pm$}0.025 & \llap{$\pm$}0.026 & \llap{$\pm$}0.013 & \llap{$\pm$}0.020 & \llap{$\pm$}0.020 &       &       &       &       &       \\
NGC   613 & 0.483 & 0.395 & 0.346 & 0.780 & 0.449 &  63.0 &  48.0 &  75.0 &  88.5 &  81.5 \\
          & \llap{$\pm$}0.096 & \llap{$\pm$}0.064 & \llap{$\pm$}0.089 & \llap{$\pm$}0.078 & \llap{$\pm$}0.045 &       &       &       &       &       \\
NGC   986 & 0.586 & 0.436 & 0.497 & 1.220 & 0.733 &  46.5 &  35.0 &  63.0 &  62.5 &  63.5 \\
          & \llap{$\pm$}0.106 & \llap{$\pm$}0.064 & \llap{$\pm$}0.132 & \llap{$\pm$}0.061 & \llap{$\pm$}0.037 &       &       &       &       &       \\
NGC  1300 & 0.573 & 0.502 & 0.232 & 0.789 & 0.346 &  60.5 &  59.0 & 123.5 & 117.5 & 118.5 \\
          & \llap{$\pm$}0.090 & \llap{$\pm$}0.074 & \llap{$\pm$}0.063 & \llap{$\pm$}0.039 & \llap{$\pm$}0.017 &       &       &       &       &       \\
NGC  1566 & 0.234 & 0.153 & 0.230 & 0.611 & 0.316 &  75.0 &  27.0 &  74.5 &  68.0 &  66.5 \\
          & \llap{$\pm$}0.041 & \llap{$\pm$}0.022 & \llap{$\pm$}0.046 & \llap{$\pm$}0.031 & \llap{$\pm$}0.016 &       &       &       &       &       \\
NGC  4593 & 0.343 & 0.341 & 0.064 & 0.332 & 0.200 &  45.5 &  45.5 &  60.0 &  98.5 &  75.5 \\
          & \llap{$\pm$}0.055 & \llap{$\pm$}0.051 & \llap{$\pm$}0.024 & \llap{$\pm$}0.088 & \llap{$\pm$}0.059 &       &       &       &       &       \\
NGC  5101 & 0.233 & 0.227 & 0.043 & 0.216 & 0.066 &  40.0 &  40.0 & 100.0 &  84.5 &  59.5 \\
          & \llap{$\pm$}0.033 & \llap{$\pm$}0.033 & \llap{$\pm$}0.026 & \llap{$\pm$}0.011 & \llap{$\pm$}0.017 &       &       &       &       &       \\
NGC  5335 & 0.425 & 0.422 & 0.049 & 0.107 & 0.136 &  20.0 &  20.0 &  46.5 &  47.5 &  47.5 \\
          & \llap{$\pm$}0.061 & \llap{$\pm$}0.062 & \llap{$\pm$}0.010 & \llap{$\pm$}0.005 & \llap{$\pm$}0.007 &       &       &       &       &       \\
NGC  5365 & 0.102 & 0.106 & 0.013 & 0.063 & 0.018 &  38.0 &  37.0 &  43.0 &  67.5 &  65.5 \\
          & \llap{$\pm$}0.015 & \llap{$\pm$}0.016 & \llap{$\pm$}0.005 & \llap{$\pm$}0.015 & \llap{$\pm$}0.018 &       &       &       &       &       \\
NGC  6221 & 0.421 & 0.400 & 0.261 & 0.455 & 0.197 &  28.0 &  28.0 &  48.0 &  49.5 &  66.5 \\
          & \llap{$\pm$}0.076 & \llap{$\pm$}0.059 & \llap{$\pm$}0.080 & \llap{$\pm$}0.046 & \llap{$\pm$}0.020 &       &       &       &       &       \\
NGC  6384 & 0.112 & 0.127 & 0.106 & 0.255 & 0.106 &  18.5 &  19.0 &  50.0 &  53.5 &  79.5 \\
          & \llap{$\pm$}0.019 & \llap{$\pm$}0.020 & \llap{$\pm$}0.019 & \llap{$\pm$}0.038 & \llap{$\pm$}0.016 &       &       &       &       &       \\
NGC  6782 & 0.183 & 0.180 & 0.037 & 0.283 & 0.112 &  23.0 &  23.0 &  57.0 &  42.5 &  38.5 \\
          & \llap{$\pm$}0.026 & \llap{$\pm$}0.027 & \llap{$\pm$}0.009 & \llap{$\pm$}0.119 & \llap{$\pm$}0.068 &       &       &       &       &       \\
NGC  6907 & 0.384 & 0.272 & 0.349 & 0.714 & 0.371 &  27.0 &  24.0 &  44.5 &  37.5 &  32.5 \\
          & \llap{$\pm$}0.084 & \llap{$\pm$}0.042 & \llap{$\pm$}0.085 & \llap{$\pm$}0.070 & \llap{$\pm$}0.029 &       &       &       &       &       \\
NGC  7155 & 0.185 & 0.185 & 0.031 & 0.333 & 0.089 &  22.0 &  22.0 &  61.0 &  67.5 &  54.5 \\
          & \llap{$\pm$}0.027 & \llap{$\pm$}0.027 & \llap{$\pm$}0.009 & \llap{$\pm$}0.017 & \llap{$\pm$}0.004 &       &       &       &       &       \\
NGC  7329 & 0.322 & 0.313 & 0.153 & 0.486 & 0.109 &  27.0 &  27.0 &  55.5 &  84.5 &  72.5 \\
          & \llap{$\pm$}0.049 & \llap{$\pm$}0.048 & \llap{$\pm$}0.035 & \llap{$\pm$}0.073 & \llap{$\pm$}0.011 &       &       &       &       &       \\
NGC  7513 & 0.660 & 0.666 & 0.114 & 0.132 & 0.214 &  27.0 &  27.0 &  43.5 &  63.5 &  75.5 \\
          & \llap{$\pm$}0.117 & \llap{$\pm$}0.115 & \llap{$\pm$}0.023 & \llap{$\pm$}0.017 & \llap{$\pm$}0.011 &       &       &       &       &       \\
NGC  7552 & 0.409 & 0.401 & 0.100 & 0.406 & 0.309 &  41.5 &  41.0 &  62.0 &  85.5 &  73.5 \\
          & \llap{$\pm$}0.065 & \llap{$\pm$}0.062 & \llap{$\pm$}0.037 & \llap{$\pm$}0.100 & \llap{$\pm$}0.100 &       &       &       &       &       \\
NGC  7582 & 0.427 & 0.417 & 0.110 & 0.291 & 0.157 &  57.0 &  56.0 &  69.0 & 108.0 &  83.5 \\
          & \llap{$\pm$}0.069 & \llap{$\pm$}0.064 & \llap{$\pm$}0.036 & \llap{$\pm$}0.100 & \llap{$\pm$}0.050 &       &       &       &       &       \\
IC  1438 & 0.125 & 0.124 & 0.052 & 0.285 & 0.156 &  19.5 &  19.5 &  74.0 &  62.5 &  64.5 \\
          & \llap{$\pm$}0.018 & \llap{$\pm$}0.018 & \llap{$\pm$}0.014 & \llap{$\pm$}0.014 & \llap{$\pm$}0.008 &       &       &       &       &       \\
IC  4290 & 0.532 & 0.522 & 0.079 & 0.635 & 0.309 &  20.0 &  20.0 &  50.0 &  51.5 &  56.5 \\
          & \llap{$\pm$}0.077 & \llap{$\pm$}0.077 & \llap{$\pm$}0.018 & \llap{$\pm$}0.032 & \llap{$\pm$}0.015 &       &       &       &       &       \\
IC  5092 & 0.508 & 0.506 & 0.151 & 0.324 & 0.111 &  17.0 &  17.0 &  27.5 &  32.5 &  29.5 \\
          & \llap{$\pm$}0.086 & \llap{$\pm$}0.075 & \llap{$\pm$}0.031 & \llap{$\pm$}0.039 & \llap{$\pm$}0.036 &       &       &       &       &       \\
UGC  10862 & 0.826 & 0.796 & 0.152 & 0.279 & 0.145 &  12.0 &  12.0 &  17.0 &  23.5 &  20.5 \\
          & \llap{$\pm$}0.117 & \llap{$\pm$}0.117 & \llap{$\pm$}0.030 & \llap{$\pm$}0.098 & \llap{$\pm$}0.052 &       &       &       &       &       \\
\enddata
\tablenotetext{a}{Col. 1: galaxy name; 
2: total nonaxisymmetric
maximum relative torque; 3: bar strength; 4: spiral strength; 5: spiral maximum $m$=2 contrast; 
6: spiral maximum $m$=4 contrast; 7-11: radii of $Q_g$, $Q_b$, $Q_s$, $A_{2s}$, and $A_{4s}$
maxima in arcseconds.}
\end{deluxetable}

%% file: t08.tex
\makeatletter
\def\jnl@aj{AJ}
\ifx\revtex@jnl\jnl@aj\let\tablebreak=\nl\fi
\makeatother
 
\begin{deluxetable}{lcc}
\tablenum{8}
\tablewidth{0pc}
\tablecaption{Mean Parameters for Sample\tablenotemark{a}}
\tablehead{
\colhead{Parameter}&
\colhead{SG} &
\colhead{DG} 
\\
\colhead{1} &
\colhead{2} &
\colhead{3} 
} 
\startdata
$<A_{2b}>$         &  0.59$\pm$0.07  & 0.82$\pm$0.05 \\
$<A_{4b}>$         &  0.34$\pm$0.05  & 0.46$\pm$0.04 \\
$<Q_g>$         &  0.34$\pm$0.09  & 0.34$\pm$0.05 \\
$<Q_b>$         &  0.32$\pm$0.08  & 0.32$\pm$0.05 \\
$<r_{b}/h_r>$ &  1.17$\pm$0.14  & 1.56$\pm$0.17 \\
$<B/T>$         &  0.14$\pm$0.05  & 0.26$\pm$0.04 \\
$<n>$           &  1.71$\pm$0.37  & 1.92$\pm$0.24 \\
No. of galaxies &       7         &     11        \\
\enddata
\tablenotetext{a}{Col. (1) Parameter; 
(2) mean values for single gaussian bar Fourier profile galaxies; 
(3) mean values for double gaussian bar Fourier profile galaxies}
\end{deluxetable}